\documentclass[aps,prx,twocolumn,superscriptaddress,nobalancelastpage]{revtex4-2}
\usepackage{preamble}

\begin{document}
\title{Force-current structure in Markovian open quantum systems and its applications: geometric housekeeping-excess decomposition and thermodynamic trade-off relations}
\author{Kohei Yoshimura}
\email{kyoshimura@ubi.s.u-tokyo.ac.jp}
\affiliation{Department of Physics, The University of Tokyo, 7-3-1 Hongo, Bunkyo-ku, Tokyo 113-0033, Japan}
\author{Yoh Maekawa}
\affiliation{Department of Physics, The University of Tokyo, 7-3-1 Hongo, Bunkyo-ku, Tokyo 113-0033, Japan}
\author{Ryuna Nagayama}
\affiliation{Department of Physics, The University of Tokyo, 7-3-1 Hongo, Bunkyo-ku, Tokyo 113-0033, Japan}
\author{Sosuke Ito}
\affiliation{Department of Physics, The University of Tokyo, 7-3-1 Hongo, Bunkyo-ku, Tokyo 113-0033, Japan}
\affiliation{Universal Biology Institute, The University of Tokyo, 7-3-1 Hongo, Bunkyo-ku, Tokyo 113-0033, Japan}
\date{\today}

\begin{abstract}
    Thermodynamic force and irreversible current are the foundational concepts of classical nonequilibrium thermodynamics. 
    Entropy production rate is provided by their product in classical systems, ranging from mesoscopic to macroscopic systems. 
    However, there is no complete quantum extension of such a structure that respects quantum mechanics. 
    In this paper, we propose anti-Hermitian operators that represent currents and forces accompanied by a gradient structure in open quantum systems described by the quantum master equation.
    We prove that the entropy production rate is given by the product of the force and current operators, which extends the canonical expression of the entropy production rate in the classical systems. 
    The framework constitutes a comprehensive analogy with the nonequilibrium thermodynamics of discrete classical systems. 
    We also show that the structure leads to the extensions of some results in stochastic thermodynamics: the geometric housekeeping-excess decomposition of entropy production and thermodynamic trade-off relations such as the thermodynamic uncertainty relation and the dissipation-time uncertainty relation. 
    In discussing the trade-off relations, we will introduce a measure of fluctuation, which we term the quantum diffusivity. 
\end{abstract}

\maketitle

\section{Introduction}

The thermodynamic theory of nonequilibrium systems has been explored since the last century, both in classical and quantum systems~\cite{de1962non,breuer2002theory,strasberg2022quantum}. 
The significant development of stochastic thermodynamics in this century~\cite{seifert2012stochastic,shiraishi2023introduction} has been pushing forward our understanding of not only classical but also quantum systems~\cite{campisi2011colloquium,landi2021irreversible,strasberg2022quantum}. 
The second law of thermodynamics is believed not to be violated even in quantum mechanics, and extensions of modern thermodynamic laws unveiled by stochastic thermodynamics, such as the fluctuation theorem~\cite{jarzynski1997nonequilibrium} or the thermodynamic uncertainty relations~\cite{barato2015thermodynamic}, to quantum systems have been actively studied~\cite{campisi2011colloquium,hasegawa2020quantum,hasegawa2021thermodynamic}.

However, there is a structure that has been underexplored in quantum thermodynamics, i.e., the structure of irreversible currents and thermodynamic forces~\cite{onsager1931reciprocal1,onsager1931reciprocal2,de1962non,schnakenberg1976network}. 
In classical systems, from mesoscopic stochastic systems to macroscopic nonlinear systems, it is known that the entropy production rate $\dot{\Sigma}$ is provided as the product between currents $J_\alpha$ and forces $F_\alpha$ as
\begin{align}
    \dot{\Sigma}=\sum_\alpha J_\alpha F_\alpha. \label{eq:JF}
\end{align}
Although this is regarded as a fundamental law of classical nonequilibrium thermodynamics, there are few studies on quantum extensions. 
One important methodology was established in Ref.~\cite{esposito2006fluctuation} and similar results have repeatedly been presented in the literature~\cite{horowitz2012quantum,horowitz2013entropy,funo2019speed}.
While their approach captures an aspect of the profound connection between quantum and classical nonequilibrium thermodynamics, it still has some problems. First, since the expression relies on a specific basis, it lacks not only formal elegance but also theoretical transparency. 
We may be unable to identify what is quantum and what is not. 
In addition, as discussed in the main text, the formulation can encounter a problem when defining conservative and nonconservative forces. 
This is crucial because, in short, this decomposition of forces corresponds to the separation between equilibrium and nonequilibrium systems. 

\begin{figure}
    \centering
    \includegraphics[width=\linewidth]{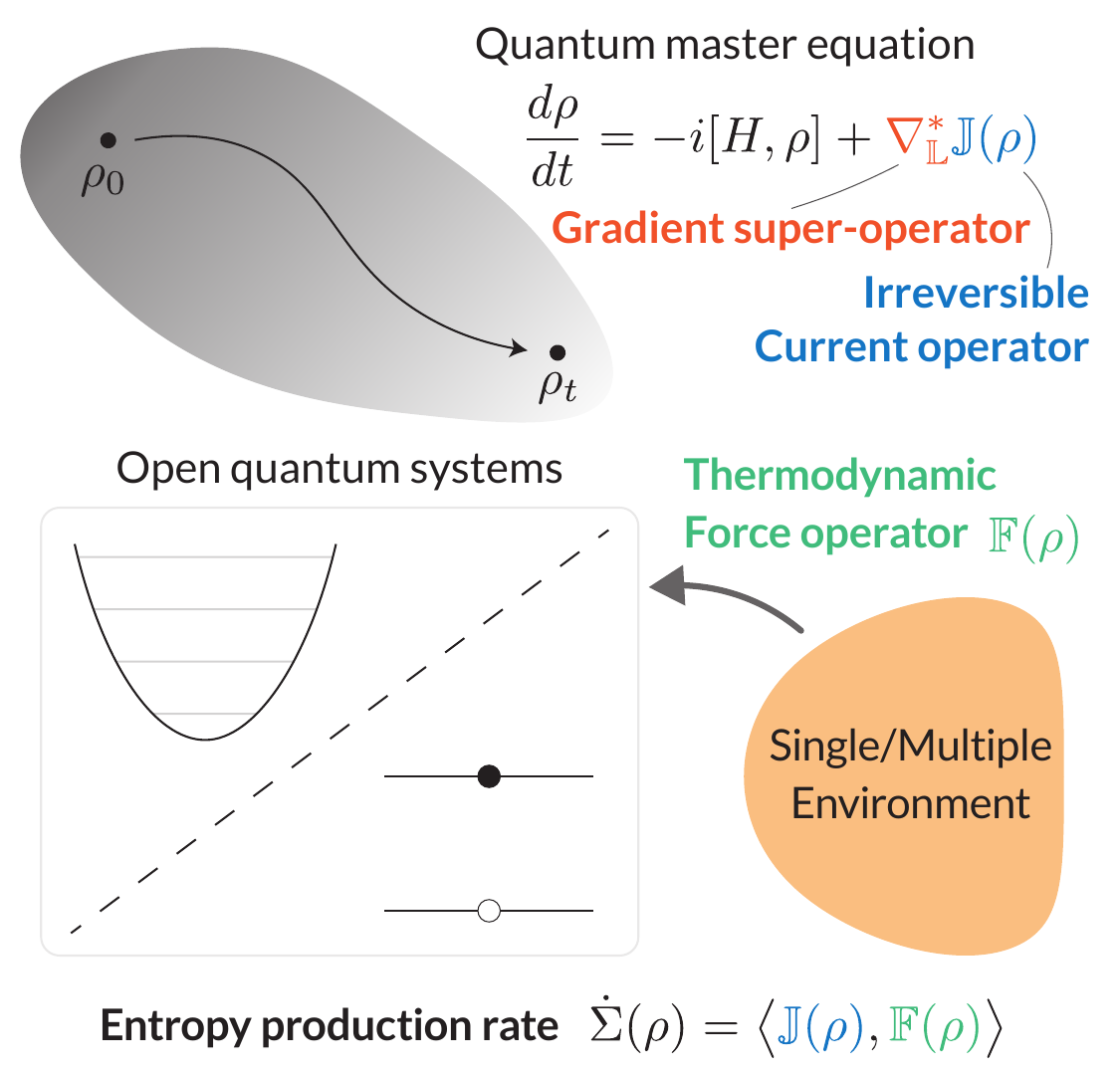}
    \caption{Summary of the fundamental results. We consider open quantum systems described by the quantum master equation. We introduce the (irreversible) current operator $\curr(\rho)$ and the gradient super-operator $\nabla_\strc$, which rewrite the dissipator into the form of the continuity equation [Eq.~\eqref{eq:eq7}]. We also define the (thermodynamic) force operator $\force(\rho)$, representing the thermodynamic force exerted on the system. The gradient super-operator distinguishes whether the force is conservative or non-conservative, i.e., whether the system is detailed balanced [Sec.~\ref{subsec:connect}]. The current and force operators jointly provide the entropy production rate as $\epr(\rho)=\langle\curr(\rho),\force(\rho)\rangle$, which is the quantum extension of a formula widely found in the classical nonequilibrium systems. }
    \label{fig:concept}
\end{figure}

In this paper, we provide a force-current structure for quantum thermodynamics of Markovian open quantum systems described by the quantum master equation (also known as the Gorini--Kossakowski--Sudarshan--Lindblad equation)~\cite{gorini1976completely,lindblad1976generators}, 
defining force and current \textit{operators} so that we can treat them as quantum objects (see Fig.~\ref{fig:concept}). 
We also consider an accompanying gradient structure by introducing a gradient super-operator, which enables us to discuss the relationship between dynamics and thermodynamics clearly. 
Consequently, we obtain a comprehensive correspondence between stochastic and quantum thermodynamics, as summarized in Table.~\ref{tab:QuantumClassical}. 

Further, we apply the developed framework to extend results obtained for classical systems in recent years.
We first derive the geometric housekeeping-excess decomposition of the entropy production rate~\cite{oono1998steady,hatano2001steady,esposito2010three,spinney2012nonequilibrium,spinney2012entropy,ge2016nonequilibrium,rao2016nonequilibrium,maes2014nonequilibrium,dechant2022geometric1,dechant2022geometric2,yoshimura2023housekeeping,kolchinsky2022information,kolchinsky2024generalized,nagayama2023geometric,yoshimura2024two,ito2023geometric,kobayashi2022hessian,horowitz2014equivalent,manzano2018quantum}. 
The housekeeping-excess decomposition~\cite{oono1998steady} splits the total dissipation into one part to keep the system out of equilibrium (the housekeeping dissipation) and the other stemming from the system's dynamics (the excess dissipation).
As it can tighten thermodynamic inequalities such as the second law, it enables us to obtain a more precise understanding of irreversibility and fundamental limitations in the dynamics~\cite{oono1998steady,hatano2001steady,maes2014nonequilibrium,shiraishi2018speed,tuan2020unified,dechant2022geometric1,dechant2022geometric2,yoshimura2023housekeeping,kolchinsky2022information,kolchinsky2024generalized,ito2023geometric,nagayama2023geometric,yoshimura2024two}. 
The geometric approach~\cite{ito2023geometric}, which we adopt, does not require any particular steady state, unlike the conventional one~\cite{horowitz2014equivalent,manzano2018quantum}, so it provides a more general way of decomposition. 

In addition, we show that the framework reveals thermodynamic trade-off relations in quantum dynamics. 
Thermodynamic trade-off relations such as the thermodynamic uncertainty relations~\cite{barato2015thermodynamic,pietzonka2016universal,dechant2018multidimensional,liu2020thermodynamic,horowitz2020thermodynamic,yoshimura2021thermodynamic,hasegawa2020quantum,van2023thermodynamic} or the thermodynamic speed limits~\cite{shiraishi2018speed,tuan2020unified,van2021geometrical,dechant2022minimum,dechant2022geometric1,dechant2022geometric2,yoshimura2023housekeeping,kolchinsky2022information,kolchinsky2024generalized,van2023thermodynamic}, involving the entropy production, provide the dynamics with fundamental limitations tighter than the second law of thermodynamics. 
We derive trade-off relations between the entropy production rate and fluctuations or a typical time scale from the force-current structure. 
The former one is the quantum extension of the short-time thermodynamic uncertainty relation~\cite{otsubo2020estimating}. We will explain that this generic inequality leads to some specific results reported recently~\cite{funo2019speed,tajima2021superconducting}. 
The latter is similar to the dissipation-time uncertainty relation~\cite{falasco2020dissipation} and the time-energy uncertainty relation~\cite{mandelstam1945uncertainty}. 
In discussing these trade-offs, we will introduce a measure of fluctuation, which we term the quantum diffusivity. 
We show that the quantum diffusivity coincides with a classical counterpart when the observable is classical. 

Finally, we demonstrate the results numerically and analytically with several examples: a two-level system attached to thermal baths at different temperatures, a toy model of superradiance, and the damped harmonic oscillator. 
We numerically integrate a two-level system and show how the housekeeping-excess decomposition works. We also provide an approximation formula for the excess dissipation. 
The superradiance model will be used to check the thermodynamic uncertainty relation; we will discuss the association between the trade-off and coherence through nonuniform relaxation. 
We will also compute the quantum diffusivity analytically for the damped harmonic oscillator, which has an infinite dimension.

\section{Preliminary}
\label{sec:Preliminary}

\subsection{Open quantum systems}
\label{sec:qme}
We consider an open quantum system described by a Hilbert space $\hilb$. 
Let $\opr(\hilb)$ denote the linear operators on $\hilb$, and $\herm(\hilb)$ and $\anti(\hilb)$ the subsets of Hermitian and anti-Hermitian operators in $\opr(\hilb)$. 
The density operator $\rho\in\herm(\hilb)$ ($\rho>0$, $\tr(\rho)=1$) is assumed to obey the quantum master equation (also known as the Gorini--Kossakowski--Sudarshan--Lindblad equation)~\cite{gorini1976completely,lindblad1976generators}
\begin{align}
    &\frac{d\rho}{dt}
    =-i[H,\rho]+\mathcal{D}(\rho),\\
    &\mathcal{D}(\rho)\coloneqq\sum_{k\in K_{\mathrm{all}}}\gamma_k\left(L_k\rho L_k^\dagger-\frac{1}{2}\{L_k^\dagger L_k,\rho\}\right). \label{eq:qme}
\end{align}
Here, $H\in\herm(\hilb)$ is the Hamiltonian, the reduced Planck constant $\hbar$ is set to one, $K_{\mathrm{all}}$ is the set of jumps, and $\{\cdot,\cdot\}$ is the anti-commutator. 
Each $k\in K_{\mathrm{all}}$ has a positive rate $\gamma_k>0$ and a jump operator $L_k\in\opr(\hilb)$, which represents state changes. 
We assume that every $k\in K_{\mathrm{all}}$ has a unique counterpart $-k\in K_{\mathrm{all}}$ such that $L_{-k}=L_k^\dagger$. 
Let $K$ be the set of indexes that only includes either of each pair. 
On the other hand, we do not assume the uniqueness of the steady state of the quantum master equation, and the system can have nontrivial conservative quantities.
For later convenience, we define for each $k\in K$
\begin{equation}
    \begin{split}
        \mathcal{D}_k(\rho)
        &\coloneqq\gamma_k\left(L_k\rho L_k^\dagger-\frac{1}{2}\{L_k^\dagger L_k,\rho\}\right)\\
        &\phantom{\coloneqq}+\gamma_{-k}\left(L_k^\dagger\rho L_k-\frac{1}{2}\{L_k L_k^\dagger,\rho\}\right). 
    \end{split}
\end{equation}
Thus, we have $\mathcal{D}(\rho)=\sum_{k\in K}\mathcal{D}_k(\rho)$. 

Thermodynamics is introduced to the model by imposing the local detailed balance condition $\ln(\gamma_k/\gamma_{-k})=s_k$ for each pair $(k,-k)$, where $s_k$ is the entropy flux into the environment ($s_{-k}=-s_k$)~\cite{horowitz2013entropy}. 
The entropy production rate (EPR) of the total system  (system + environment) is given as~\cite{strasberg2022quantum}
\begin{align}
    \epr(\rho)=-k_{\mathrm{B}}\tr\big(\mathcal{D}(\rho)\ln\rho\big)+\sum_{k\in K_{\mathrm{all}}} s_k\gamma_k\tr(L_k^\dagger L_k\rho). \label{eq:EPRdef}
\end{align}
The first term gives the von Neumann entropy change $-(d/dt)k_\mathrm{B}\tr(\rho\ln\rho)$, and the second provides the entropy change in the environment. 
As shown later, the EPR is always non-negative owing to the local detailed balance condition.
Hereafter, we set the Boltzmann constant $k_\mathrm{B}$ to one for simplicity. 

Especially when process $k$ is mediated by a thermal bath at inverse temperature $\beta_k$, we may assume the thermodynamic consistency conditions~\cite{horowitz2013entropy}
\begin{align}
    s_k = \beta_k\omega_k,\quad [L_k, H] = \omega_k L_k, \label{eq:TC}
\end{align}
where $\omega_k$ is the energy gap corresponding to $k$ and $\beta_k\omega_k$ provides the entropy change $s_k$ in the bath. 
In this case, the second term in the right-hand side of Eq.~\eqref{eq:EPRdef} is given by $-\sum_k\beta_k\tr(H\mathcal{D}_k(\rho))$ and $-\tr(H\mathcal{D}_k(\rho))$ is interpreted as the heat flux into bath $k$~\cite{strasberg2022quantum}.
In addition, if $\beta_k=\beta$ for every $k$ with a single parameter $\beta$ holds, the system can have an equilibrium state $\rho^{\mathrm{eq}}=e^{-\beta H}/Z$ with $Z=\tr(e^{-\beta H})$, where the EPR vanishes. 
We call systems satisfying the conditions in Eq.~\eqref{eq:TC} with $\beta_k = \beta$ \textit{detailed balanced}. 

Throughout the paper, we always assume the Hamiltonian, the jump operators, and the rates are time-independent; namely, we only consider autonomous systems. However, this does not restrict the generality of the results, and they can easily be extended to non-autonomous systems because we only deal with quantities defined at each time, and time derivatives are not generally involved.

We should emphasize the fact that $\gamma_k$ and $L_k$ in Eq.~\eqref{eq:qme} are not uniquely separable because the quantum master equation is invariant under the transformation $(\gamma_k, L_k)\to (\gamma_k/|c_k|^2, c_kL_k)$ with any $c_k\in \mathbb{C}$ such that $c_k\neq 0$ and $c_{-k}=c_k^*$, where the star indicates the complex conjugate.
For example, when considering a two-level system, $L_k$ can be chosen as $\sigma_-$ or $\sigma_-/2$. 
Although this indefiniteness can be inherited by operators that we introduce later, it will be seen that it does not cause any crucial issues. 

\subsection{Classical expression of the quantum master equation}
The quantum master equation can be reformed into a classical master equation~\cite{funo2019speed}. 
We assume the spectral decomposition $\rho=\sum_i p_i\ketbras{i}$. 
Then, the diagonal elements obey the classical master equation
\begin{align}
    \frac{dp_i}{dt}=\sum_{j,k\in K_{\mathrm{all}}}R_{ij}^k p_j,
\end{align}
where $R_{ij}^k = \gamma_k|\bra{i} L_k \ket{j}|^2$ for $i\neq j$ and $R_{ii}^k = -\sum_{j\neq i} R_{ij}^k$. 
This equation means the equivalence between the dynamics of diagonal elements and that of an occupation probability in the continuous-time Markov chain (Markov jump process) with transition rates $\{R_{ij}^k\}$. Label $k$ now indicates the ``route'' used in the transition from $j$ to $i$. 

Further, ``currents'' and ``forces'' are defined by $J_{ij}^k=R_{ji}^kp_i-R_{ij}^{-k}p_j$, $F_{ij}^k=\ln(R_{ji}^kp_i/(R_{ij}^{-k}p_j))$. 
As a consequence, we can prove 
\begin{align}
    \dot{\Sigma}(\rho)=\sum_{i,j}\sum_{k\in K} J_{ij}^k F_{ij}^k,
\end{align}
which is the standard form known in classical systems. 

These classical expressions have been utilized in exploring quantum thermodynamics (e.g.,~\cite{funo2019speed,van2021geometrical,van2023thermodynamic}); however, we can find some issues in this method. 
First, the transition rates change in time, even if $\gamma_k$ and $L_k$ do not because the eigenbasis $\ket{i}$ is time-dependent. 
Second, the current and force have nothing to do with any operator expression, so the connection to quantum mechanics is lost. 
The first point can be attributed to the Hamiltonian dynamics, which should not be directly associated with the dissipation because it is supposed to govern the reversible dynamics. 
In fact, our definition will never refer to the Hamiltonian. 
The second problem is also resolved in our formulation; we will introduce operators expressing currents and forces. 
We will provide a method to treat quantum thermodynamics as a quantum theory without relying on any basis. 
In addition to these problems, another one exists in the conventional definition of force and current, as discussed in the next section. 

\subsection{Brief review of classical systems}
\label{subsec:reviewCL}
Before considering the quantum extension of the force-current structure, reviewing some preliminary on classical systems would be better. 
More details and examples can be found in Ref.~\cite{yoshimura2023housekeeping}. 

Generally, nonequilibrium thermodynamics of classical discrete systems such as Markov jump processes (MJPs) and chemical reaction networks (CRNs) can be treated simultaneously in a single framework.
We consider a distribution $\bm{x} = (x_i)$, corresponding to the occupation probability in MJPs or the concentration distribution in CRNs. Label $i$ indicates microscopic states or chemical species then. 
The distribution changes as processes occur, such as state transitions or chemical reactions. The processes are labeled by $e\in E$, where $E$ is the set of processes. 
The influence of process $e$ is encoded in an integer matrix called the stoichiometric matrix ${\nabla}^\mathrm{T}$ whose $(i,e)$-element provides how $x_i$ increases/decreases in process $e$ (superscript $\mathrm{T}$ means transposition). 
For each $e$, we define current $J_e$ as the process's occurrence rate, which is probability current in MJPs and reaction rate in CRNs. 
Finally, the equation of motion (master equation/reaction rate equation) reads
\begin{align}
    \frac{d\bm{x}}{dt}=\nabla^\mathrm{T}\bm{J}(\bm{x}), \label{eq:clMaster}
\end{align}
where $\bm{J}$ is the vector of $J_e$'s and $\bm{J}(\bm{x})$ indicates the dependence of $\bm{J}$ on $\bm{x}$. 

We can equip the dynamics with a thermodynamic structure by assuming that the currents are split into forward and backward fluxes $J_e^+>0$ and $J_e^->0$, as $J_e=J_e^+-J_e^-$ and the local detailed balance $F_e = \ln(J_e^+/J_{e}^-)$. Here, $F_e$ is the thermodynamic force on process $e$ and provides the EPR $\dot{\Sigma} = \bm{J}^\mathrm{T}\bm{F} = \sum_{e\in E} J_eF_e$, where the EPR is in advance defined by the system's entropy change and environmental entropy increase~\cite{de1962non,seifert2012stochastic}. 
Let us examine these quantities in a little more depth. 

First, we can characterize the detailed balance in terms of the thermodynamic force. 
We define a force $\bm{F}$ to be conservative if it is provided as $-\nabla\bm{\phi}$ with a potential vector $\bm{\phi}$, which has the same dimension as $\bm{x}$. 
Then, if we assume the so-called mass-action form of $\bm{J}^\pm=(J_e^\pm)$, the following two statements can be proved equivalent: 1) $\bm{F}(\bm{x})$ is conservative 2) Eq.~\eqref{eq:clMaster} has an equilibrium solution $\bm{x}^\mathrm{eq}$, where the detailed balance $\bm{J}(\bm{x}^\mathrm{eq})=\bm{0}$ holds~\cite{schuster1989generalization}. 
The mass action form of fluxes is widely accepted in chemical thermodynamics~\cite{kondepudi2014modern} and holds automatically in MJPs (for details, see Appendix~\ref{app:mak}). 
Because the gradient operator usually has nontrivial null space, potential $\bm{\phi}(\bm{x})$ such that $\bm{F}(\bm{x})=-\nabla\bm{\phi}(\bm{x})$ is not unique. Nonetheless, we can provide one of them by $\bm{\phi}(\bm{x})=\ln\bm{x}-\ln\bm{x}^\mathrm{eq}$ with an equilibrium solution $\bm{x}^\mathrm{eq}$ (where the log of a vector, e.g. $\ln\bm{x}$, denotes the vector of logs, like $(\ln x_i)$). 
Therefore, asking whether or not the force is conservative is equivalent to asking whether or not the system is detailed balanced in a broad class of systems. 

Second, the EPR has an additional geometric interpretation, whose utility will be shown in later sections. 
We define a diagonal matrix $\mathsf{M}=(M_e\delta_{ee'})$ by $M_{e}=(J_{e}^+-J_{e}^-)/\ln(J_e^+/J_e^-)$ ($\delta_{ee'}$ is the Kronecker delta), which connects the force and current as $\bm{J}=\mathsf{M}\bm{F}$. 
The diagonal elements of $\mathsf{M}$ are the so-called logarithmic means between $\bm{J}^+$ and $\bm{J}^-$~\cite{carlson1972logarithmic}. Since the log mean is always positive as long as the two numbers are positive, $\mathsf{M}$ is a positive definite matrix. 
Therefore, the EPR can be seen as the squared norm of $\bm{F}$ with $\mathsf{M}$ being the metric, $\dot{\Sigma}=\bm{F}^\mathrm{T}\mathsf{M}\bm{F}$. 
Consequently, we can decompose the EPR into the housekeeping and excess parts in general by projecting the force onto the conservative subspace orthogonally regarding this metric (if interested, see also~\cite{dechant2022geometric1,dechant2022geometric2,kobayashi2022hessian,kolchinsky2022information,kolchinsky2024generalized,yoshimura2024two}). 

\bigskip

Even though the classical representation of the quantum master equation in the previous section fits in this framework from the dynamical point of view, it has a crucial problem regarding thermodynamics. 
Now, we could regard the eigenvalues of $\rho$ as the distribution. The processes are labeled by $e=(i,j,k)$ and currents read $J_e = R_{ij}^k p_j - R_{ji}^{-k} p_i$. 
Since $e=(i,j,k)$ corresponds to a jump from $j$ to $i$, $[\nabla^{\mathrm{T}}]_{ie}=1$, $[\nabla^{\mathrm{T}}]_{je}=-1$ and $0$ otherwise. 
However, when reflecting on the thermodynamic force, we encounter an issue; in the light of the classical case, we expect that the force is given by a potential as $F_e=-\sum_i[\nabla]_{ei}\phi_i$ if there is an equilibrium state $\rhoeq$, and that the potential determines the equilibrium state. 
However, $\rho$ and $\rhoeq$ do not generally commute, so how the potential vector and the equilibrium state are connected is unclear. 
Then, we wonder if there is a potential \textit{operator} that defines conservative forces and has a natural connection to $\rhoeq$. 
In the following, we answer this question affirmatively.

\section{Reformulating quantum thermodynamics}
\label{sec:CFstrcture}

\subsection{Current operator and force operator}
First of all, we introduce an auxiliary Hilbert space $\hilbb\coloneqq\mathbb{C}^{2|K|}$ and operators $\strc\in\herm(\hilbb\otimes\hilb)$ and $\Gamma\in\herm(\hilbb)$ as
\begin{align}
    \strc\coloneqq\bigoplus_{k\in K}\begin{pmatrix}
        0&L_k^\dagger\\L_k&0
    \end{pmatrix},\quad
    \Gamma\coloneqq\bigoplus_{k\in K}\begin{pmatrix}
        \gamma_k/2&0\\0&\gamma_{-k}/2
    \end{pmatrix},
\end{align}
where $\otimes$ means the tensor product and $\oplus$ the direct sum of operators
(the direct sum can be seen as a diagonal alignment of the blocks of operators/real numbers).
These operators are supposed to separately represent the transition structure and the transition rates in the dynamics. 
Hereafter, we do not distinguish the number zero and the zero operators and write them as $0$. 

The introduction of an auxiliary space is essential; in classical systems, a jump can be labeled by a pair of states $i,j$ without introducing a special label for jumps. On the other hand, in quantum dynamics, a jump can involve more than two (eigen)states. In our formulation, an auxiliary space is necessary to treat quantum jumps apart from specific states, and it enables us to discuss connections between dynamics and thermodynamics more efficiently. 
As will be discussed later, the characterization of conservative systems in terms of thermodynamic forces and the definition of the geometric housekeeping-excess decomposition are only possible with the introduction of an auxiliary space.

Next, regarding $\Gamma\otimes\rho$ as the fluxes ($\bm{J}^\pm$) in the classical case, we introduce the current and the force operator as follows:
\begin{align}
    \curr(\rho)&\coloneqq[\strc,\Gamma\otimes\rho], \label{eq:defCurr}\\
    \force(\rho)&\coloneqq [\strc,\ln(\Gamma\otimes\rho)]. \label{eq:defForce}
\end{align}
They are anti-Hermitian operators on $\hilbb\otimes\hilb$. 
We can express them alternatively as
\begin{equation}
    \begin{split}
        &\curr(\rho)=\bigoplus_{k\in K}
        \begin{pmatrix}
            0&J_{-k}\\
            J_k&0
        \end{pmatrix}\\
        &\text{with}\quad J_k = \frac{1}{2}(\gamma_k L_k\rho-\gamma_{-k}\rho L_k)
    \end{split}
    \label{eq:curr2}
\end{equation}
and
\begin{equation}
    \begin{split}
        &\force(\rho) = 
        \bigoplus_{k\in K}
        \begin{pmatrix}
            0&F_{-k}\\
            F_k&0
        \end{pmatrix}\\
        &\text{with}\quad F_k = \ln\left(\frac{\gamma_k}{\gamma_{-k}}\right)L_k + [L_k,\ln\rho].
    \end{split}
    \label{eq:force2}
\end{equation}
Note that, in particular, the thermodynamic consistency conditions~\eqref{eq:TC} with the local detailed balance condition allow us to rewrite $F_k$ into
\begin{align}
    F_k = [L_k,\ln\rho+\beta_k H]. \label{eq:forceTC}
\end{align}
The reason we regard these operators as currents and forces is that they satisfy the dynamical and thermodynamic relationships that classical currents and forces fulfill, as will be shown in the following. A few of them are listed in Table~\ref{tab:QuantumClassical}. Some correspondence to the classical notion is also suggested in Example~\ref{sec:tls}.
We comment here that an expression similar to Eq.~\eqref{eq:curr2} is provided recently in Ref.~\cite{de2023quantum}. 

The current and force operators are connected as 
\begin{align}
    \curr(\rho)=\lm_{\Gamma\otimes\rho}(\force(\rho)) \label{eq:forceToCurrent} 
\end{align}
with a linear super-operator $\lm_{\Gamma\otimes\rho}$, which is defined as
\begin{align}
    \lm_{B}(A)\coloneqq\int_0^1 B^s AB^{1-s}ds
\end{align}
for an operator $A$ and a positive operator $B$. 
Given the spectral decomposition $B=\sum_n b_n \ketbras{n}$, we find the explicit formula~\cite{ottinger2010nonlinear}
\begin{align}
    \lm_{B}(A)=\sum_{n,m}\frac{b_n-b_m}{\ln(b_n/b_m)}\bra{n}A\ket{m}\ket{n}\bra{m}. \label{eq:lmExplicit}
\end{align}
That is, the force and current operators are associated by a quantum generalization of the logarithmic mean. 
Despite its nontrivial look, Eq.~\eqref{eq:forceToCurrent} is proved easily; first notice the identity $-(d/ds)\big(e^{sQ}Ae^{(1-s)Q}\big)=e^{sQ}[A,Q]e^{(1-s)Q}$. Inserting $Q=\ln B$ and integrating both sides over $s\in[0,1]$, we get
\begin{align*}
    \lm_B([A,\ln B])=-BA+AB=[A,B].
\end{align*}
Setting $A=\strc$ and $B=\Gamma\otimes\rho$ finally leads to Eq.~\eqref{eq:forceToCurrent}. 

In the following, the inner product given by $\lm_{\Gamma\otimes\rho}$, 
\begin{align}
    \langle \force,\force'\rangle_{\Gamma\otimes\rho}
    \coloneqq \langle \force,\lm_{\Gamma\otimes\rho}(\force')\rangle,
\end{align}
and the induced norm 
\begin{align}
    \lVert\force\rVert_{\Gamma\otimes\rho}\coloneqq
    \sqrt{\langle \force,\force\rangle_{\Gamma\otimes\rho}} \label{eq:lmnorm}
\end{align}
will play a central role. Hereafter, the plain inner product $\langle\cdot,\cdot\rangle$ represents the Hilbert--Schmidt inner product $\langle A, B\rangle=\tr(A^\dagger B)$ for the appropriate operator space. 
To see that $\lm_{\Gamma\otimes\rho}$ actually induces an inner product, we show some crucial properties of $\lm_B$ for an arbitrary positive operator $B$. 

Let $A$ and $C$ also be arbitrary (possibly not positive) operators on the same Hilbert space as the one on which $B$ acts. 
First, $\lm_B$ is linear, which is evident from the definition. Second, it is a symmetric super-operator regarding the Hilbert--Schmidt inner product because
\begin{align*}
    \langle \lm_{B}(A),C \rangle
    &=\sum_{n,m}\frac{b_n-b_m}{\ln(b_n/b_m)}
    \bra{m}A^\dagger\ket{n}\bra{n}C\ket{m}\\
    &=\langle A, \lm_{B}(C) \rangle. 
\end{align*}
Third, it is positive definite because, for an arbitrary operator, 
\begin{align*}
    \langle \lm_{B}(A),A \rangle
    &=\sum_{n,m}\frac{b_n-b_m}{\ln(b_n/b_m)}
    |\bra{n}A\ket{m}|^2\geq 0, 
\end{align*}
and this quantity is zero only if $A$ is a zero operator. 
As a result, the induced inner product $\langle A, C\rangle_{B}\coloneqq\langle A, \lm_{B}(C) \rangle$ satisfies the axioms of inner product, and we can define the norm $\lVert A\rVert_B\coloneqq \sqrt{\langle A,A\rangle_B}$. 

In addition, we note that the inverse $\lm_B^{-1}$ can be defined by inverting the log-mean factor in Eq.~\eqref{eq:lmExplicit} as
\begin{align}
    \lm_B^{-1}(A)\coloneqq \sum_{n,m}\frac{\ln(b_n/b_m)}{b_n-b_m}
    \bra{n}A\ket{m}\ket{n}\bra{m}.
\end{align}
Since it shares all three properties with $\lm_B$, it also leads to an inner product.

Now that the three properties of $\lm_{\Gamma\otimes\rho}$, linearity, symmetry, and positive definiteness, are proved, we can connect Eq.~\eqref{eq:forceToCurrent} to the renowned Onsager reciprocal relation~\cite{onsager1931reciprocal1,onsager1931reciprocal2}. 
Let us consider physical currents, e.g., electric currents or heat currents, $j_\alpha(\rho)\in\mathbb{R}$ ($\alpha=1,2,\dots$) given by 
\begin{align}
    j_\alpha(\rho)=\langle \mathbb{A}_\alpha,\curr(\rho)\rangle 
\end{align}
with anti-Hermitian operators $\mathbb{A}_\alpha\in\anti(\hilbb\otimes\hilb)$.
We assume that the force operator is given in the dual form 
\begin{align}
    \force(\rho)=\sum_\alpha f_\alpha(\rho)\mathbb{A}_\alpha \label{eq:force_assumption}
\end{align}
by physical forces $f_\alpha(\rho)\in\mathbb{R}$, such as voltage or temperature gradient. 
We further define a matrix $L(\rho)$ by
\begin{align}
    L_{\alpha\beta}(\rho)\coloneqq\langle\mathbb{A}_\alpha,\mathbb{A}_\beta\rangle_{\Gamma\otimes\rho}=\langle \mathbb{A}_\alpha,\lm_{\Gamma\otimes\rho}(\mathbb{A}_\beta)\rangle.  
\end{align}
If $\rho=\rhoeq=e^{-\beta H}/Z$, this definition resembles the Green--Kubo formula~\cite{green1954markoff,kubo2012statistical} as 
\begin{align}
    &L_{\alpha\beta}(\rhoeq)\notag\\
    &= \frac{1}{Z}\int_0^1 \tr[\mathbb{A}_\alpha(\Gamma^s\otimes e^{-s\beta H}) \mathbb{A}_\beta (\Gamma^{1-s}\otimes e^{-(1-s)\beta H})]ds. 
\end{align}
Then, Eq.~\eqref{eq:forceToCurrent} leads to the relation 
\begin{align}
    j_{\alpha}(\rho)=\sum_\beta L_{\alpha\beta}(\rho)f_\beta(\rho) \label{eq:classical_onsager0}
\end{align}
because $\sum_\beta L_{\alpha\beta}(\rho)f_\beta(\rho)=\sum_\beta \langle \mathbb{A}_\alpha,\lm_{\Gamma\otimes\rho}(f_\beta(\rho)\mathbb{A}_\beta)\rangle=\langle \mathbb{A}_\alpha,\lm_{\Gamma\otimes\rho}(\force(\rho))\rangle$ and $\curr(\rho)=\lm_{\Gamma\otimes\rho}(\force(\rho))$. 
Let us discuss that Eq.~\eqref{eq:classical_onsager0} is a nonequilibrium extension of the Onsager reciprocal relation. 
When $\rho$ is close to equilibrium $\rhoeq$, $j_\alpha(\rho)$ and $f_\beta(\rho)$ will be of the order of the difference between $\rho$ and $\rhoeq$, which we write $\varepsilon$. 
Then, $L(\rho)$ can be replaced by $L(\rhoeq)$ because $[L_{\alpha\beta}(\rho)-L_{\alpha\beta}(\rhoeq)]f_\beta(\rho)=O(\varepsilon^2)$. 
Thus, near the equilibrium state $\rhoeq$, to leading order, we obtain  
\begin{align}
    j_{\alpha}(\rho)=\sum_\beta L_{\alpha\beta}(\rhoeq)f_\beta(\rho). \label{eq:classical_onsager}
\end{align}
Moreover, since $\mathbb{A}_\alpha$ are anti-Hermitian, the symmetry of the inner product $\langle\cdot,\cdot\rangle_{\Gamma\otimes\rho}$ proves the reciprocal relation 
\begin{align}
    L_{\alpha\beta}(\rho)=L_{\beta\alpha}(\rho). 
\end{align}
Therefore, Eq.~\eqref{eq:classical_onsager} is the Onsager reciprocal relation~\cite{onsager1931reciprocal1,onsager1931reciprocal2} and Eq.~\eqref{eq:classical_onsager0} is seen as its nonequilibrium extension~\cite{vroylandt2018degree}. 
If $\mathbb{A}_\alpha$ are linearly independent, $L(\rho)$ becomes positive definite because for any $\bm{\xi}=(\xi_\alpha)$, we have 
\begin{align}
    \bm{\xi}^{\mathrm{T}}L(\rho)\bm{\xi}
    =\Big\lVert\sum_\alpha \xi_\alpha\mathbb{A}_\alpha\Big\rVert_{\Gamma\otimes\rho}^2,
\end{align}
and it is positive unless $\bm{\xi}=0$. 
In general, we have the following expression of the EPR 
\begin{align}
    \epr(\rho)=\sum_{\alpha,\beta}f_{\alpha}(\rho)L_{\alpha\beta}(\rho)f_{\beta}(\rho). 
\end{align}
We finally mention that Eq.~\eqref{eq:force_assumption} is a crucial assumption and should be proved generally or verified in concrete cases.

\subsection{Connection to dynamics}
\label{subsec:connect}
Next, we explain how the operators are related to the dynamics. 
First, we define the gradient super-operator $\nabla_{\strc}:\opr(\hilb)\to\opr(\hilbb\otimes\hilb)$ by
\begin{align}
    \nabla_{\strc} A\coloneqq[(I_\hilbb\otimes A),\strc]
    =\bigoplus_{k\in K}
    \begin{pmatrix}
        0&[A,L_{k}^\dagger]\\
        [A,L_k]&0
    \end{pmatrix},
\end{align}
where $I_\hilbb$ is the identity matrix of $\hilbb$. 
Specifically, it maps $\herm(\hilb)$ into $\anti(\hilbb\otimes\hilb)$. As it is defined with a commutator, the Leibniz rule holds; $\nabla_\strc(AB)=\nabla_\strc A(I_\hilbb\otimes B)+(I_\hilbb\otimes A)\nabla_\strc B$. 
The divergence super-operator $\nabla_{\strc}^*$, i.e., the adjoint of $\nabla_\strc$, is provided as
\begin{align}
    \nabla_{\strc}^* \mathbb{B}=\tr_\hilbb [\mathbb{B},\strc]. 
\end{align} 
It can be seen as follows: for $A\in\opr(\hilb)$ and $\mathbb{B}\in\opr(\hilbb\otimes\hilb)$, we have $\langle \nabla_\strc A,\mathbb{B}\rangle=\tr([\strc,(I_\hilbb\otimes A^\dagger)]\mathbb{B})=\tr((I_\hilbb\otimes A^\dagger)[\mathbb{B},\strc])=\tr_\hilb(A^\dagger\tr_\hilbb[\mathbb{B},\strc])=\langle A,\tr_\hilbb[\mathbb{B},\strc]\rangle$, where $\tr_\hilb$ and $\tr_\hilbb$ are the partial traces over $\hilb$ and $\hilbb$. When $\mathbb{B}$ is in the form
\begin{align}
    \mathbb{B}=\bigoplus_{k\in K}
    \begin{pmatrix}
        0&B_{-k}\\
        B_k&0
    \end{pmatrix},
\end{align}
the divergence reads
\begin{align}
    \nabla_{\strc}^* \mathbb{B}
    = \sum_{k\in K}\big([B_k,L_k^\dagger]+[B_{-k},L_k]\big). \label{eq:divexplicit}
\end{align}

With the gradient super-operator, we can define conservative forces; define a force operator $\force(\rho)$ to be \textit{conservative} if it is provided as $-\nabla_\strc\phi(\rho)$ with a potential operator $\phi(\rho)\in\herm(\hilb)$. 
Moreover, the divergence super-operator allows us to rewrite the dissipator in a continuity-equation form
\begin{align}
    \mathcal{D}(\rho)=\nabla_{\strc}^*\curr(\rho). \label{eq:eq7}
\end{align}
This expression was proved for detailed-balanced systems in the form $\mathcal{D}(\rho)=-\tr_\hilbb([\strc,[\strc,\Gamma\otimes\rho]])$ in Ref.~\cite{mittnenzweig2017entropic}, but we stress that Eq.~\eqref{eq:eq7} holds whether or not we impose detailed balance. 
Keeping the relation $L_{-k}=L_k^\dagger$ in mind, one can confirm Eq.~\eqref{eq:eq7} by a straightforward calculation combining Eqs.~\eqref{eq:curr2} and \eqref{eq:divexplicit}. 

When the system is detailed balanced, the system has an equilibrium solution $\rhoeq=e^{-\beta H}/Z$, where the current vanishes $\curr(\rhoeq)=0$. 
Note that, in a general steady state $\rho^{\mathrm{ss}}$, we only have $-i[H,\rho^{\mathrm{ss}}]+\nabla_\strc^*\curr(\rho^{\mathrm{ss}})=0$ and the current $\curr(\rho^{\mathrm{ss}})$ can be nonzero. 
To show the statement, first note that the commutation relation $[L_k,H]=\omega_k L_k$ leads to $L_k H^n = (H+\omega_k)^n L_k$, which further brings $L_ke^{-\beta H}=e^{-\beta(H+\omega_k)} L_k$~\footnote{The first equality is proved by the mathematical induction; $L_kH = (H+\omega_k)L_k$ holds and assuming $L_kH^n=(H+\omega_k)^nL_k$ leads to $L_kH^{n+1}=(H+\omega_k)^nL_kH =(H+\omega_k)^{n+1}L_k$. The second one is derived by applying this equality to each term of the Taylor expansion of $e^{-\beta H}$. }.
Applying this to Eq.~\eqref{eq:curr2}, we get
\begin{align}
    J_k(\rhoeq)=\frac{1}{2}(\gamma_ke^{-\beta\omega_k}-\gamma_{-k})\rhoeq L_k=0 \label{eq:detailedbalance}
\end{align}
for any $k$, where the last equality comes from the local detailed balance $\gamma_k/\gamma_{-k}=e^{\beta\omega_k}$. 

As in the classical case, whether or not the force is conservative can be proved to be equivalent to whether or not the system has an equilibrium solution. 
Assume there is an equilibrium state $\pi$ such that $\curr(\pi)=0$. Then, we can rewrite the force operator as
\begin{align}
    \force(\rho)&=\force(\rho)-\force(\pi)
    =[\strc,\ln(\Gamma\otimes\rho)-\ln(\Gamma\otimes\pi)]\notag\\
    &=-[I_\hilbb\otimes(\ln\rho-\ln\pi),\strc]
    =-\nabla_\strc\phi(\rho)
\end{align}
with $\phi(\rho)\coloneqq\ln\rho-\ln\pi$, where we used the fact $\force(\pi)=0$, which is because $\force(\pi)=\lm_{\Gamma\otimes\pi}^{-1}(\curr(\pi))$ and $\lm_{\Gamma\otimes\pi}^{-1}$ is linear, and the relation
\begin{align*}
    &\ln(\Gamma\otimes\rho)-\ln(\Gamma\otimes\pi)\\
    &=\ln\Gamma\otimes I_{\hilb} + I_{\hilbb}\otimes\ln\rho
    -(\ln\Gamma\otimes I_{\hilb} + I_{\hilbb}\otimes\ln\pi)\\
    &=I_{\hilbb}\otimes(\ln\rho-\ln\pi),
\end{align*} 
where $I_\hilb$ is the identity operator of $\hilb$. 
On the other hand, assume the force operator is given as $\force(\rho)=-\nabla_\strc\phi$. If we introduce $\pi=e^{-\psi}/\tr(e^{-\psi})$ ($\psi = \phi-\ln\rho$), then we have $\force(\pi)=0$ because
\begin{align}
    \force(\pi)
    &=[\strc,\ln\Gamma\otimes I_\hilb]
    +[\strc,I_\hilbb\otimes\ln\pi]\notag\\
    &=[\strc,\ln\Gamma\otimes I_\hilb]
    +[\strc,I_\hilbb\otimes\ln\rho]
    -[\strc,I_\hilbb\otimes\phi]\notag\\
    &=\nabla_\strc\phi + \force(\rho) = 0,
\end{align}
where the contribution from $\tr(e^{-\psi})$ disappears in the commutator in the second line. 
Therefore, $\curr(\pi)=0$ immediately follows from the relation $\curr(\pi)=\lm_{\Gamma\otimes\pi}(\force(\pi))$ and the linearity of $\lm_{\Gamma\otimes\pi}$. 
In summary, $\nabla_\strc$ enables one to characterize the detailed balance in terms of force in quantum thermodynamics as in the classical case and resolves the problem of the conventional classical expression discussed in Sec.~\ref{subsec:reviewCL}.

In addition, the gradient operator gives a sufficient condition of conserved quantities in autonomous systems. If a time-independent observable $X$ satisfies $[H,X]=0$, the standard condition of conserved quantities in quantum mechanics, and 
\begin{align}
    \nabla_{\strc}X=0,
\end{align}
then we find $\langle X\rangle=\tr(\rho X)$ is conserved because
\begin{align*}
    \frac{d}{dt}\langle X\rangle 
    &= -i\tr(X[H,\rho])+\tr(X\mathcal{D}(\rho))\\
    &=i\tr(\rho[H,X])+\langle X,\nabla_{\strc}^*\curr(\rho)\rangle\\
    &= i\tr(\rho[H,X])+\langle \nabla_{\strc} X,\curr(\rho)\rangle=0. 
\end{align*}
This is analogous to the situation in CRNs, where the stoichiometric matrix plays the same role as $\nabla_{\strc}^*$ and no counterpart to the Hamiltonian exists~\cite{polettini2014irreversible,rao2016nonequilibrium}. 
An example of such a conserved quantity can be found in the system discussed in Sec.~\ref{subsec:TF}. 

\subsection{Entropy production rate}
\begin{table}
    \centering
    \renewcommand{\arraystretch}{2}
    \setlength{\tabcolsep}{0pt} 
    \begin{tabular}{c|c|c}
         & Classical & Quantum \\\hline\hline
        \rowcolor{gray!10}
        Gradient & $\nabla$ & $\nabla_\strc$ \\
        Continuity & $\;\; d\bm{x}/dt=\nabla^\mathrm{T}\bm{J}(\bm{x})\;\;$ & $\mathcal{D}(\rho)=\nabla_\strc^*\curr(\rho)$\\
        \rowcolor{gray!10}
        Current & $J_{e}=J_e^+-J_e^-$ & $\curr(\rho)=[\strc,\Gamma\otimes\rho]$\\
        Force & $\;\;F_{e}=\ln(J_e^+/J_e^-)\;\;$ & $\;\;\force(\rho)=[\strc,\ln(\Gamma\otimes\rho)]\;\;$\\
        \rowcolor{gray!10}
        Conservative & $\bm{F}=-\nabla\bm{\phi}$ & $\force(\rho)=-\nabla_\strc\phi$ \\
        \hspace{6pt}Detailed balance\hspace{6pt} & $\bm{J}(\bm{x}^\mathrm{eq})=0$ & $\curr(\rhoeq)=0$ \\
        \rowcolor{gray!10}
        EPR & $\bm{J}^{\mathrm{T}}\bm{F}$ & $\tr\big(\curr(\rho)^\dagger\force(\rho)\big)$\\\hline
    \end{tabular}
    \caption{Classical-quantum correspondence in nonequilibrium thermodynamics. Quantum counterparts are given by operators or super-operators. }
    \label{tab:QuantumClassical}
\end{table}

Finally, we arrive at the following expression of the EPR:
\begin{align}
    \epr(\rho) = \langle \curr(\rho),\force(\rho)\rangle. \label{eq:canonicalEPR}
\end{align}
Not only in MJPs and CRNs, this form of current times force widely appears in nonequilibrium systems~\cite{de1962non}. 
It is fair to note that expressions essentially close to ours can be found in Refs.~\cite{ottinger2010nonlinear,de2023quantum}; however, the result in Ref.~\cite{ottinger2010nonlinear} is only valid under the assumption of detailed balance, and none of them could identify the force-current structure. 
This time, the clear and widely applicable definition of the force-current structure becomes available for the first time by considering the auxiliary space $\hilbb$ and operator $\Gamma$. 
Therefore, Eq.~\eqref{eq:canonicalEPR} is the first complete extension of the canonical formula of the EPR to the open quantum systems. 

Before proving Eq.~\eqref{eq:canonicalEPR}, let us see a few facts. 
In this expression, $\curr$ and $\force$ can be swapped as they are both anti-Hermitian. Moreover, since $\curr(\rho)=\lm_{\Gamma\otimes\rho}(\force(\rho))$, we have 
\begin{align}
    \epr(\rho)=\lVert\force(\rho)\rVert_{\Gamma\otimes\rho}^2, \label{eq:EPRandSquaredForce}
\end{align}
where the norm is defined in Eq.~\eqref{eq:lmnorm}. 
Up to here, this is just a rewriting, but the geometric representation will be useful for deriving inequalities in later sections. 

Let us prove Eq.~\eqref{eq:canonicalEPR}. 
First, referring to expressions in Eqs.~\eqref{eq:curr2} and \eqref{eq:force2}, we can get
\begin{align}
    \langle \curr(\rho),\force(\rho)\rangle 
    = \sum_{k\in K} [\tr(J_k^\dagger F_k)+\tr(J_{-k}^\dagger F_{-k})]. 
\end{align}
Considering the first term of $F_k=s_k L_k + [L_k,\ln\rho]$ and using the definition of $J_k$, we have
\begin{align*}
    &s_k\tr(J_k^\dagger L_k)+s_{-k}\tr(J_{-k}^\dagger L_{-k})\\
    &=s_k\gamma_k\tr(L_k^\dagger L_k\rho)+s_{-k}\gamma_{-k}\tr(L_{-k}^\dagger L_{-k}\rho).
\end{align*}
Summing over $k\in K$, we obtain the second term in Eq.~\eqref{eq:EPRdef}. 
The second term of $F_k$ leads to
\begin{align*}
    &\tr(J_k^\dagger[L_k,\ln\rho])
    +\tr(J_{-k}^\dagger[L_{-k},\ln\rho])\\
    &=-\tr(\{[J_{-k},L_k]+[J_{k},L_{-k}]\}\ln\rho)\\
    &=-\tr_\hilb\left(\tr_{\mathbb{C}^2}\left\{\left[
        \begin{pmatrix}
            0&J_{-k}\\J_k&0
        \end{pmatrix},
        \begin{pmatrix}
            0&L_{-k}\\L_k&0
        \end{pmatrix}
    \right]\right\}\ln\rho\right),
\end{align*}
where we used $J_k^\dagger = -J_{-k}$ in the second equality. 
By summing over $k\in K$, because of the continuity-equation representation of $\mathcal{D}(\rho)$, we obtain $-\tr(\mathcal{D}(\rho)\ln\rho)$, which is the first term in Eq.~\eqref{eq:EPRdef}, and complete the proof. 
Note that we only assumed the local detailed balance condition, so Eq.~\eqref{eq:canonicalEPR} is valid even without the thermodynamic consistency condition~\eqref{eq:TC}. 

In this proof, the calculations are done with respect to each $k$. This is because the dissipative term of the quantum master equation is essentially separable; $\mathcal{D}(\rho)$ is given as a sum of $\mathcal{D}_k(\rho)$'s and the EPR can be split into partial EPRs
\begin{equation}
    \begin{split}
        \epr_k(\rho)\coloneqq
        &-\tr(\mathcal{D}_k(\rho)\ln\rho)\\
        &+s_k\gamma_k\tr(L_k^\dagger L_k\rho)+s_{-k}\gamma_{-k}\tr(L_kL_k^\dagger\rho). 
    \end{split}
\end{equation}
The direct sums appearing in the definitions of $\curr$, $\force$, $\strc$, and $\Gamma$ can be understood to reflect this structure.
Moreover, if we define $\curr_k(\rho)$ and $\force_k(\rho)$ by
\begin{align}
    \curr_k(\rho)
    =\begin{pmatrix}
        0&J_{-k}\\
        J_k&0
    \end{pmatrix},\quad
    \force_k(\rho)
    =\begin{pmatrix}
        0&F_{-k}\\
        F_k&0
    \end{pmatrix},
\end{align}
we get not only $\curr(\rho)=\bigoplus_{k\in K}\curr_k(\rho)$, $\force(\rho)=\bigoplus_{k\in K}\force_k(\rho)$, but also
\begin{align}
    \epr_k(\rho)=\langle \curr_k(\rho),\force_k(\rho)\rangle, \label{eq:partialEPR}
\end{align}
which is evident from the proof of Eq.~\eqref{eq:canonicalEPR}. 
The positivity of partial EPRs follows from the fact that positive super-operator $\lm_{\Gamma_k\otimes\rho}$ connects $\curr_k(\rho)$ and $\force_k(\rho)$, where 
\begin{align}
    \Gamma_k=\begin{pmatrix}
        \gamma_k/2&0\\0&\gamma_{-k}/2
    \end{pmatrix}. 
\end{align}
As in the preceding global discussion, we can define an inner product and a norm induced by $\lm_{\Gamma_k\otimes\rho}$ and derive the expression $\epr_k(\rho)=\lVert\force_k(\rho)\rVert_{\Gamma_k\otimes\rho}^2$. 
For later convenience, we further define
\begin{align}
    \strc_k
    =\begin{pmatrix}
        0&L_k^\dagger\\L_k&0
    \end{pmatrix},
\end{align}
which yields $\nabla_{\strc_k}A=[I_{\mathbb{C}^2}\otimes A,\strc_k]$ and $\nabla_{\strc_k}^*\mathbf{X}=\tr_{\mathbb{C}^2}[\mathbf{X},\strc_k]$ for $A\in\opr(\hilb)$ and $\mathbf{X}\in\opr(\mathbb{C}^2\otimes \hilb)$. 

\bigskip
In summary, up to this section, we have defined current and force operators for the quantum master equation, which constitute a clear analogy with the classical framework. 
The correspondence is summarized in Table.~\ref{tab:QuantumClassical}. 

We note that under the transformation $(\gamma_k, L_k)\to (\gamma_k/|c_k|^2, c_kL_k)$, which does not change the quantum master equation as mentioned, the operators we introduced are \textit{not} invariant. 
Nevertheless, this fact does not make those ``variant'' operators meaningless but provides a criterion as to whether or not the result we obtain with them is physically relevant.
For example, the geometric housekeeping-excess decomposition, which we will derive in the next section extensively utilizing the geometric structure developed in this section, can be proved invariant under the transformation. 

In what follows, we discuss what we can learn from the analogy. 
In particular, we will obtain two applications:
First, we define a geometric housekeeping-excess decomposition of the EPR. It extends the geometric decomposition studied in classical systems and resolves a problem in the conventional decomposition in quantum thermodynamics. 
Second, we derive some thermodynamic trade-off relations. 
We extend the so-called short-time thermodynamic uncertainty relation known in classical systems and obtain a trade-off between time and dissipation. 
There, we will introduce a quantity that can be useful in measuring the fluctuations in quantum thermodynamics. 

\section{Applications}
\label{sec:Applications}

\subsection{Geometric housekeeping-excess decomposition}
\label{subsec:Decomposition}

We first consider the geometric housekeeping-excess decomposition in the open quantum systems as an application of the developed framework. 
The housekeeping-excess decomposition, also known as the adiabatic-nonadiabatic decomposition, was first proposed to study irreversibility more precisely by dividing entropy production into distinct contributions~\cite{oono1998steady}: 
The housekeeping part quantifies inevitable dissipation to keep the system out of equilibrium, while the excess is dissipation due to transient dynamics. 
It allows us to obtain tighter inequalities, focusing on physically distinct aspects of the dynamics~\cite{shiraishi2018speed,tuan2020unified,dechant2022geometric1,dechant2022geometric2,yoshimura2023housekeeping,kolchinsky2022information,kolchinsky2024generalized,ito2023geometric,nagayama2023geometric,yoshimura2024two,kolchinsky2024generalized}. 
In classical systems, essentially two concrete approaches have been proposed: one uses the long time limit of the dynamics (thus, the steady state)~\cite{hatano2001steady,komatsu2008steady}, and the other focuses on the gradient structure of the dynamics~\cite{maes2014nonequilibrium,ito2023geometric}. 

Some studies have revealed that we can establish such a decomposition for quantum systems~\cite{horowitz2014equivalent,manzano2018quantum} based on the former strategy. However, their method requires a steady state that satisfies a particular condition so that the decomposed EPRs are positive. 
Although this problem also exists in the classical method~\cite{ge2016nonequilibrium,rao2016nonequilibrium}, it was recently overcome by using the geometric method, which involves the instantaneous gradient structure rather than the steady state~\cite{dechant2022geometric1,dechant2022geometric2,yoshimura2023housekeeping,kolchinsky2022information,kolchinsky2024generalized,kobayashi2022hessian,nagayama2023geometric}.
Still, the geometric decomposition has yet to be obtained for the quantum master equation; thus, a generally applicable decomposition of EPR is still lacking. 

In what follows, we show that this issue can be resolved in quantum systems by using the geometric framework obtained in the previous section.
We derive the geometric housekeeping-excess decomposition for the open quantum systems by relying on the geometric expression of the EPR~\eqref{eq:EPRandSquaredForce} and the gradient super-operator introduced in the previous section. 
Before that, we begin with a brief review of the conventional way of decomposing the EPR in quantum thermodynamics. 

\subsubsection{Conventional decomposition and its problem}
In the context of quantum thermodynamics, the decomposition is often referred to as the adiabatic-nonadiabatic decomposition~\cite{horowitz2013entropy,horowitz2014equivalent,manzano2018quantum}. 
The adiabatic EPR corresponds to the housekeeping EPR in our terminology, while the nonadiabatic to the excess. 
We call the decomposition adopted in Refs.~\cite{horowitz2013entropy,horowitz2014equivalent,manzano2018quantum} the \textit{adiabatic-nonadiabatic} decomposition to distinguish it from our geometric decomposition explained later. 

The adiabatic-nonadiabatic decomposition is defined as follows~\cite{manzano2018quantum}: 
Assume there is a steady state $\pi$, which satisfies $-i[H,\pi]+\mathcal{D}(\pi)=0$. 
We define $\Phi\coloneqq -\ln\pi$ and further assume the condition
\begin{align}
    [\Phi,L_k] = \Delta\varphi_k L_k \label{eq:priviledged}
\end{align}
for all $k\in K$ with $\Delta\varphi_k$ being a difference between two eigenvalues of $\Phi$. 
This condition is valid if $\pi=e^{-\beta H}/Z$ because
then $\Phi = \beta H+\ln Z$ and Eq.~\eqref{eq:priviledged} turns into the thermodynamic consistency condition~\eqref{eq:TC}. 
The nonadiabatic EPR is then defined by
\begin{align}
    \dot{\Sigma}^{\mathrm{na}}(\rho)\coloneqq-\frac{d}{dt}D(\rho(t)\Vert\pi), \label{eq:na}
\end{align}
where $D(\rho\Vert\pi)=\tr\big(\rho(\ln\rho-\ln\pi)\big)$ is the relative entropy. 
The adiabatic EPR can be obtained by subtracting the nonadiabatic EPR from the total EPR~\footnote{Though we defined the decomposition for autonomous systems, where $H$ and $\gamma_k$ are constant, the decomposition can be applied to non-autonomous cases with a slight modification.}. 

Unfortunately, the condition in Eq.~\eqref{eq:priviledged} is not always guaranteed, and the adiabatic EPR can be negative, as shown in Ref.~\cite{manzano2018quantum}. 
Because the entropy production must be associated with the second law (the nonnegativity) and the decomposition's purpose is to gain a more detailed understanding of the second law and irreversibility, the violation of positivity is a crucial issue for the decomposition. 
In the following, we propose another way of decomposition, namely, the geometric housekeeping-excess decomposition; it does not even refer to a steady state and is always non-negative, so it can provide more general insights into open quantum systems. 

\subsubsection{Housekeeping EPR}
\begin{figure}
    \centering
    \includegraphics[width=\linewidth]{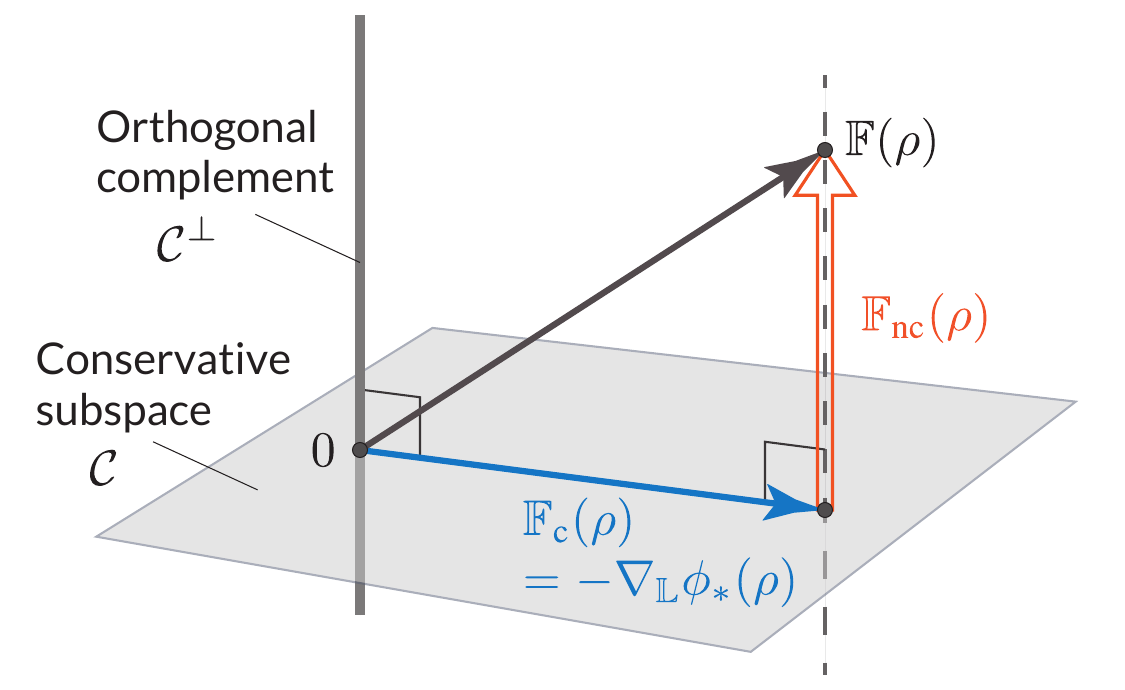}
    \caption{Schematics of the geometric housekeeping-excess decomposition. 
    We have two subspaces in the entire space of thermodynamic forces: $\mathcal{C}$, the space of conservative forces $\{-\nabla_{\strc}\phi\mid\phi\in\herm(\hilb)\}$, and the orthogonal complement $\mathcal{C}^\perp$ with respect to the inner product $\langle\cdot,\cdot\rangle_{\Gamma\otimes\rho}$. If two forces $\force'$ and $\force''$ satisfy $\force'-\force''\in\mathcal{C}^\perp$, then they provide the same dissipator as $\nabla_\strc^*\curr'=\nabla_\strc^*\curr''$, where $\curr'=\lm_{\Gamma\otimes\rho}(\force')$ and $\curr''=\lm_{\Gamma\otimes\rho}(\force'')$. 
    In the decomposition, the actual force $\force(\rho)$ is projected onto the conservative subspace $\mathcal{C}$ orthogonally. 
    Given the projected force be $\force_\mathrm{c}(\rho)$,
    the distance between $\force(\rho)$ and $\mathcal{C}$ is obtained as the norm of $\force_\mathrm{nc}(\rho)=\force(\rho)-\force_\mathrm{c}(\rho)$. 
    The squared norm of $\force_\mathrm{nc}(\rho)$ is interpreted as the housekeeping EPR, while that of $\force_\mathrm{c}(\rho)$ is the excess EPR. 
    As suggested in the schematics, the projected force $\force_\mathrm{c}(\rho)$ is unique, and its norm can be identified with the distance between the zero force $0$ and the affine subspace $\force(\rho)+\mathcal{C}^\perp$.}
    \label{fig:decomposition}
\end{figure}

We first define the housekeeping EPR in a geometric manner. 
We assume that the Hilbert space $\hilb$ has a finite dimension for technical reasons, but we expect that it can be generalized to infinite-dimensional cases with appropriate functional analytic treatment.
In the geometric decomposition, the housekeeping EPR is defined as the squared ``distance'' between the actual force and the subspace of conservative forces (see Fig.~\ref{fig:decomposition}). 
Now, we have the squared-norm expression of EPR $\epr(\rho)=\lVert\force(\rho)\rVert_{\Gamma\otimes\rho}^2$ and the definition of conservative force, i.e., the form $-\nabla_\strc\phi$. 
Therefore, we can define the housekeeping EPR as
\begin{align}
    \eprHK(\rho) \coloneqq \min_{\force'\in\mathcal{C}}\lVert\force(\rho)-\force'\rVert_{\Gamma\otimes\rho}^2, \label{eq:hk}
\end{align}
where $\mathcal{C}\coloneqq\{-\nabla_{\strc}\phi\mid\phi\in\herm(\hilb)\}\subset\mathcal{F}$ is the subspace of conservative forces and $\mathcal{F}=\anti(\hilbb\otimes\hilb)$ denotes the space of forces. 

The housekeeping EPR vanishes when the system is conservative, i.e., detailed balanced. 
On the other hand, when $\force(\rho)$ contains a non-conservative contribution, it has a positive value. 
Therefore, the housekeeping EPR quantifies how the external driving that makes the system non-conservative causes dissipation. 

To understand those facts in more depth, we consider the properties of the minimizer. 
We write the minimizer of Eq.~\eqref{eq:hk} as $\force_\mathrm{c}(\rho)=-\nabla_\strc\phi_*(\rho)$. 
The minimizer $\force_\mathrm{c}(\rho)$ is the conservative force closest to the actual force in terms of the metric determined by $\Gamma\otimes\rho$. 
From standard facts in linear algebra, we have the following facts~\cite{puntanen2011matrix} (see also Fig.~\ref{fig:decomposition}):
\begin{enumerate}
    \item $\force_\mathrm{c}(\rho)$ is unique,
    \item $\langle\force(\rho)-\force_\mathrm{c}(\rho),\force_\mathrm{c}(\rho)\rangle_{\Gamma\otimes\rho}=0$,
    \item $\force(\rho)-\force_\mathrm{c}(\rho)\in \mathcal{C}^\perp$,
\end{enumerate}
where $\mathcal{C}^\perp$ is the orthogonal complement of $\mathcal{C}$, given as
\begin{align}
    \mathcal{C}^\perp=\{\force'\in\mathcal{F}\mid \nabla_{\strc}^*\lm_{\Gamma\otimes\rho}(\force')=0\}. 
\end{align}
If we define the non-conservative force and the conservative and non-conservative currents as $\force_\mathrm{nc}(\rho)=\force(\rho)-\force_\mathrm{c}(\rho)$, $\curr_\mathrm{c}(\rho)=\lm_{\Gamma\otimes\rho}(\force_\mathrm{c}(\rho))$ and $\curr_\mathrm{nc}(\rho)=\lm_{\Gamma\otimes\rho}(\force_\mathrm{nc}(\rho))$, we have $\curr(\rho) = \curr_\mathrm{c}(\rho)+\curr_\mathrm{nc}(\rho)$ and  
\begin{align}
    \nabla_\strc^*\curr(\rho)
    =\nabla_\strc^*\curr_\mathrm{c}(\rho),\quad
    \nabla_\strc^*\curr_\mathrm{nc}(\rho)=0. \label{eq:jcjnc}
\end{align}
Therefore, by the definition of the housekeeping EPR, we can divide the total current $\curr(\rho)$ into the ``futile'' part $\curr_\mathrm{nc}(\rho)$ that does not contribute to the dynamics and the essential one $\curr_\mathrm{c}(\rho)$. 
The non-conservative motion is only used to keep the system out of equilibrium and does not affect state change. 

Since $\eprHK = \lVert\force_\mathrm{nc}(\rho)\rVert_{\Gamma\otimes\rho}^2$, we have the expression
\begin{align}
    \eprHK(\rho) = \langle \curr_\mathrm{nc}(\rho),\force_\mathrm{nc}(\rho)\rangle, \label{eq:cyclic}
\end{align}
where the inner product is the Hilbert--Schmidt one. 
This expression represents the fact that the housekeeping dissipation stems from the ``futile'' motion in the dynamics. 
The formula has been obtained for steady-state dissipation in Schnakenberg's seminal work~\cite{schnakenberg1976network} and extended to the housekeeping dissipation in classical systems in and out of steady states in Ref.~\cite{yoshimura2023housekeeping}. 
Equation~\eqref{eq:cyclic} can be regarded as a quantum extension of those results. 

\subsubsection{Excess EPR}
Next, we define the excess EPR by $\eprEX\coloneqq\epr-\eprHK$. 
It has other expressions
\begin{align}
    \eprEX = \lVert\force_\mathrm{c}(\rho)\rVert_{\Gamma\otimes\rho}^2
    =\min_{\force'\in\force(\rho)+\mathcal{C}^\perp}
    \lVert\force'\rVert_{\Gamma\otimes\rho}^2, \label{eq:ex}
\end{align}
where $\force(\rho)+\mathcal{C}^\perp=\{\force'\mid\exists \force''\in \mathcal{C}^\perp \;\mathrm{s.t.}\;\force'=\force(\rho)+\force''\}$. 
These equations are derived from the standard facts in linear algebra
\begin{itemize}
    \item[4.] $\lVert\force(\rho)\rVert_{\Gamma\otimes\rho}^2
    =\lVert\force_\mathrm{nc}(\rho)\rVert_{\Gamma\otimes\rho}^2
    +\lVert\force_\mathrm{c}(\rho)\rVert_{\Gamma\otimes\rho}^2$
    \item[5.] $\displaystyle\lVert\force_\mathrm{c}(\rho)\rVert_{\Gamma\otimes\rho}^2=\min_{\force'\in\force(\rho)+\mathcal{C}^\perp}\lVert\force'\rVert_{\Gamma\otimes\rho}^2$.
\end{itemize} 
The implication of the condition $\force'\in \force(\rho)+\mathcal{C}^\perp$ becomes clearer if we focus on currents rather than forces; we can rewrite Eq.~\eqref{eq:ex} as follows: 
\begin{equation}
    \begin{split}
        &\eprEX(\rho) = \min_{\curr'}\langle \curr',\lm_{\Gamma\otimes\rho}^{-1}(\curr')\rangle
        \\
        &\mathrm{s.t.}\quad
        \nabla_{\strc}^*\curr'=\nabla_{\strc}^*\curr(\rho)
        \equiv\mathcal{D}(\rho).
    \end{split} \label{eq:exj}
\end{equation}
Therefore, the excess EPR can be understood as the minimum dissipation to induce the given dissipative dynamics $\mathcal{D}(\rho)$. 

We remark on a connection to the optimal transport theory~\cite{villani2009optimal}.
We can derive the classical version of Eqs.~\eqref{eq:ex} and $\eqref{eq:exj}$~\cite{yoshimura2023housekeeping,nagayama2023geometric}, which directly provides a discrete generalization of the 2-Wasserstein distance~\cite{maas2011gradient,liero2013gradient} through the Benamou--Brenier formula~\cite{benamou2000computational}. 
While further consideration is required on how to treat the unitary term $-i[H,\rho]$ in the dynamics, the expressions in Eqs.~\eqref{eq:ex} and \eqref{eq:exj} will likely serve as an essential stepping stone in considering optimal transport theory in quantum mechanics. 

\subsubsection{Maximization formula}
The housekeeping and excess EPRs can be formulated by maximization:
\begin{align}
    \eprHK(\rho)&=\max_{\force'\in\mathcal{C}^\perp}\frac{|\langle\force',\force(\rho)\rangle_{\Gamma\otimes\rho}|^2}{\lVert\force'\rVert_{\Gamma\otimes\rho}^2}, \label{eq:maxHK}\\
    \eprEX(\rho)&=\max_{\force'\in\mathcal{C}}\frac{|\langle\force',\force(\rho)\rangle_{\Gamma\otimes\rho}|^2}{\lVert\force'\rVert_{\Gamma\otimes\rho}^2}. \label{eq:maxEX}
\end{align}
Note that the difference between the two equations is in the range of maximization. 
These are proved at once as follows: we let $\alpha$ stand for $\mathrm{c}$ or $\mathrm{nc}$. In the Cauchy--Schwarz inequality
\begin{align}
    \lVert\force'\rVert_{\Gamma\otimes\rho}^2
    \lVert\force_\alpha(\rho)\rVert_{\Gamma\otimes\rho}^2
    \geq |\langle\force',\force_\alpha(\rho)\rangle_{\Gamma\otimes\rho}|^2, \label{eq:cs48}
\end{align}
we have 
\begin{align}
    \langle\force',\force_\alpha(\rho)\rangle_{\Gamma\otimes\rho}
    =\langle\force',\force(\rho)\rangle_{\Gamma\otimes\rho}
\end{align}
for $\force'\in\mathcal{C}$ if $\alpha=\mathrm{c}$, and for $\force'\in\mathcal{C}^\perp$ if $\alpha=\mathrm{nc}$. 
With the corresponding subspace denoted by $\mathcal{C}_\alpha$, rearranging Eq.~\eqref{eq:cs48} leads to
\begin{align}
    \lVert\force_\alpha(\rho)\rVert_{\Gamma\otimes\rho}^2
    \geq \frac{|\langle\force',\force(\rho)\rangle_{\Gamma\otimes\rho}|^2}{\lVert\force'\rVert_{\Gamma\otimes\rho}^2}
\end{align}
for any $\force'\in \mathcal{C}_\alpha$. The equality can be achieved by setting $\force'=\force_\alpha(\rho)\in\mathcal{C}_\alpha$. 

\subsubsection{Invariance}
\label{subsec:invariance}
We show the invariance of the decomposition under the transformation $(\gamma_k, L_k)\to (\gamma_k/|c_k|^2, c_kL_k)$ discussed in Sec.~\ref{sec:qme}. 
We prove this statement in two steps: first, we show that the potential(s) $\phi$ that provides the conservative force is independent of the transformation. Next, we show that the excess EPR is invariant, which implies the invariance of the decomposition.

First, we note that the conservative force $\force_\mathrm{c}(\rho)$ is the unique element that satisfies the conditions $\force_\mathrm{c}(\rho)\in\mathcal{C}$ and $\force(\rho)-\force_\mathrm{c}\in\mathcal{C}^\perp$ simultaneously (c.f.\ Fig.~\ref{fig:decomposition}). 
Thus, any $\phi$ that solves the equation
\begin{align}
    -\nabla_\strc^*\lm_{\Gamma\otimes\rho}(\nabla_\strc\phi)=\mathcal{D}(\rho) \label{eq:51}
\end{align}
can give the conservative force as $\force_\mathrm{c}(\rho)=-\nabla_{\strc}\phi$ because then $-\nabla_\strc\phi$ satisfies both the conditions. Note that $\mathcal{D}(\rho)$ appeared from $\nabla_\strc^*\lm_{\Gamma\otimes\rho}(\force(\rho))$ and is invariant under the transformation. 

Importantly, the left-hand side of Eq.~\eqref{eq:51} is also invariant regardless of $\phi$;
For any $X\in\opr(\hilb)$, we can show
\begin{align}
    &\nabla_\strc^*\lm_{\Gamma\otimes\rho}(\nabla_\strc X)\notag\\
    &=\sum_{k\in K}\bigg(\int_0^1 ds\frac{\gamma_k^{s}\gamma_{-k}^{1-s}}{2}\big[\rho^s[X,L_k^\dagger]\rho^{1-s},L_k\big]\notag\\
    &\phantom{=\sum_{k\in K}\Big(}
    +\int_0^1 ds\frac{\gamma_k^{1-s}\gamma_{-k}^{s}}{2}\big[\rho^s[X,L_k]\rho^{1-s},L_k^\dagger\big]\bigg), \label{eq:52}
\end{align}
and this is invariant under $(\gamma_k, L_k)\to (\gamma_k/|c_k|^2, c_kL_k)$ with $c_{-k}$ being the complex conjugate of $c_k$ because the pair of $L_k$ and $L_k^\dagger$ yields $|c_k|^2$, while both $\gamma_k^{s}\gamma_{-k}^{1-s}$ and $\gamma_{k}^{1-s}\gamma_{-k}^s$ result in the canceling factor $1/|c_k|^2$ (the derivation of Eq.~\eqref{eq:52} is given in Appendix~\ref{app:invariance}). 
Therefore, we can define the \textit{Laplacian} $\mathcal{L}$ as
\begin{align}
    \mathcal{L}(X)\coloneqq\nabla_\strc^*\lm_{\Gamma\otimes\rho}(\nabla_\strc X)
\end{align}
and it becomes invariant. 

Consequently, Eq.~\eqref{eq:51} reads
\begin{align}
    -\mathcal{L}(\phi)=\mathcal{D}(\rho) \label{eq:GFE}
\end{align}
and it is clear that the potentials that can give the conservative force are insensitive to the transformation. This means that $\eprEX(\rho)= \lVert\force_\mathrm{c}(\rho)\rVert_{\Gamma\otimes\rho}^2$ is also invariant under the transformation because 
\begin{align}
    \lVert\force_\mathrm{c}(\rho)\rVert_{\Gamma\otimes\rho}^2
    =\lVert-\nabla_\strc\phi\rVert_{\Gamma\otimes\rho}^2
    =\langle \phi,\mathcal{L}(\phi)\rangle
\end{align}
for any $\phi$ that solves $-\mathcal{L}(\phi)=\mathcal{D}(\rho)$. 
Therefore, the excess and housekeeping EPRs are well-defined regardless of the indefiniteness of $\gamma_k$ and $L_k$. 

We note that Eq.~\eqref{eq:52} further implies that the Laplacian is separable and the decomposed parts are also invariant: 
\begin{equation}
    \begin{split}
        &\mathcal{L}(X)=\sum_{k\in K}\mathcal{L}_k(X)\\
        &\text{with}\quad
        \mathcal{L}_k(X)\coloneqq\nabla_{\strc_k}^*\lm_{\Gamma_k\otimes\rho}(\nabla_{\strc_k} X).
    \end{split}
    \label{eq:partialLaplacian}
\end{equation}

\subsubsection{Unseparability}

We discuss that the decomposed EPRs are \textit{not} separable, unlike the total EPR; if we do the minimization~\eqref{eq:hk} for each $k$, we will get a force of the form $\bigoplus_{k\in K}(-\nabla_{\strc_k}\phi_k)$, which is not a \textit{globally} conservative force but a \textit{fully-locally} conservative force. 
In fact, this fully local conservativeness is automatically satisfied if the dissipations are mediated by thermal baths; then, the expression in Eq.~\eqref{eq:forceTC} implies $\force_k(\rho) = -\nabla_{\strc_k}\phi_k(\rho)$, where $\phi_k(\rho)=\beta_k H+\ln\rho$. 
On the other hand, global conservativeness, namely, expression $\force(\rho)=-\nabla_{\strc}\phi$ by a single potential operator $\phi$, is available only when the system is free from external driving so that it relaxes to the equilibrium. 
Therefore, in the decomposition, it is crucial that the minimization is implemented simultaneously to quantify the external effect on dissipation. 

Formally, we may define partial elements because the conservative force and current keep the direct-sum structure as $\force_\mathrm{c}(\rho)=\bigoplus_{k\in K}(-\nabla_{\strc_k}\phi_*(\rho))$ and $\curr_\mathrm{c}(\rho)=\bigoplus_{k\in K}\lm_{\Gamma_k\otimes\rho}(-\nabla_{\strc_k}\phi_*(\rho))$. 
However, in addition to the point that conservativeness is crucial only globally, we should be aware that the first equation in Eq.~\eqref{eq:jcjnc}, $\nabla_\strc\curr_\mathrm{c}(\rho)=\mathcal{D}(\rho)$, does not mean that each current component $\lm_{\Gamma_k\otimes\rho}(\nabla_{\strc_k}\phi_*(\rho))$ gives $\mathcal{D}_k(\rho)$ when operated by $\nabla_{\strc_k}^*$; therefore, the fragments of the housekeeping and excess EPRs would not be associated with conservativeness or minimum dissipation in the global sense. 
We note that, however, it could be interesting to consider \textit{partially-local} conservativeness if we have a reasonable partition of $K$ into a few parts, as in the bipartite systems considered in information thermodynamics~\cite{parrondo2015thermodynamics}. 

\subsection{Thermodynamic trade-off relations}
\label{subsec:tradeoff}

Next, we discuss trade-off relations derived from the geometric structure. 
Let us begin with a few general results. 
The first one is a lower bound on the excess EPR derived from the maximization representation of the excess EPR~\eqref{eq:maxEX}. For any $X\in\herm(\hilb)$, the maximization implies 
\begin{align}
    \eprEX(\rho)\geq \frac{|\tr (X\mathcal{D}(\rho))|^2}{\lVert\nabla_\strc X\rVert_{\Gamma\otimes\rho}^2}, \label{eq:preTUR}
\end{align}
where the numerator is derived from
\begin{align*}
    \langle\nabla_\strc X,\force(\rho)\rangle_{\Gamma\otimes\rho}
    &=\langle X, \nabla_\strc^*\lm_{\Gamma\otimes\rho}(\force(\rho))\rangle\\
    &=\langle X, \mathcal{D}(\rho)\rangle
    =\tr(X\mathcal{D}(\rho)). 
\end{align*}
Since the excess EPR $\eprEX(\rho)$ is smaller than or equal to the total EPR $\epr(\rho)$, a weaker inequality is obtained by replacing $\eprEX(\rho)$ with $\epr(\rho)$ in Eq.~\eqref{eq:preTUR}. 
When $X$ is time-independent, the time derivative of the expectation value $\langle X\rangle=\tr(X\rho)$ is given by
\begin{align}
    \frac{d}{dt}\langle X\rangle 
    = -i\tr(X[H,\rho])+\tr(X\mathcal{D}(\rho)). \label{eq:expTimeDerivative}
\end{align}
If $X$ further commutes with $\rho$ or $H$, we obtain $\tr(X[H,\rho])=0$; then, Eq.~\eqref{eq:preTUR} reads
\begin{align}
    \eprEX(\rho)\geq \frac{|\frac{d}{dt}\langle X\rangle|^2}{\lVert\nabla_\strc X\rVert_{\Gamma\otimes\rho}^2}. 
\end{align}

The second one is a lower bound on the partial EPRs: 
\begin{align}
    \epr_k(\rho)\geq\frac{|\tr(X\mathcal{D}_k(\rho))|^2}{\lVert \nabla_{\strc_k}X\rVert_{\Gamma_k\otimes\rho}^2}. \label{eq:preTURex}
\end{align}
It is proved by applying the Cauchy--Schwarz inequality as
\begin{align}
    \epr_k(\rho)=\lVert\force_k(\rho)\rVert_{\Gamma_k\otimes\rho}^2
    \geq \frac{|\langle \nabla_{\strc_k}X,\force_k(\rho)\rangle_{\Gamma_k\otimes\rho}|^2}{\lVert\nabla_{\strc_k}X\rVert_{\Gamma_k\otimes\rho}^2}
\end{align}
and rewriting the numerator. 

\subsubsection{Quantum diffusivity}
In the two bounds in Eqs.~\eqref{eq:preTUR} and \eqref{eq:preTURex}, the denominators can be upper bound as
\begin{align}
    \lVert\nabla_{\strc_k} X\rVert_{\Gamma_k\otimes\rho}^2\leq \mathscr{D}_X^k,\quad
    \lVert\nabla_\strc X\rVert_{\Gamma\otimes\rho}^2\leq \mathscr{D}_X \label{eq:bound}
\end{align}
with
\begin{align}
    \mathscr{D}_X^k
    &\coloneqq\frac{1}{2}\Big[\tr\big(X^2\mathcal{D}_k(\rho)\big)-\tr\big(X\mathcal{D}_k(\{X,\rho\})\big)\Big]\\
    \mathscr{D}_X
    &\coloneqq\sum_{k\in K}\mathscr{D}_X^k\\
    &=\frac{1}{2}\Big[\tr\big(X^2\mathcal{D}(\rho)\big)-\tr\big(X\mathcal{D}(\{X,\rho\})\big)\Big]. \label{eq:DX}
\end{align}
As discussed later, $\mathscr{D}_X$ can be regarded as a measure of fluctuation of observable $X$; so we call $\mathscr{D}_X$ the \textit{quantum diffusivity} of $X$. 
Technically, because of the separable structure, we have $\lVert\nabla_\strc X\rVert_{\Gamma\otimes\rho}^2=\sum_{k\in K}\lVert\nabla_{\strc_k} X\rVert_{\Gamma_k\otimes\rho}^2$; thus, the second inequality in Eq.~\eqref{eq:bound} follows from the first one. 
Yet, since the proof is complicated, we skip it to Appendix~\ref{app:diffusion}. 
Instead, we prove here that $\mathscr{D}_X$ coincides with the classical definition of diffusivity when $X$ commutes with $\rho$. 

When $[X,\rho]=0$, they are diagonalized simultaneously as $\rho=\sum_ip_i\ketbras{i}$ and $X=\sum_i x_i\ketbras{i}$.
Then, we can show the formula 
\begin{align}
    \mathscr{D}_X=\frac{1}{2}\sum_{i,j,k\in K_\mathrm{all}}(x_i-x_j)^2R_{ij}^k p_j \label{eq:STV}
\end{align}
(recall $R_{ij}^k=\gamma_k|\bra{i}L_k\ket{j}|^2$).
The meaning of this quantity becomes clear if we consider a classical variable $\hat{x}$ that takes value $x_i$ at probability $p_i$ whose dynamics is described as an MJP with rates $\{R_{ij}^k\}$. 
Then, we obtain the relation
\begin{align}
    \lim_{dt\to 0}\frac{\mathrm{Var}(\Delta \hat{x})}{2dt} = \frac{1}{2}\sum_{i,j,k\in K_\mathrm{all}}(x_i-x_j)^2R_{ij}^k p_j, \label{eq:classicalDiffusivity}
\end{align}
where $\mathrm{Var}$ indicates the variance and $\Delta \hat{x} = \hat{x}(t+dt) - \hat{x}(t)$. 
Equation~\eqref{eq:classicalDiffusivity} is proved as follows: in an infinitesimal time interval $dt$, change in state $j\to i(\neq j)$ occurs at probability $R_{ij}p_jdt$ ($R_{ij}=\sum_k R_{ij}^k$), so $\hat{x}$'s change $\Delta\hat{x}$ has moments $\sum_{i,j}(x_i-x_j)^n R_{ij}p_j dt$. The equation immediately follows by considering the first and second moments. 
To show Eq.~\eqref{eq:STV}, we first note that for any $A=\sum_i a_i\ketbras{i}$, 
\begin{align}
    \tr\big(\ketbras{i} \mathcal{D}(A)\big)=\sum_{j,k\in K_\mathrm{all}}(R_{ij}^k a_j-R_{ji}^k a_i).
\end{align}
Applying this equality to each term in Eq.~\eqref{eq:DX} leads to
\begin{align*}
    \tr(X^2\mathcal{D}(\rho))&=\sum_{i,j,k\in K_\mathrm{all}}x_i^2(R_{ij}^kp_j-R_{ji}^kp_i),\\
    \tr(X\mathcal{D}(\{X,\rho\}))
    &=2\sum_{i,j,k\in K_\mathrm{all}}x_i(R_{ij}^kp_jx_j-R_{ji}^kp_ix_i).
\end{align*}
Therefore, we obtain
\begin{align}
    \mathscr{D}_X
    &=\frac{1}{2}\sum_{i,j,k\in K_\mathrm{all}}R_{ij}^kp_j(x_i^2-x_j^2-2x_ix_j+2x_j^2)\notag\\
    &=\frac{1}{2}\sum_{i,j,k\in K_\mathrm{all}}(x_i-x_j)^2R_{ij}^k p_j.
\end{align}

We have seen that $\mathscr{D}_X$ coincides with the classical diffusivity when $X$ commutes with $\rho$. 
In addition, $\mathscr{D}_X$ is defined for any observable $X$ and always nonnegative, which is evident if we admit Eq.~\eqref{eq:bound}.
Thus, we can regard $\mathscr{D}_X$ as a quantum extension of diffusivity. 
This should be important because the quantum extension of the variance of change (i.e., the mean square displacement) is not obvious as $X$ at time $t$ and $t+dt$ do not necessarily commute. 
We discuss analytical results on the quantum diffusivity with the damped harmonic oscillator in Sec.~\ref{sec:ex3}. 

Finally, we remark that with the adjoint super-operators $\mathcal{D}_k^*$ and $\mathcal{D}^*$, the quantum diffusivity is rewritten as
\begin{align}
    \mathscr{D}_X^k
    &=\frac{1}{2}\tr\Big[\rho\big(\mathcal{D}_k^*(X^2)-\{\mathcal{D}_k^*(X),X\}\big)\Big],\\
    \mathscr{D}_X
    &=\frac{1}{2}\tr\Big[\rho\big(\mathcal{D}^*(X^2)-\{\mathcal{D}^*(X),X\}\big)\Big]. \label{eq:QD2}
\end{align}
Here, the adjoint super-operators are given as
\begin{equation}
    \begin{split}
        \mathcal{D}_k^*(X)
        &=\gamma_k\left(L_k^\dagger X L_k-\frac{1}{2}\{L_k^\dagger L_k,X\}\right)\\
        &\phantom{=}+\gamma_{-k}\left(L_kX L_k^\dagger-\frac{1}{2}\{L_k L_k^\dagger,X\}\right)
    \end{split}
\end{equation}
and $\mathcal{D}^*=\sum_{k\in K}\mathcal{D}_k^*$. 

\subsubsection{Thermodynamic uncertainty relation}
Defining the quantum diffusivity, we obtain the inequalities
\begin{align}
    \epr_{k}(\rho)\frac{\mathscr{D}_X^k}{|\tr (X\mathcal{D}_k(\rho))|^2}&\geq 1, \label{eq:tur}\\
    \eprEX(\rho)\frac{\mathscr{D}_X}{|\tr (X\mathcal{D}(\rho))|^2}&\geq 1, \label{eq:turEX}
\end{align}
which can be seen as a quantum generalization of the thermodynamic uncertainty relation (TUR)~\cite{barato2015thermodynamic,horowitz2020thermodynamic}; 
the inequalities show universal trade-off relations between dissipation and the short-time fluctuation relative to the changing rate. 
As shown soon later, these inequalities are generalizations of known quantum trade-off relations~\cite{funo2019speed,tajima2021superconducting}. 

The first inequality further leads to
\begin{align}
    \epr(\rho)\mathscr{D}_X\geq \left(\sum_{k\in K}\big|\tr (X\mathcal{D}_k(\rho))\big|\right)^2. \label{eq:ineq1}
\end{align}
From the triangle inequality, we see that $\sum_{k\in K}\big|\tr (X\mathcal{D}_k(\rho))|$ is larger than $|\tr (X\mathcal{D}(\rho))|$, which appears in Eq.~\eqref{eq:turEX}; 
thus, if $\epr(\rho)=\eprEX(\rho)$, hence the system is conservative, the inequality in Eq.~\eqref{eq:ineq1} is stronger than that of Eq.~\eqref{eq:turEX}, where $\eprEX(\rho)$ reads $\epr(\rho)$. 
If the system is not conservative ($\eprHK(\rho)\neq 0$), there is no such strict order since $\eprEX(\rho)$ is smaller than $\epr(\rho)$. 
Equation~\eqref{eq:ineq1} is shown as
\begin{align*}
    &\left(\sum_{k\in K}\big|\tr (X\mathcal{D}_k(\rho))\big|\right)^2
    \leq\left(\sum_{k\in K}\sqrt{\epr_k(\rho)\mathscr{D}_X^k}\right)^2\\
    &\leq \sum_{k\in K}\epr_k(\rho)\times
    \sum_{k\in K}\mathscr{D}_X^k
    =\epr(\rho)\mathscr{D}_X,
\end{align*}
where we used the Cauchy--Schwarz inequality in the second line. 

Let us see that the TURs generalize previously known results:
First, we show Eq.~\eqref{eq:ineq1} leads to a key inequality in Ref.~\cite{tajima2021superconducting}. Take $X=H$ and assume $\beta_k=\beta$ for all $k$. Then, 
\begin{align}
    Q_k(\rho)\coloneqq |\tr (H\mathcal{D}_k(\rho))|
\end{align}
provides the magnitude of the heat flow between the system and bath $k$. 
Using the formula~\eqref{eq:QD2} and the commutation relation $[L_k,H]=\omega_k L_k$, we can derive the relation
\begin{align}
    \mathscr{D}_H = \frac{1}{2}\sum_{k\in K_{\mathrm{all}}} \gamma_{k}\omega_k^2\tr(L_k^\dagger L_k\rho) \label{eq:TF}
\end{align}
(for details, see Appendix~\ref{app:TF}). 
By equation~\eqref{eq:ineq1}, we obtain the inequality
\begin{align}
    \frac{Q_\mathrm{tot}(\rho)^2}{\dot{\Sigma}(\rho)}\leq \mathscr{D}_H, \label{eq:76}
\end{align}
where $Q_\mathrm{tot}(\rho)=\sum_{k\in K}Q_k(\rho)$. 
This is the central inequality in the proof of the trade-off relation provided as Eq.~(4) in Ref.~\cite{tajima2021superconducting}.
In Sec.~\ref{subsec:TF}, we further demonstrate that the bound can be tightened (if we regard Eq.~\eqref{eq:76} as a lower bound on the EPR) by changing the observable we focus on from the Hamiltonian. 

Second, we discuss that Eq.~\eqref{eq:turEX} tightens the main inequality provided in Ref.~\cite{funo2019speed}. 
Let $\rho$ be eigendecomposed as $\sum_i p_i\ketbras{i}$. We define $X$ by $X=\sum_ix_i\ketbras{i}$ with
\begin{align}
    x_i \coloneqq \frac{1}{2}\frac{dp_i/dt}{|dp_i/dt|}
\end{align}
if $|dp_i/dt|\neq0$; otherwise, we set $x_i=0$.  
Then, we obtain
\begin{align}
    \tr(X\mathcal{D}(\rho)) &= \frac{1}{2}\sum_i \left|\frac{dp_i}{dt}\right|= \lVert\mathcal{D}(\rho)\rVert_\mathrm{tr},\\
    \mathscr{D}_X &\leq \frac{1}{2}\sum_{i,j(\neq i),k\in K_{\mathrm{all}}}R_{ij}^kp_j=:A(\rho), 
\end{align}
where $\lVert \cdot\rVert_\mathrm{tr}$ indicates the trace distance and $A(\rho)$ is the so-called dynamical activity~\cite{maes2008canonical}. 
The inequality in the second line comes from $(x_i-x_j)^2\leq (|x_i|+|x_j|)^2\leq 1$. 
Therefore, equation~\eqref{eq:turEX} leads to 
\begin{align}
    \lVert\mathcal{D}(\rho)\rVert_\mathrm{tr}\leq \sqrt{\frac{1}{2}\eprEX(\rho)A(\rho)},
\end{align}
which generalizes a part of the main result presented as Eq.~(39) in Ref.~\cite{funo2019speed}.
In the original result, $\epr(\rho)$ is used instead of $\eprEX(\rho)$, so the bound is tightened by using the excess EPR. 

\subsubsection{Time-dissipation uncertainty relation}
Another insight can be extracted from the inequalities if we define time $\tau_X$ by 
\begin{align}
    \tau_X\coloneqq
    \frac{2\mathscr{D}_X}{|\tr(X\mathcal{D}(\rho))|^2}. \label{eq:timeDissipation}
\end{align}
If $X$ is time-independent and commutes with $H$ or $\rho$, then $\tr(X\mathcal{D}(\rho))=d\langle X\rangle/dt$ and the expectation value of $X$ approximately changes $\tau_X\tr(X\mathcal{D}(\rho))$ in the interval $\tau_X$. 
On the other hand, the ``standard deviation'' of the change would approximately be given by $\sqrt{2\mathscr{D}_X\tau_X}$ once we admit $\mathscr{D}_X$ is a quantum extension of diffusivity. By definition of $\tau_X$, we have
\begin{align}
    \tau_X\big|\tr(X\mathcal{D}(\rho))\big| = \sqrt{2\mathscr{D}_X\tau_X}.
\end{align} 
Therefore, $\tau_X$ provides a typical time scale for $\langle X\rangle$ to change as much as its fluctuation. 

A similar notion was introduced in Ref.~\cite{mandelstam1945uncertainty} as
\begin{align}
    \tilde{\tau}_X \coloneqq \frac{\sigma_X}{|\tr(X[H,\rho])|}
\end{align}
to obtain the time-energy uncertainty relation (also known as the quantum speed limit (QSL)~\cite{pires2016generalized})
\begin{align}
    \tilde{\tau}_X\cdot\sigma_H\geq \frac{1}{2}, \label{eq:teu}
\end{align}
where $\sigma_X$ and $\sigma_H$ indicate the standard deviation, $\sigma_X = \sqrt{\langle X^2\rangle - \langle X\rangle^2}$ and $\sigma_H = \sqrt{\langle H^2\rangle - \langle H\rangle^2}$.
If there are no dissipators, the denominator $|\tr(X[H,\rho])|$ becomes $|d\langle X\rangle/dt|$ and $\tilde{\tau}_X$ provides the time it takes $\langle X\rangle$ to change as much as the standard deviation, similarly to $\tau_X$. 

As we have the time-energy uncertainty relation for $\tilde{\tau}_X$, we can get an inequality for $\tau_X$, 
\begin{align}
    \tau_X\cdot \eprEX(\rho)\geq 2, \label{eq:tdu}
\end{align}
which is derived by combining the definition of $\tau_X$ and the inequality in Eq.~\eqref{eq:turEX}. 
Since this inequality holds for any $X\in\herm(\hilb)$, we can further obtain the symbolic form 
\begin{align}
    \tau_*\cdot\eprEX(\rho)\geq 2
\end{align}
by defining the minimum time $\tau_*\coloneq\min_{X\in\herm(\hilb)}\tau_X$.
Those inequalities represent the universal trade-off relation between the typical time of the observable or the system itself and the excess component of the dissipation; namely, it shows the fact that changing faster requires more dissipation. 
It is reasonable that the excess component appears because it is related to the minimum dissipation, as discussed in Sec.~\ref{subsec:Decomposition}. 
We may regard the inequalities as a variety of the time-dissipation uncertainty relation~\cite{falasco2020dissipation}. 

While time-energy uncertainty relation (QSL)~\eqref{eq:teu} focuses on the unitary term in the dynamics (cf. Eq.~\eqref{eq:expTimeDerivative}), we have derived another uncertainty relation (QSL)~\eqref{eq:tdu} by focusing on the dissipative contribution.
Yet, it should be noted that neither of the denominators in the definitions of $\tau_X$ and $\tilde{\tau}_X$ unconditionally provide the time derivative. Thus, interpreting them as the time required to overcome the fluctuation is not always valid. 
Nonetheless, they can complement each other since the interpretations are exclusively reasonable; $d\langle X\rangle/dt\simeq \tr(X\mathcal{D}(\rho))$ holds if $-i\tr(X[H,\rho])$ is negligible, and vice versa.

\section{Examples}
\label{sec:Example}
\subsection{Two-level system attached to two heat baths}
\label{sec:tls}
\begin{figure}
    \centering
    \includegraphics[width=\linewidth]{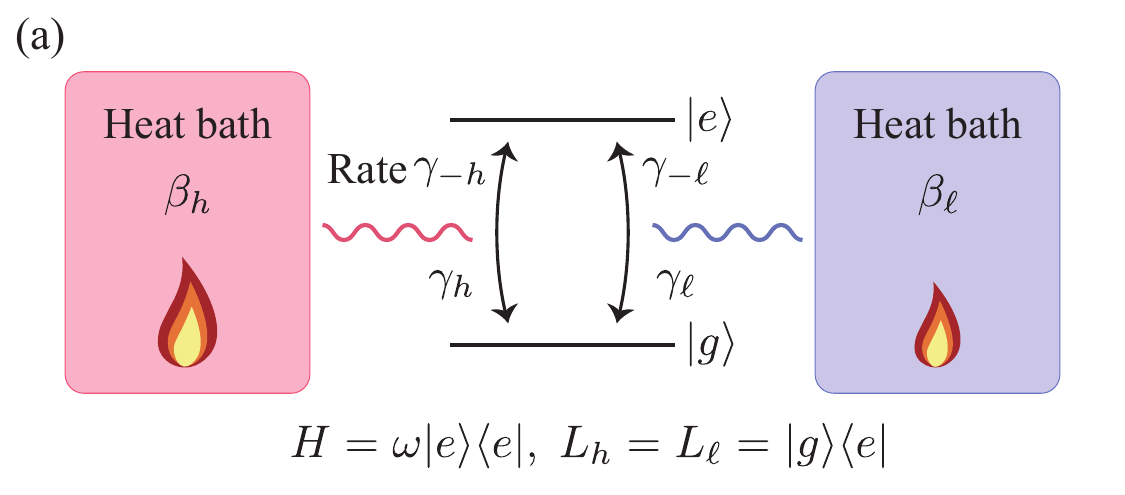}
    \includegraphics[width=\linewidth]{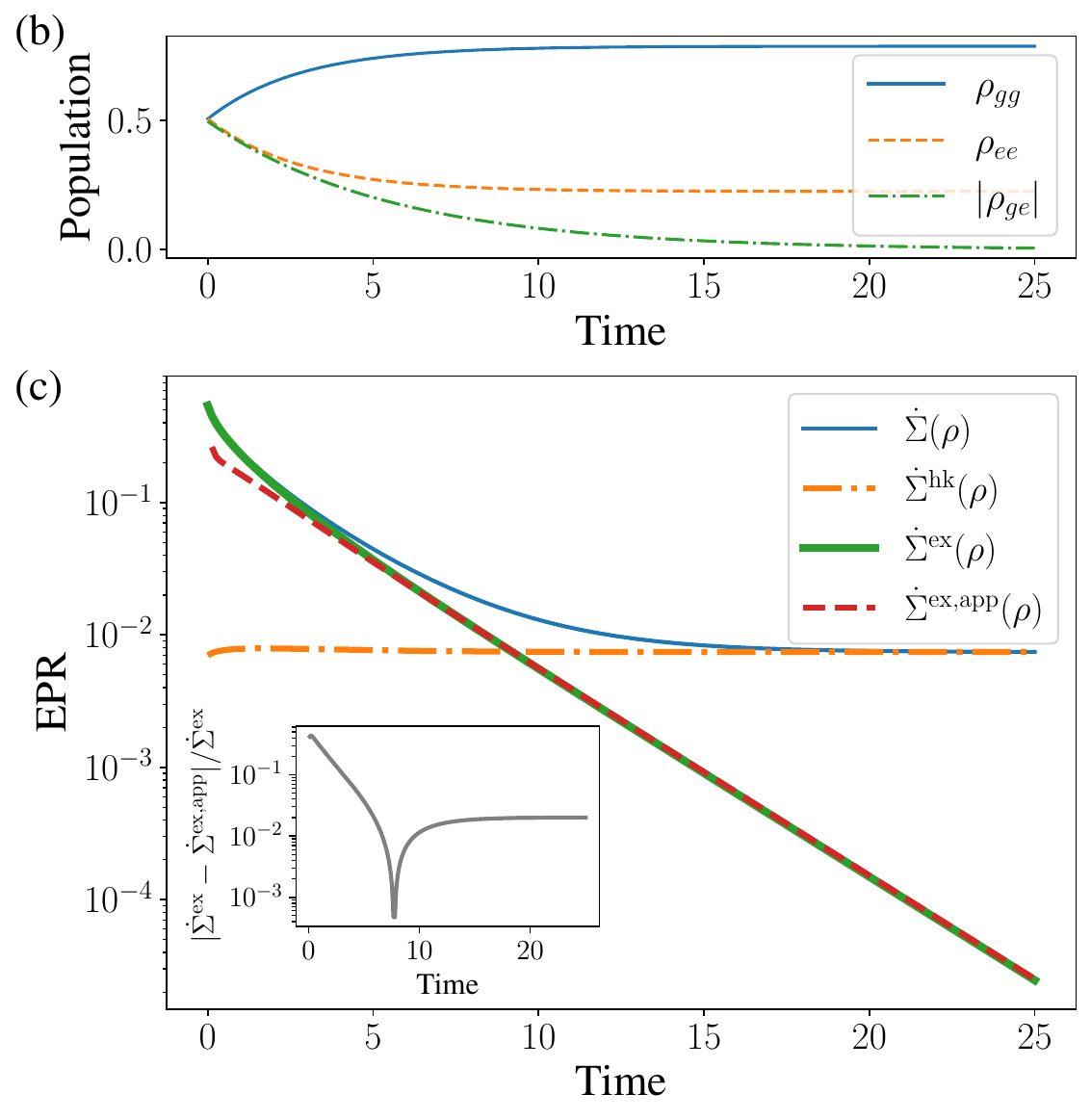}
    \caption{Summary of the results in the first example. (a) We simulate the dissipative dynamics of a two-level system coupled to two heat baths at different temperatures. (b) We begin with a pure state $(\ket{g}+\ket{e})/\sqrt{2}$ added a small noise to avoid divergence of $\ln\rho$. The coherence $\rho_{ge}$ vanished while the classical population converges to the stationary distribution. We do the simulation with parameters $\omega=1$, $\beta_h=0.58$, $\beta_\ell=1$, and $\gamma_+=\gamma_-=0.1$. (c) We plot the EPRs. The excess and housekeeping EPRs are computed by solving Eq.~\eqref{eq:GFE}.} As the system relaxes, the total and housekeeping EPRs converge to a single nonzero value, and the excess EPR becomes zero. The approximation formula reproduces the excess EPR well with a small error (shown in the inset). 
    \label{fig:example1}
\end{figure}
First, let us illustrate the general framework and the decomposition through a minimal model of the nonequilibrium quantum system.
Consider a two-level system that is equipped with Hamiltonian $H=\omega\ketbras{e}$ and attached to two heat baths at inverse temperatures $\beta_h$ and $\beta_\ell$ ($\beta_h<\beta_\ell$). 
The two ways of dissipation induced by the baths are represented by jump operators $L_h=L_\ell=\ket{g}\bra{e}$. 
The rates obey the local detailed balance 
\begin{align}
    \ln\frac{\gamma_h}{\gamma_{-h}}=\beta_h\omega,\quad
    \ln\frac{\gamma_\ell}{\gamma_{-\ell}}=\beta_\ell\omega.
\end{align}
The system possesses the steady state
\begin{align}
    \rho^\mathrm{ss}
    =\frac{\Gamma_+}{\Gamma_++\Gamma_-}\ketbras{g}+\frac{\Gamma_-}{\Gamma_++\Gamma_-}\ketbras{e} \label{eq:ex1ss}
\end{align}
where $\Gamma_{\pm}=\gamma_{\pm h}+\gamma_{\pm\ell}$. 
The situation is summarized in Fig.~\ref{fig:example1} (a). 

Now, the components of the force and current operators, defined in Eqs.~\eqref{eq:defCurr} and \eqref{eq:defForce}, read
\begin{align}
    J_k &= \frac{\gamma_k\rho_{ee}-\gamma_{-k}\rho_{gg}}{2}L_k+\frac{\rho_{eg}}{2}(\gamma_k\ketbras{g}-\gamma_{-k}\ketbras{e}), \label{eq:ex1J}\\
    F_k &= \beta_k\omega L_k + \big[L_k,\ln\rho\big]. 
\end{align}
for $k=h,\ell$. Here, $\rho_{ij}$ denotes the matrix elements of $\rho$ in basis $\{\ket{g},\ket{e}\}$. 
We see $J_k$ has a classical component proportional to $\gamma_k\rho_{ee}-\gamma_{-k}\rho_{gg}$ and a non-classical part depending on $\rho_{eg}$. 
Since $\beta_k$ depends on $k$, $F_k$ cannot be expressed as $[\phi,L_k]$ (thus, the system is non-conservative due to the temperature gap). 

For this simple case, we can analytically obtain the potential operator $\phi\in\herm(\hilb)$ such that $\force_\mathrm{c}(\rho)=-\nabla_\strc\phi$. 
Because $\nabla_\strc$ is insensible to the identity operator ($\nabla_\strc I_{\hilb}=0$), we can assume the form
\begin{align}
    \phi=\begin{pmatrix}
        0&\psi\\\psi^*&\Delta
    \end{pmatrix}, \label{eq:phi}
\end{align}
where $\Delta$ is a real number corresponding to the potential difference between energy eigenstates, while $\psi$ are non-classical terms. 
As discussed in Sec.~\ref{subsec:invariance}, such a $\phi$ is obtained by solving Eq.~\eqref{eq:GFE}. 
The excess and housekeeping EPRs can be computed by using the potential as $\eprHK=\lVert\force(\rho)-(-\nabla_{\strc}\phi)\rVert_\rho^2$ and $\eprEX=\lVert-\nabla_{\strc}\phi\rVert_\rho^2$.
For the present model, moreover, it is possible to solve the equation analytically. 
However, the exact form of the solution, given in Appendix~\ref{app:rigorous}, is so complicated that it is little tractable. 
Instead, here we consider some extreme cases. 

First, when the system has no coherence, $\rho_{ge}=\rho_{eg}=0$, we recover the classical result provided in Ref.~\cite{yoshimura2023housekeeping} (the sign is flipped since the ``positive'' direction of the jump is opposite),
\begin{align}
    \Delta = \Delta_\mathrm{cl}(\rho) \coloneqq 
    \frac{J_\mathrm{cl}(\rho)}{L_\mathrm{cl}(\rho)},\quad \psi = 0, \label{eq:phiclassical}
\end{align}
where
\begin{align}
    J_\mathrm{cl}(\rho)&=\sum_{k=h,\ell} (\gamma_{k}\rho_{ee}-\gamma_{-k}\rho_{gg}),\\
    L_\mathrm{cl}(\rho)&=\sum_{k=h,\ell}\frac{\gamma_{k}\rho_{ee}-\gamma_{-k}\rho_{gg}}{\ln(\gamma_{k}\rho_{ee}/(\gamma_{-k}\rho_{gg}))}. \label{eq:onsager}
\end{align}
In addition, if $\beta_h=\beta_\ell=\beta$, the potential difference will be $f_{ee}-f_{gg}$ with $f = \ln\rho-\ln\rho^\mathrm{ss}$; this is because the isothermality leads to
\begin{align*}
    \ln\frac{\gamma_{k}\rho_{ee}}{\gamma_{-k}\rho_{gg}}
    =\beta\omega+\ln\frac{\rho_{ee}}{\rho_{gg}}
\end{align*}
and $\rho^\mathrm{ss}_{gg}/\rho^\mathrm{ss}_{ee}=\Gamma_+/\Gamma_-=e^{\beta\omega}$. Thus, we get 
\begin{align}
    \frac{J_\mathrm{cl}(\rho)}{L_\mathrm{cl}(\rho)}
    =\beta\omega+\ln\frac{\rho_{ee}}{\rho_{gg}}
    =\ln\frac{\rho_{ee}}{\rho_{ee}^\mathrm{ss}}-\ln\frac{\rho_{gg}}{\rho_{gg}^\mathrm{ss}}. 
\end{align}

Second, when $\rho_{ge}$ is small and $\rho$ is close to steady state $\rho^\mathrm{ss}$, we obtain the approximation formulas
\begin{align}
    \tilde{\Delta} &= \Delta_\mathrm{cl}(\rho) + \frac{2A_\mathrm{cl}(u)}{\delta\rho}\left(\frac{1}{L_\mathrm{cl}(u)}-\frac{1}{L_\mathrm{cl}(\rho)}\right) |\rho_{ge}|^2, \label{eq:approx1}\\ 
    \tilde{\psi} &= \frac{A_\mathrm{cl}(u)}{L_\mathrm{cl}(u)}\rho_{ge}, \label{eq:approx2}
\end{align}
where $u= I_\hilb/2$ is the maximally mixed state, $\delta\rho = \rho_{gg}-\rho_{ee}$ is the population difference, and $A_\mathrm{cl}(\rho)=\sum_{k=h,\ell}(\gamma_k\rho_{ee}+\gamma_{-k}\rho_{gg})$ is the classical dynamical activity. 
With the approximated potential $\tilde{\phi}$ given by $\tilde{\Delta}$ and $\tilde{\psi}$, we can compute the approximated value $\dot{\Sigma}^\mathrm{ex,app}(\rho)$ by
\begin{align}
    \dot{\Sigma}^\mathrm{ex,app}(\rho)
    =\lVert-\nabla_{\strc}\tilde{\phi}\rVert_{\Gamma\otimes\rho}^2.
\end{align}

In Fig.~\ref{fig:example1} (b) and (c), we show numerical results of the two-level system demonstrating the accuracy of the approximation. 
The numerical results here and in the next example are obtained with the Python quantum toolbox, QuTiP~\cite{johansson2012qutip}. 
In the time evolution, the system approaches the nonequilibrium steady state where heat flows from the hot bath to the cold bath. 
At the same time, the total EPR approaches the value of the housekeeping EPR, and the decomposition splits the net dissipation into the two contributions. 
The housekeeping EPR is almost at a constant value corresponding to the stationary EPR, whereas the excess EPR is vanishing. We can see that the excess EPR is well approximated by $\dot{\Sigma}^\mathrm{ex,app}(\rho)$ even when $\rho$ is not so close to the steady state. 

\subsection{Relaxation of a superradiant system}
\label{subsec:TF}
\begin{figure}
    \centering
    \includegraphics[width=\linewidth]{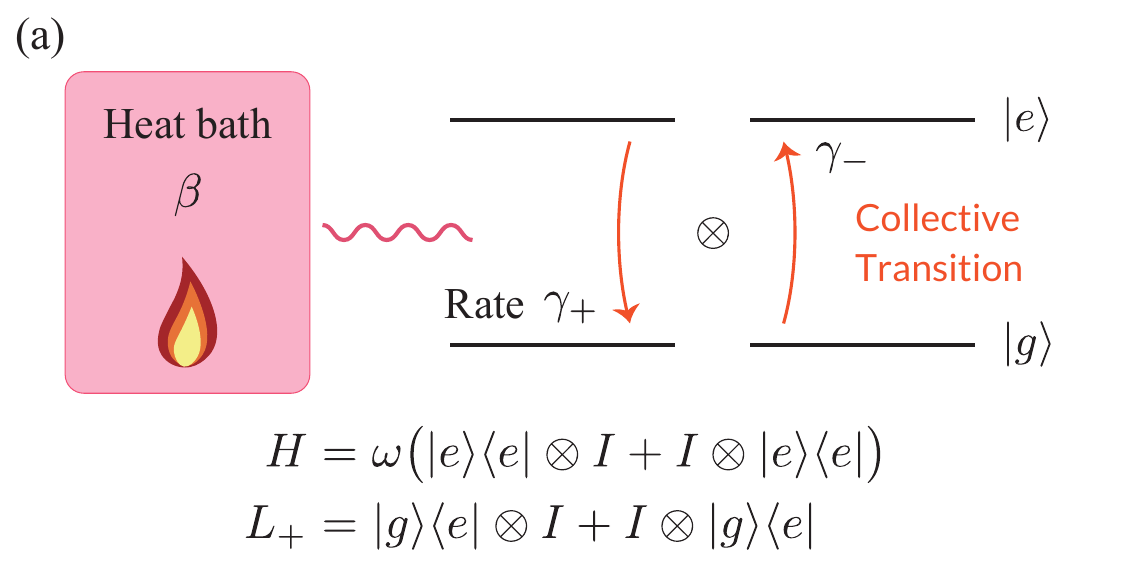}
    \includegraphics[width=\linewidth]{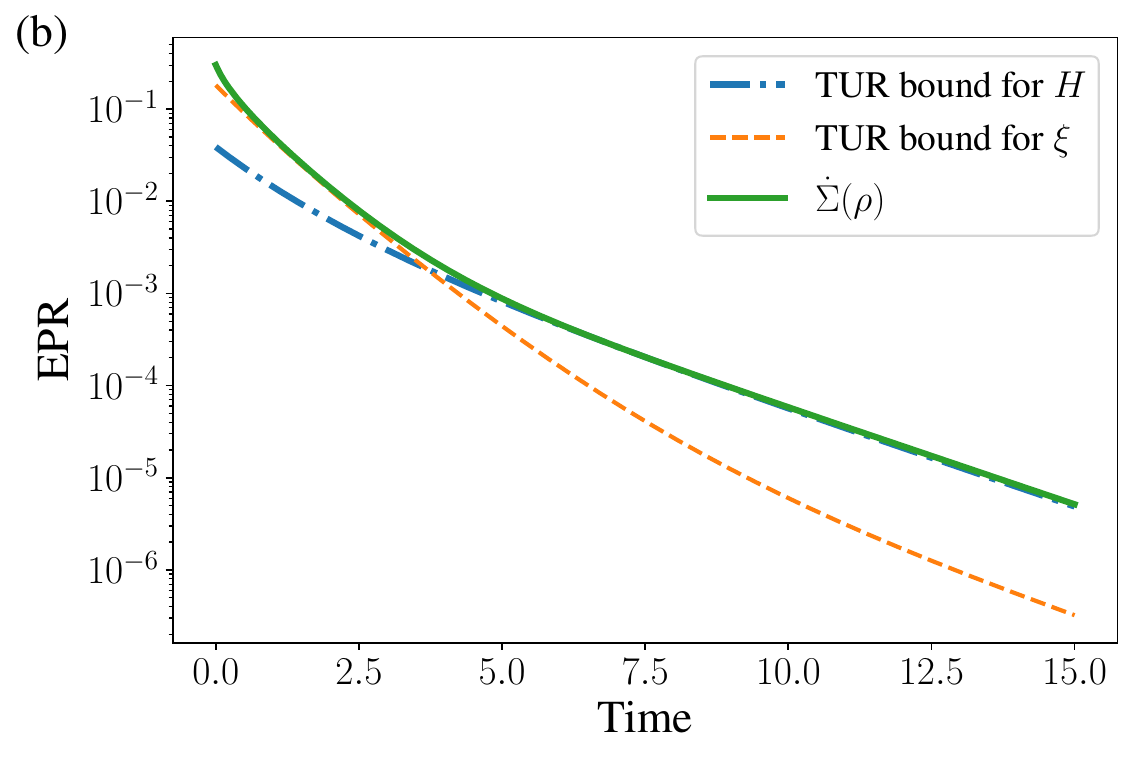}
    \caption{(a) We consider a model of superradiance, where two two-level systems are simultaneously influenced by a heat bath. We simulate the time evolution with parameters $\omega=1$, $\beta=1$, $\gamma_+=\Gamma(n_\mathrm{th}+1)$ and $\gamma_-=\Gamma n_\mathrm{th}$, where $\Gamma=0.1$ and $n_\mathrm{th}=1/(e^{\beta\omega}-1)$. (b) We can observe two stages of relaxation; in the earlier region (before $t\approx 3$), the TUR bound given by $\xi=\ket{ge}\bra{eg}+\ket{eg}\bra{ge}$ is tighter than that given by $H$, and the inverse holds in the later regime. }
    \label{fig:tur}
\end{figure}
Next, we check how the TUR works by analyzing the model of superradiance used in Ref.~\cite{tajima2021superconducting}. 
As depicted in Fig.~\ref{fig:tur} (a), the system consists of two two-level systems with Hamiltonian
\begin{align}
    H = \omega\big(\ketbras{e}\otimes I+I\otimes \ketbras{e}\big),
\end{align}
which has degeneracy; $\ket{g}\ket{e}$ and $\ket{e}\ket{g}$ both have the eigenvalue $\omega$. 
We also have a single collective dissipation ($|K|=1$) represented by
\begin{align}
    L_+&=\ket{g}\bra{e}\otimes I+I\otimes \ket{g}\bra{e}=L_-^\dagger. 
\end{align}
With the system coupled to a single bath at inverse temperature $\beta$, the local detailed balance $\gamma_+/\gamma_-=\exp(\beta\omega)$ is assumed. 
Considering this model, the authors of Ref.~\cite{tajima2021superconducting} exemplified that the coherence between the degenerate states enables the reduction of the dissipation beyond the classical limit.

We note that the steady state is not unique; the system has a conserved quantity in addition to the trace,
\begin{align}
    Q=\frac{1}{2}\big(\ket{ge}\bra{ge}+\ket{eg}\bra{eg}-\ket{ge}\bra{eg}-\ket{eg}\bra{ge}\big),
    \label{eq:qdef}
\end{align}
where $\ket{ge}=\ket{g}\ket{e}$ and $\ket{eg}=\ket{e}\ket{g}$. 
Namely, if initially $\tr(Q\rho_{t=0})=q$, we have $\tr(Q\rho_t)=q$ for any $t$. 
This is because $[H,Q]=0$ and $L_\pm Q=QL_{\pm}=0$, thus $(d/dt)\tr(Q\rho_t)=0$. 
As a result, we can expect that the system relaxes to the so-called generalized Gibbs ensemble~\cite{rigol2007relaxation}
\begin{align}
    \rho_{\beta,\lambda}=\frac{e^{-\beta H - \lambda Q}}{\tr(e^{-\beta H - \lambda Q})},
\end{align}
rather than the thermal equilibrium state $e^{-\beta H}/\tr(e^{-\beta H})$. 
Here, $\beta$ is the bath inverse temperature, and $\lambda$ is a parameter determined by the initial condition. 
In Appendix~\ref{app:convergence}, we show that (i) $\rho_{\beta,\lambda}$ is a steady state of the quantum master equation and (ii) $\tr(Q\rho_{\beta,\lambda})=q$ is solved uniquely by 
\begin{align}
    \lambda(\beta,q)=-\ln\left[\frac{q}{1-q}(e^{\beta\omega}+e^{-\beta\omega}+1)\right]. \label{eq:lambda}
\end{align} 
Although these facts do not directly mean that $\rho_t$ starting from $\rho_0$ such that $\tr(Q\rho_0)=q$ converges to $\rho_{\beta,\lambda(\beta,q )}$, 
we obtain numerical evidence of the convergence to $\rho_{\beta,\lambda}$ in the Appendix. 

In illustrating the TUR, we take an initial state that has significant coherence between $\ket{ge}$ and $\ket{eg}$ as
\begin{align}
    \rho_0 = \frac{e^{-\beta H}}{\tr(e^{-\beta H})} + \alpha\frac{e^{-\beta\omega}}{\tr(e^{-\beta H})} \xi
\end{align}
with $\xi=\ket{ge}\bra{eg}+\ket{eg}\bra{ge}$. 
To keep the positivity of $\rho_0$, $\alpha$ must satisfy $-1<\alpha <1$. 
We choose $\alpha = -0.9$ and plot the time course of the EPR and the TUR bounds~\eqref{eq:tur} for $X=H$ and $X=\xi$ in Fig.~\ref{fig:tur} (b) with parameter values described in the caption. 

In Fig.~\ref{fig:tur} (b), EPR's time evolution shows that there are two regimes of relaxation. 
We can estimate that the earlier stage is the decay of coherence, and the latter is the usual thermalization. 
In fact, in the earlier step, the TUR bound for $X=\xi$ is much tighter than that for $X=H$, while $H$'s bound gets closer to the EPR in the later step. 
Therefore, in the first stage, conflict with the ``fluctuation'' of $\xi$ is more crucial for the EPR, and later, the TUR associated with the energy will be more important. 
We expect that our general TUR bounds can generally provide limitations suited to the situation by properly choosing the observable. 

\subsection{Damped harmonic oscillator}
\label{sec:ex3}
Finally, we examine the quantum diffusivity in the damped harmonic oscillator. 
The system is described by Hamiltonian $H=\hbar\omega (a^\dagger a+1/2)$, jump operators $L_+=a=L_-^\dagger$ and decay rates $\gamma_+=\Gamma(\nth+1)$ and $\gamma_-=\Gamma\nth$, where $a$ is the annihilation operator, which satisfies $[a,a^\dagger]=1$, and $\nth=1/(e^{\beta\hbar\omega}-1)$~\cite{breuer2002theory}. 
Here, we recover the reduced Planck constant to clarify the dimension later. 
Even though the Hilbert space is not finite-dimensional, our theory can be applied to this system. 

We can compute the quantum diffusivity with the formula~\eqref{eq:QD2}, for example. 
Let us consider the position operator
\begin{align}
    Q = \sqrt{\frac{\hbar}{2m\omega}}(a+a^\dagger),
\end{align}
where $m$ is the mass of the particle. 
Because of the equalities
\begin{align}
    \mathcal{D}^*(Q^2)&=-\frac{\hbar\Gamma}{2m\omega}(a^2+(a^\dagger)^2+2a^\dagger a-2\nth), \label{eq:106}\\
    \{\mathcal{D}^*(Q),Q\}&=-\frac{\hbar\Gamma}{2m\omega}(a+a^\dagger)^2, \label{eq:107}
\end{align}
the quantum diffusivity is given as
\begin{align}
    \mathscr{D}_Q = \frac{\hbar\Gamma}{m\omega}\Big(\nth+\frac{1}{2}\Big) \label{eq:diffHO}
\end{align}
regardless of $\rho$ (the derivation of Eqs.~\eqref{eq:106} and \eqref{eq:107} are given in Appendix~\ref{app:HO}). 
It is easy to see that $\mathscr{D}_Q$ has the same dimension of $(\text{Length})^2/\text{Time}$ as the classical diffusion coefficient. 
Moreover,  $\mathscr{D}_Q$ has a direct connection to the equilibrium fluctuation of $Q$
\begin{align}
    (\sigma_Q^{\mathrm{eq}})^2
    &=\tr\big(Q^2\rhoeq\big)
    -\big(\tr\big(Q\rhoeq\big)\big)^2\\
    &=\frac{\hbar}{m\omega}\Big(\nth+\frac{1}{2}\Big).
\end{align}
Therefore, we obtain the equality 
\begin{align}
    \mathscr{D}_Q=\Gamma(\sigma_Q^{\mathrm{eq}})^2. \label{eq:diffandsigma}
\end{align}
Note that whereas the expression of $(\sigma_Q^{\mathrm{eq}})^2$ is only valid in equilibrium, the quantum diffusivity is always provided in the form of Eq.~\eqref{eq:diffHO}. 
This indicates that the quantum diffusivity is a more intrinsic measure of fluctuation. 
Still, we should be aware that $\mathscr{D}_X$ generally depends on the state $\rho$, and Eq.~\eqref{eq:diffandsigma} could be a specific result in this simple model; this is similar to the classical case, where the constant diffusivity is an idealization that is invalid in complex systems. 

We note that the quantum diffusivity may be seen as a generalization of the quantum diffusion coefficient investigated in the study of the damped quantum harmonic oscillators from the dynamical point of view~\cite{isar1994open}. 
Although the detailed comparison is a subject for future research, we emphasize that our diffusivity can be defined for any Hermitian operators in arbitrary open systems if they are described by the quantum master equation, not restricted to the coordinate and momentum of the harmonic oscillator.

\section{Conclusion}
We have studied the thermodynamics of the open quantum systems described by the quantum master equation to find the force-current structure, which ubiquitously exists in classical nonequilibrium systems. 
The force and current are provided as anti-Hermitian operators, accompanied by a consistent gradient structure characterized by the gradient super-operator. 
The developed framework constitutes a comprehensive analogy with the classical counterpart without resorting to any method that bluntly uses the information of eigenbases.

Not only does it have mathematical elegance, but the force-current structure also provides several physical insights. 
Its geometric structure lets us decompose the entropy production rate (EPR) into the housekeeping and excess parts. 
The geometric decomposition overcomes the problem 
of the conventional approach of decomposition, which requires a special steady state. 
Moreover, the geometric form of EPRs naturally leads to fundamental restrictions residing in the dynamics, such as the thermodynamic uncertainty relation (TUR) and the time-dissipation trade-off. 
Our inequalities generalize some previous results and enable us to discuss trade-offs in a generic manner. 

In deriving the TURs, we have found a quantity that we term the quantum diffusivity.
We have discussed that it can be a quantum measure of thermal fluctuation.
Revealing its more detailed theoretical properties should be an important future task; for example, it may be possible to connect the diffusivity to the variance of a short-time displacement. 

We would like to end this paper with some more comments on future directions: In this paper, we focused on the Euclidean geometry of forces. On the other hand, in classical systems, we can develop another type of theory that has more to do with optimization and the variational principle by using a non-Euclidean geometry called information geometry~\cite{kolchinsky2022information,kolchinsky2024generalized}. 
It would be an interesting avenue to consider the connection between thermodynamics and information geometry. 

In addition, we expect the thermodynamic uncertainty relations (TURs)~\eqref{eq:tur} and \eqref{eq:turEX} can be practically useful in examining infinite-dimensional systems. 
As discussed in Sec.~\ref{sec:ex3}, our theory does not require the finite dimension of the Hilbert space. 
However, the EPR is hard to compute in such systems because it is generally challenging to calculate $\ln\rho$~\cite{weiderpass2020neumann}. 
Although we should carefully confirm whether the bound can be tight or not, the TURs may be utilized to estimate the EPR in infinite-dimensional systems (as is actually done in classical systems~\cite{otsubo2020estimating,manikandan2020inferring}). 

\begin{acknowledgments}
K.Y.\ thanks Ken Hiura, Naruo Ohga, and Artemy Kolchinsky for their suggestive comments. 
K.Y.\ is supported by Grant-in-Aid for JSPS Fellows (Grant No.~22J21619). 
S.I.\ is supported by JSPS KAKENHI Grants No.~21H01560, No.~22H01141, No.~23H00467, and No.~24H00834, 
JST ERATO Grant No.~JPMJER2302, 
and UTEC-UTokyo FSI Research Grant Program. 
This research was supported by JSR Fellowship, the University of Tokyo.
\end{acknowledgments}

\appendix
\section{Mass action kinetics}
\label{app:mak}
Consider reactions
\begin{align}
    \sum_i \nu_{ei}^\mathrm{R}X_i\rightleftharpoons\sum_i \nu_{ei}^\mathrm{P}X_i
\end{align}
labeled by $e$. Here, $\nu_{ei}^\mathrm{R}$ and $\nu_{ei}^\mathrm{P}$ are non-negative integers representing the stoichiometry of reactants and products.
Then, the stoichiometric matrix is given by
\begin{align}
    \nabla_{ei}=\nu_{ei}^\mathrm{P}-\nu_{ei}^\mathrm{R}. 
\end{align}
Mass action kinetics is defined by
\begin{align}
    J_e^+=k_e^+ \prod_i x_i^{\nu_{ei}^\mathrm{R}},\quad
    J_e^-=k_e^- \prod_i x_i^{\nu_{ei}^\mathrm{P}},
\end{align}
where $k_e^\pm$ are constants. 
If it is assumed, the force reads
\begin{align}
    F_e = \ln\frac{k_e^+}{k_e^-}
    -\sum_{i}\nabla_{ei}\ln x_i. 
\end{align}
While $F$ is not generally written as an image of $\nabla$, if
\begin{align}
    \ln\frac{k_e^+}{k_e^-} = \sum_{i}\nabla_{ei}\psi_i
\end{align}
holds, $F=-\nabla\phi$ with $\phi = \ln x-\psi$ and $F$ vanishes at $\pi$ proportional to $e^{\psi}$. 

Markov jump processes are an example of this framework. 
A state change from $i$ to $j$ can be regarded as a reaction with $\nu_{le}^\mathrm{R}=\delta_{il}$ and $\nu_{le}^\mathrm{P}=\delta_{jl}$. 
The forward and backward fluxes read
\begin{align}
    J_e^+ = k_e^+ x_i,\quad J_e^- = k_e^- x_j,
\end{align}
where we can interpret $k_e^\pm$ as the transition rates of the jump. 
Since the dynamics conserves $\sum_i x_i=1$, the equilibrium state $\pi$ is given so that $\sum_i \pi_i = 1$. 

\section{Derivation of Eq.~\eqref{eq:52}}
\label{app:invariance}
We first compute
\begin{align}
    \lm_{\Gamma\otimes\rho}(\nabla_\strc X)
    =\int_0^1 ds(\Gamma\otimes\rho)^{s} [I_\hilbb\otimes X,\strc](\Gamma\otimes\rho)^{1-s}. 
\end{align}
Since
\begin{align}
    (\Gamma\otimes\rho)^s = \bigoplus_{k\in K}
    \begin{pmatrix}
        (\gamma_k/2)^s\rho^s & 0\\
        0 & (\gamma_{-k}/2)^s\rho^s
    \end{pmatrix}
\end{align}
and 
\begin{align}
    [I_\hilbb\otimes X,\strc]
    =\bigoplus_{k\in K}
    \begin{pmatrix}
        0 & [X,L_k^\dagger]\\
        [X,L_k] & 0
    \end{pmatrix},
\end{align}
we have
\begin{align}
    &(\Gamma\otimes\rho)^{s} [I_\hilbb\otimes X,\strc]\notag\\
    &= \bigoplus_{k\in K}
    \begin{pmatrix}
        0 & \Big(\frac{\gamma_k}{2}\Big)^s\rho^s[X,L_k^\dagger]\\
        \Big(\frac{\gamma_{-k}}{2}\Big)^s\rho^s[X,L_k] & 0
    \end{pmatrix}
\end{align}
and thus
\begin{align}
    &(\Gamma\otimes\rho)^{s} [I_\hilbb\otimes X,\strc](\Gamma\otimes\rho)^{1-s}\notag\\
    &= \bigoplus_{k\in K}
    \begin{pmatrix}
        0 & \frac{\gamma_k^s\gamma_{-k}^{1-s}}{2}\rho^s[X,L_k^\dagger]\rho^{1-s}\\
        \frac{\gamma_k^{1-s}\gamma_{-k}^{s}}{2}\rho^s[X,L_k]\rho^{1-s} & 0
    \end{pmatrix}.
\end{align}

By swapping the integral and the commutator, we have
\begin{align}
    &\nabla_\strc^*\lm_{\Gamma\otimes\rho}(\nabla_\strc X)\notag\\
    &=\int_0^1ds\tr_\hilbb
    \big[(\Gamma\otimes\rho)^{s} [I_\hilbb\otimes X,\strc](\Gamma\otimes\rho)^{1-s},\strc\big]. 
\end{align}
The integrand then reads
\begin{align}
    &\tr_\hilbb
    \big[(\Gamma\otimes\rho)^{s} [I_\hilbb\otimes X,\strc](\Gamma\otimes\rho)^{1-s},\strc\big]\notag\\
    &=\sum_{k\in K}\bigg(\frac{\gamma_k^s\gamma_{-k}^{1-s}}{2}\big[\rho^s[X,L_k^\dagger]\rho^{1-s},L_k\big]\notag\\
    &\phantom{=\sum_{k\in K}\bigg(}
    +\frac{\gamma_k^{1-s}\gamma_{-k}^{s}}{2}\big[\rho^s[X,L_k]\rho^{1-s},L_k^\dagger\big]\bigg).
\end{align}

\section{Derivation of Eq.~\eqref{eq:bound}}
\label{app:diffusion}
We first prove the equality
\begin{align}
    \mathscr{D}_X^k= 
    \langle \nabla_{\strc_k}X,\mathcal{A}_{\Gamma_k\otimes\rho}(\nabla_{\strc_k}X)\rangle \label{eq:DX2}
\end{align}
for any $X\in\herm(\hilb)$, where $\mathcal{A}_B(C)\coloneqq (BC+CB)/2$ is the quantum generalization of the multiplication of the arithmetic mean. 
Without loss of generality, we assume $|K|=1$ and do not write $k$ hereafter. 
The right-hand side of Eq.~\eqref{eq:DX2} is transformed as
\begin{align}
    &\langle \nabla_{\strc}X,\mathcal{A}_{\Gamma\otimes\rho}(\nabla_{\strc}X)\rangle\notag\\
    &=\frac{1}{2}\tr\big((\nabla_\strc X)^\dagger(\Gamma\otimes\rho)(\nabla_\strc X)\big)\notag\\
    &\phantom{=}+\frac{1}{2}\tr\big((\nabla_\strc X)^\dagger(\nabla_\strc X)(\Gamma\otimes\rho)\big)\\
    &=\tr((\nabla_\strc X)^\dagger(\nabla_\strc X)\Gamma\otimes\rho), 
\end{align}
where we used $(\nabla_\strc X)^\dagger=-\nabla_\strc X$ in the last line. 
We also have 
\begin{align}
    &(\nabla_\strc X)^\dagger(\nabla_\strc X)\notag\\
    &=-\begin{pmatrix}
        0 & [L^\dagger,X]\\
        [L,X] & 0
    \end{pmatrix}
    \begin{pmatrix}
        0 & [L^\dagger,X]\\
        [L,X] & 0
    \end{pmatrix}\\
    &= 
    \begin{pmatrix}
        [L^\dagger, X][X,L] & 0 \\
        0 & [X,L][L^\dagger, X]
    \end{pmatrix}.
\end{align}
Thus, 
\begin{align}
    &\langle \nabla_{\strc}X,\mathcal{A}_{\Gamma\otimes\rho}(\nabla_{\strc}X)\rangle\notag\\
    &=\frac{\gamma_+}{2}\tr_\hilb\big([L^\dagger, X][X,L]\rho\big)
    +\frac{\gamma_-}{2}\tr_\hilb\big([X,L][L^\dagger, X]\rho\big). \label{eq:C6}
\end{align}
Each term can be expanded as
\begin{align}
    &[L^\dagger, X][X,L]\notag\\
    &=L^\dagger X^2 L-\frac{1}{2}\{L^\dagger L,X^2\}
    +\frac{1}{2}\{L^\dagger L,X^2\}\notag\\
    &\phantom{=}-L^\dagger XL X-XL^\dagger X L
    +XL^\dagger L X\\
    &=L^\dagger X^2 L-\frac{1}{2}\{L^\dagger L,X^2\}\notag\\
    &\phantom{=}-X\Big(-\frac{1}{2}XL^\dagger L+L^\dagger X L -\frac{1}{2}L^\dagger LX\Big)\notag\\
    &\phantom{=}-\Big(-\frac{1}{2}L^\dagger LX+L^\dagger X L -\frac{1}{2}XL^\dagger L\Big)X\\
    &=L^\dagger X^2 L-\frac{1}{2}\{L^\dagger L,X^2\}\notag\\
    &\phantom{=}-\Big\{X,L^\dagger X L-\frac{1}{2}\{L^\dagger L,X\}\Big\}
\end{align}
and similarly, 
\begin{align}
    [X,L][L^\dagger, X]
    &=L X^2 L^\dagger-\frac{1}{2}\{L L^\dagger,X^2\}\notag\\
    &\phantom{=}-\Big\{X,L X L^\dagger-\frac{1}{2}\{L L^\dagger,X\}\Big\}.
\end{align}
Therefore, plugging them into Eq.~\eqref{eq:C6}, we obtain
\begin{align}
    &\langle \nabla_{\strc}X,\mathcal{A}_{\Gamma\otimes\rho}(\nabla_{\strc}X)\rangle\notag\\
    &=\frac{1}{2}\tr_\hilb\big((\mathcal{D}^*(X^2)-\{\mathcal{D}^*(X),X\})\rho\big),
\end{align}
which shows Eq.~\eqref{eq:DX2} even if $|K|>1$. 

Next, we show the general inequality for positive Hermitian operator $B$ and arbitrary operator $C$ 
\begin{align}
    \langle C,\lm_{B}(C)\rangle\leq \langle C,\mathcal{A}_{B}(C)\rangle. \label{eq-app:c12}
\end{align}
Choosing $B=\Gamma_k\otimes\rho$ and $C=\nabla_{\strc_k}X$ leads to the inequalities in Eq.~\eqref{eq:bound}. 
Equation~\eqref{eq-app:c12} is proved with the eigendecomposition $B=\sum_i b_i\ketbras{i}$; 
then, we have
\begin{align}
    \mathcal{A}_B(C)
    =\sum_i \frac{b_i+b_j}{2}\bra{i}C\ket{j}\ket{i}\bra{j} 
\end{align}
because
\begin{align}
    \bra{i}\frac{BC+CB}{2}\ket{j}
    =\frac{b_i\bra{i}C\ket{j}+\bra{i}C\ket{j}b_j}{2}. 
\end{align}
From the inequality between log mean and arithmetic mean
\begin{align}
    \frac{x-y}{\ln(x/y)}\leq \frac{x+y}{2}, \label{eq-app:logmeanineq}
\end{align}
Eq.~\eqref{eq-app:c12} follows as 
\begin{align}
    &\langle C,\lm_{B}(C)\rangle
    =\sum_{i,j}\frac{b_i-b_j}{\ln(b_i/b_j)}|\bra{i}C\ket{j}|^2\notag\\
    &\leq \sum_{i,j}\frac{b_i+b_j}{2}|\bra{i}C\ket{j}|^2
    =\langle C,\mathcal{A}_{B}(C)\rangle. 
\end{align}

In case, we leave proof of Eq.~\eqref{eq-app:logmeanineq}; using the Cauchy--Schwarz inequality, we see
\begin{align}
    (x-y)^2
    =\Big(\int_y^x\sqrt{s}\frac{1}{\sqrt{s}}ds\Big)^2\\
    \leq
    \int_y^x s ds\int_y^x \frac{1}{s} ds\\
    =\frac{x^2-y^2}{2}\ln\Big(\frac{x}{y}\Big). 
\end{align}
We obtain Eq.~\eqref{eq-app:logmeanineq} by dividing both sides by $(x-y)\ln(x/y)$, which is always positive because $x-y$ and $\ln(x/y)$ have the same sign.

\section{Derivation of Eq.~\eqref{eq:TF}}
\label{app:TF}
From the linearity, it is sufficient to show
\begin{align}
    \tr\Big[\rho\big(\tilde{\mathcal{D}}_k(H^2)-\{\tilde{\mathcal{D}}_k(H),H\}\big)\Big]
    =\gamma_k\omega_k^2\tr(L_k^\dagger L_k\rho), 
    \label{eq-app:TF}
\end{align}
where 
\begin{align}
    \tilde{\mathcal{D}}_k(X)=\gamma_k\Big(L_k^\dagger XL_k-\frac{1}{2}\{L_k^\dagger L_k,X\}\Big).
\end{align}
We find
\begin{align}
    &L_k^\dagger HL_k-\frac{1}{2}\{L_k^\dagger L_k,H\}\notag\\
    &=L_k^\dagger HL_k
    -\frac{1}{2}L_k^\dagger L_k H
    -\frac{1}{2}HL_k^\dagger L_k  \\
    &=-\frac{1}{2}L_k^\dagger [L_k,H]
    +\frac{1}{2}[L_k^\dagger,H]L_k\\
    &=-\omega_k L_k^\dagger L_k
\end{align}
and 
\begin{align}
    &L_k^\dagger H^2L_k-\frac{1}{2}\{L_k^\dagger L_k,H^2\}\notag\\
    &=L_k^\dagger H^2L_k
    -\frac{1}{2}L_k^\dagger L_k H^2 
    -\frac{1}{2}H^2L_k^\dagger L_k  \\
    &=-\frac{1}{2}L_k^\dagger [L_k,H^2]
    +\frac{1}{2}[L_k^\dagger,H^2]L_k \\
    &=-\frac{\omega_k}{2} L_k^\dagger L_k H
    -\frac{\omega_k}{2} L_k^\dagger HL_k \notag\\
    &\phantom{=}-\frac{\omega_k}{2} L_k^\dagger H L_k
    -\frac{\omega_k}{2} H L_k^\dagger L_k\\
    &=-\frac{\omega_k}{2}\{L_k^\dagger L_k,H\}
    -\omega_kL_k^\dagger HL_k,
\end{align}
where we used $[L_k^\dagger,H]=-[L_k,H]^\dagger = -\omega_k L_k^\dagger$ and
\begin{align}
    [L_k,H^2]=[L_k,H]H+H[L_k,H]=\omega_k \{L_k,H\}. 
\end{align}
Thus, we obtain
\begin{align}
    &\gamma_k^{-1}\big(\tilde{\mathcal{D}}_k(H^2)
    -\{\tilde{\mathcal{D}}_k(H),H\}\big)\notag\\
    &=-\frac{\omega_k}{2}\{L_k^\dagger L_k,H\}
    -\omega_kL_k^\dagger HL_k
    +\omega_k\{L_k^\dagger L_k,H\}\\
    &=\frac{\omega_k}{2}(L_k^\dagger L_k H+H L_k^\dagger L_k-2L_k^\dagger HL_k)\\
    &=\frac{\omega_k}{2}(L_k^\dagger [L_k,H]+[H, L_k^\dagger] L_k)\\
    &= \omega_k^2 L_k^\dagger L_k,
\end{align}
which shows Eq.~\eqref{eq-app:TF}. 

\section{Exact form of $\phi$ in Eq.~\eqref{eq:phi}}
\label{app:rigorous}
Let us solve the equation
\begin{align}
    -\nabla_\strc^*\lm_{\Gamma\otimes\rho}(\nabla_\strc\phi)
    =\mathcal{D}(\rho). \label{eq-app:GFE}
\end{align}
Since this equation involves a nontrivial super-operator, $\lm_{\Gamma\otimes\rho}$, turning it into an explicit linear equation for $\phi$ takes a lot of calculation, as below. 

In the current situation, the right-hand side reads
\begin{align}
    \mathcal{D}(\rho)
    &=\sum_{k=h,\ell}\bigg[\gamma_k\left(L_k\rho L_{k}^\dagger-\frac{1}{2}\{L_k^\dagger L_k,\rho\}\right)\notag\\
    &\phantom{=\sum_{k=h,\ell}}+\gamma_{-k}\left(L_{-k}\rho L_{-k}^\dagger-\frac{1}{2}\{L_{-k}^\dagger L_{-k},\rho\}\right)\bigg].
\end{align}
By the definition
\begin{align}
    L_h=L_\ell = \ket{g}\bra{e}
    =\mat{0&1\\0&0},
\end{align}
we obtain
\begin{align}
    &\mathcal{D}(\rho)\notag\\
    &=\Gamma_+\mat{\rho_{ee}&-\rho_{ge}/2\\-\rho_{eg}/2&-\rho_{ee}}
    +\Gamma_-\mat{-\rho_{gg}&-\rho_{ge}/2\\-\rho_{eg}/2&\rho_{gg}}, \label{eq-app:E-dissipator}
\end{align}
where $\Gamma_\pm\coloneqq \gamma_{\pm h}+\gamma_{\pm \ell}$ and the matrix representation is given in the energy eigenbasis. 

In computing the left-hand side of Eq.~\eqref{eq-app:GFE}, due to the separability, we first consider $\nabla_{\strc_k}^*\lm_{\Gamma_k\otimes\rho}(\nabla_{\strc_k}\phi)$ and sum up them. 
Let us assume the form
\begin{align}
    \phi=\mat{\phi_a&\phi_b^*\\\phi_b&0} \label{eq-app:phi}
\end{align}
with $\phi_a\in\mathbb{R}$ and $\phi_b\in\mathbb{C}$. 
Here, note that the parametrization is different from the main text; 
$\phi_a$ can be identified with $-\Delta$ in Eq.~\eqref{eq:phi} because adding or subtracting a matrix proportional to the identity to $\phi$ does not affect the value of $\nabla_\strc\phi$. 
Further, $\phi_b$ corresponds to $\psi^*$.
Consequently, we have 
\begin{align}
    &\nabla_{\strc_k}\phi
    =\mat{0 & \tilde{F}_{-k}\\ \tilde{F}_k & 0} \label{eq-app:E6}\\
    &\mathrm{with}\quad
    \tilde{F}_k=[\phi,L_k]=\mat{-\phi_b&\phi_a\\0&\phi_b},
\end{align}
where we can see that if we adopted $\psi$ instead of $\phi_b$, we have two complex conjugates, and that is why we parametrize the off-diagonal elements as in Eq.~\eqref{eq-app:phi}. Note that $\tilde{F}_{-k}=-\tilde{F}_k^\dagger$. 

Next, we compute $\lm_{\Gamma_k\otimes\rho}(\nabla_{\strc_k}\phi)$.
Because computing it analytically requires the eigenbasis of $\Gamma_k\otimes\rho$ (cf. Eq.~\eqref{eq:lmExplicit}), we suppose the eigendecomposition
\begin{align}
    \rho=p\ketbras{u}+(1-p)\ketbras{v}. 
\end{align}
The eigenvectors can be expressed as
\begin{align}
    \ket{u}=\mat{u_1\\u_2},\quad
    \ket{v}=\mat{-u_2^*\\u_1^*}
\end{align}
with numbers $u_1$ and $u_2$ such that $|u_1|^2+|u_2|^2=1$. 
$p$ and $1-p$ are the eigenvalues of $\rho$, so they must be in range $[0,1]$. 
Although the elements of $\ket{u}$ and $\ket{v}$ can be explicitly written down by the elements of $\rho$, we make them implicit and write just as $u_1$ and $u_2$ for simplicity. 
Without loss of generality, we can set $\ket{u}$ to the ground state $\ket{g}$ when $\rho$ is diagonal (i.e., $\rho_{ge}=\rho_{eg}=0$). 
Then, $\Gamma_k\otimes\rho$ has the following pairs of eigenvalue and eigenvector: 
\begin{equation}
    \begin{split}
        \lambda_1 = \frac{\gamma_k}{2}p
        &\quad\text{with}\quad
        \ket{\bm{u}_1}=\begin{pmatrix}
            1 \\ 0
        \end{pmatrix}
        \otimes\ket{u},\\
        \lambda_2 = \frac{\gamma_k}{2}(1-p)
        &\quad\text{with}\quad
        \ket{\bm{u}_2}=
        \begin{pmatrix}
            1 \\ 0
        \end{pmatrix}
        \otimes\ket{v},\\
        \lambda_3 = \frac{\gamma_{-k}}{2}p
        &\quad\text{with}\quad
        \ket{\bm{u}_3}=
        \begin{pmatrix}
            0 \\ 1
        \end{pmatrix}
        \otimes\ket{u},\\
        \lambda_4 = \frac{\gamma_{-k}}{2}(1-p)
        &\quad\text{with}\quad
        \ket{\bm{u}_4}=
        \begin{pmatrix}
            0 \\ 1
        \end{pmatrix}
        \otimes\ket{v}.
    \end{split}
\end{equation}
Therefore, $\lm_{\Gamma_k\otimes\rho}(\nabla_{\strc_k}\phi)$ is given by 
\begin{align}
    \sum_{i,j=1}^4
    \Lambda(\lambda_i,\lambda_j)
    \bra{\bm{u}_i}\nabla_{\strc_k}\phi
    \ket{\bm{u}_j}\ket{\bm{u}_i}\bra{\bm{u}_j}, \label{eq-app:sum1} 
\end{align}
where $\Lambda$ indicates the log mean. 

Since $\nabla_{\strc_k}\phi$ has only off-diagonal elements, we do not have to consider all pairs of $i,j$; because of Eq.~\eqref{eq-app:E6}, we have
\begin{equation}
    \begin{split}
        \nabla_{\strc_k}\phi\ket{\bm{u}_1}=\mat{0\\\tilde{F}_k\ket{u}},&\quad
        \nabla_{\strc_k}\phi\ket{\bm{u}_2}=\mat{0\\\tilde{F}_k\ket{v}},\\
        \nabla_{\strc_k}\phi\ket{\bm{u}_3}=\mat{\tilde{F}_{-k}\ket{u}\\0},&\quad
        \nabla_{\strc_k}\phi\ket{\bm{u}_4}=\mat{\tilde{F}_{-k}\ket{v}\\0}. 
    \end{split}
\end{equation}
Thus, we only need to take into account $(i,j)=(1,3),(2,3),(1,4),(2,4)$ and their reverses.
Let $\tilde{f}_{ij}^k$ denote $\bra{\bm{u}_i}\nabla_{\strc_k}\phi\ket{\bm{u}_j}$. 
Then, we have 
\begin{align}
    \tilde{f}_{31}^k
    &=-(\tilde{f}_{13}^k)^*
    =\bra{u}\tilde{F}_k\ket{u},\\
    \tilde{f}_{32}^k
    &=-(\tilde{f}_{23}^k)^*
    =\bra{u}\tilde{F}_k\ket{v},\\
    \tilde{f}_{41}^k
    &=-(\tilde{f}_{14}^k)^*
    =\bra{v}\tilde{F}_k\ket{u},\\
    \tilde{f}_{42}^k
    &=-(\tilde{f}_{24}^k)^*
    =\bra{v}\tilde{F}_k\ket{v}.
\end{align}
Further, the operator parts are given as
\begin{align}
    \ket{\bm{u}_3}\bra{\bm{u}_1}
    &=(\ket{\bm{u}_1}\bra{\bm{u}_3})^\dagger
    =\mat{0&0\\1&0}\otimes \ket{u}\bra{u},\\
    \ket{\bm{u}_3}\bra{\bm{u}_2}
    &=(\ket{\bm{u}_2}\bra{\bm{u}_3})^\dagger
    =\mat{0&0\\1&0}\otimes \ket{u}\bra{v},\\
    \ket{\bm{u}_4}\bra{\bm{u}_1}
    &=(\ket{\bm{u}_1}\bra{\bm{u}_4})^\dagger
    =\mat{0&0\\1&0}\otimes \ket{v}\bra{u},\\
    \ket{\bm{u}_4}\bra{\bm{u}_2}
    &=(\ket{\bm{u}_2}\bra{\bm{u}_4})^\dagger
    =\mat{0&0\\1&0}\otimes \ket{v}\bra{v}.
\end{align}
Therefore, the task now turns into computing
\begin{align}
    &\lm_{\Gamma_k\otimes\rho}(\nabla_{\strc_k}\phi)
    =\mat{0&\tilde{J}_{-k}\\\tilde{J}_k&0} \label{eq-app:E21}\\
    &\text{with}\quad
    \tilde{J}_k=-\tilde{J}_{-k}^\dagger=\sum_{x,y\in \{u,v\}}
    \Lambda_{xy}\bra{x}\tilde{F}_k\ket{y}
    \ket{x}\bra{y},
    \label{eq-app:sum2}
\end{align}
where $\Lambda_{xy}$ is defined by
\begin{align}
    \Lambda_{uu}
    &=\Lambda(\lambda_1,\lambda_3),\quad
    \Lambda_{uv}
    =\Lambda(\lambda_2,\lambda_3),\\
    \Lambda_{vu}
    &=\Lambda(\lambda_1,\lambda_4),\quad
    \Lambda_{vv}
    =\Lambda(\lambda_2,\lambda_4). 
\end{align}

Before examining each term in the summation of Eq.~\eqref{eq-app:sum2}, let us get closer to the final goal. 
Operating $\nabla_{\strc_k}^*$ to Eq.~\eqref{eq-app:E21} leads to
\begin{align}
    \nabla_{\strc_k}^*\lm_{\Gamma_k\otimes\rho}(\nabla_{\strc_k}\phi)
    &=[\tilde{J}_k,L_{-k}]+[\tilde{J}_{-k},L_{k}]\\
    &=[\tilde{J}_k,L_{k}^\dagger]-[\tilde{J}_{k}^\dagger,L_{k}]\\
    &=[\tilde{J}_k,L_{k}^\dagger]+[\tilde{J}_{k},L_{k}^\dagger]^\dagger.
\end{align}
If we write $\tilde{J}_k$ in the energy eigenbasis as
\begin{align}
    \tilde{J}_k
    =\mat{\tilde{J}_{11}^k&\tilde{J}_{12}^k\\
    \tilde{J}_{21}^k&\tilde{J}_{22}^k},
\end{align}
then we have
\begin{align}
    [\tilde{J}_k,L_{k}^\dagger]
    =\mat{\tilde{J}_{12}^k & 0\\
    \tilde{J}_{22}^k-\tilde{J}_{11}^k & -\tilde{J}_{12}^k},
\end{align}
and thus
\begin{align}
    \nabla_{\strc_k}^*\lm_{\Gamma_k\otimes\rho}(\nabla_{\strc_k}\phi)
    =\mat{\tilde{J}_{12}^k+(\tilde{J}_{12}^k)^* & (\tilde{J}_{22}^k)^*-(\tilde{J}_{11}^k)^*\\
    \tilde{J}_{22}^k-\tilde{J}_{11}^k & -\tilde{J}_{12}^k-(\tilde{J}_{12}^k)^*}. \label{eq-app:final0}
\end{align}
We are going to calculate $\tilde{J}_k$ and obtain a linear equation for $\phi_a$ and $\phi_b$ by comparing Eq.~\eqref{eq-app:final0} with $\mathcal{D}(\rho)$. 

After a straightforward calculation, we will obtain the following results: 
\begin{align}
    \Lambda_{uu}
    &=p\xi_k,\quad\Lambda_{uv}
    =\frac{1}{2}\eta_{k},\\
    \Lambda_{vu}
    &=\frac{1}{2}\eta_{-k},\quad\Lambda_{vv}=(1-p)\xi_k, 
\end{align}
where 
\begin{align}
    \xi_k
    &=\frac{1}{2}\frac{\gamma_k-\gamma_{-k}}{\ln(\gamma_k/\gamma_{-k})},\\
    \eta_{k}
    &=\frac{\gamma_k(1-p)-\gamma_{-k}p}{\ln(\gamma_k/\gamma_{-k})+\ln((1-p)/p)}.
\end{align}
Note that $\eta_{k}$ provides the ratio between the classical probability current and thermodynamic force (what is called the Onsager coefficient in Ref.~\cite{yoshimura2023housekeeping}). 
Further, because
\begin{align}
    \tilde{F}_k\ket{u}
    =\mat{\phi_a u_2-\phi_b u_1\\\phi_b u_2},\;\;
    \tilde{F}_k\ket{v}
    =\mat{\phi_a u_1^*+\phi_b u_2^*\\\phi_b u_1^*}, 
\end{align}
we find 
\begin{align}
    \bra{u}\tilde{F}_k\ket{u}
    &=u_1^*u_2\phi_a+(1-2|u_1|^2)\phi_b,\\
    \bra{u}\tilde{F}_k\ket{v}
    &=(u_1^*)^2\phi_a+2u_1^*u_2^*\phi_b,\\
    \bra{v}\tilde{F}_k\ket{u}
    &=-u_2^2\phi_a+2u_1u_2\phi_b,\\
    \bra{v}\tilde{F}_k\ket{v}
    &=-u_1^*u_2\phi_a-(1-2|u_1|^2)\phi_b,
\end{align}
where we used $|u_2|^2-|u_1|^2=1-2|u_1|^2$, which is because $|u_1|^2+|u_2|^2=1$. 
In addition, we can also obtain
\begin{align}
    \ket{u}\bra{u}
    =\mat{|u_1|^2 & u_1 u_2^* \\ u_1^* u_2 & |u_2|^2},&\;
    \ket{u}\bra{v}
    =\mat{-u_1 u_2 & u_1^2 \\ - u_2^2 & u_1 u_2},\notag\\
    \ket{v}\bra{u}
    =\mat{-u_1^* u_2^* & -(u_2^*)^2 \\ (u_1^*)^2 & u_1^* u_2^*},&\;
    \ket{v}\bra{v}
    =\mat{|u_2|^2 & - u_1 u_2^* \\ - u_1^* u_2 & |u_1|^2}. 
\end{align}

Finally, we calculate the summation in Eq.~\eqref{eq-app:sum2}. 
We can write down $\tilde{J}_k$ as
\begin{align}
    &\tilde{J}_k=p\xi_k\bra{u}\tilde{F}_k\ket{u}\ket{u}\bra{u}
    +\frac{1}{2}\eta_k\bra{u}\tilde{F}_k\ket{v}\ket{u}\bra{v}
    \notag\\
    &+\frac{1}{2}\eta_{-k}\bra{v}\tilde{F}_k\ket{u}\ket{v}\bra{u}
    +(1-p)\xi_k\bra{v}\tilde{F}_k\ket{v}\ketbras{v}. \label{eq-app:sum3}
\end{align}
Since we have the relations
\begin{align}
    \ket{u}\bra{u}
    &=I_\hilb-\ket{v}\bra{v},\\
    \bra{v}\tilde{F}_k\ket{v}
    &=-\bra{u}\tilde{F}_k\ket{u}, 
\end{align}
Eq.~\eqref{eq-app:sum3} can be simplified as
\begin{align}
    \tilde{J}_k&=\xi_k(pI_\hilb-\ket{v}\bra{v})\bra{u}\tilde{F}_k\ket{u}\notag\\
    &+\frac{1}{2}\eta_k\bra{u}\tilde{F}_k\ket{v}\ket{u}\bra{v}
    +\frac{1}{2}\eta_{-k}\bra{v}\tilde{F}_k\ket{u}\ket{v}\bra{u}. 
\end{align}
We only need to calculate $\tilde{J}_{11}^k, \tilde{J}_{12}^k$, and $\tilde{J}_{22}^k$, which are provided as
\begin{align}
    &\tilde{J}^{k}_{11}
    =\xi_k(p-|u_2|^2)(u_1^*u_2\phi_a+(1-2|u_1|^2)\phi_b)\notag\\
    &\phantom{=}+\frac{\eta_k}{2}(-u_1 u_2)((u_1^*)^2\phi_a+2u_1^*u_2^*\phi_b)\notag\\
    &\phantom{=}+\frac{\eta_{-k}}{2}(-u_1^* u_2^*)(-u_2^2\phi_a+2u_1u_2\phi_b)\\
    &=\Big[\xi_k(p-|u_2|^2)-\frac{\eta_k}{2}|u_1|^2+\frac{\eta_{-k}}{2}|u_2|^2\Big]u_1^*u_2\phi_a\notag\\
    &\phantom{=}+\Big[\xi_k(1-2|u_1|^2)(p-|u_2|^2)-(\eta_k+\eta_{-k})|u_1u_2|^2\Big]\phi_b,
\end{align}
\begin{align}
    &\tilde{J}_{12}^k
    =\xi_ku_1u_2^*(u_1^*u_2\phi_a+(1-2|u_1|^2)\phi_b)\notag\\
    &\phantom{=}+\frac{\eta_k}{2}u_1^2((u_1^*)^2\phi_a+2u_1^*u_2^*\phi_b)\notag\\
    &\phantom{=}+\frac{\eta_{-k}}{2}(-(u_2^*)^2)(-u_2^2\phi_a+2u_1u_2\phi_b)\\
    &=\Big[\xi_k|u_1u_2|^2+\frac{\eta_k}{2}|u_1|^4-\frac{\eta_{-k}}{2}|u_2|^4\Big]\phi_a\notag\\
    &\phantom{=}+\Big[\xi_k(1-2|u_1|^2)+\eta_k|u_1|^2-\eta_{-k}|u_2|^2\Big]u_1u_2^*\phi_b,
\end{align}
and
\begin{align}
    &\tilde{J}_{22}^k
    =\xi_k(p-|u_1|^2)(u_1^*u_2\phi_a+(1-2|u_1|^2)\phi_b)\\
    &\phantom{=}+\frac{\eta_k}{2}u_1 u_2((u_1^*)^2\phi_a+2u_1^*u_2^*\phi_b)\notag\\
    &\phantom{=}+\frac{\eta_{-k}}{2}u_1^* u_2^*(-u_2^2\phi_a+2u_1u_2\phi_b)\\
    &=\Big[\xi_k(p-|u_1|^2)+\frac{\eta_{k}}{2}|u_1|^2-\frac{\eta_{-k}}{2}|u_2|^2\Big]u_1^*u_2\phi_a\notag\\
    &+\Big[\xi_k(1-2|u_1|^2)(p-|u_1|^2)+(\eta_k+\eta_{-k})|u_1u_2|^2\Big]\phi_b. 
\end{align}

Therefore, the $(1,1)$- and $(2,1)$-elements in Eq.~\eqref{eq-app:final0} are 
\begin{align}
    &\tilde{J}_{12}^k+(\tilde{J}_{12}^k)^* 
    =\Big[2\xi_k|u_1u_2|^2+\eta_k|u_1|^4-\eta_{-k}|u_2|^4\Big]\phi_a\notag\\
    &\phantom{=}+2\Big[\xi_k(1-2|u_1|^2)+\eta_k|u_1|^2-\eta_{-k}|u_2|^2\Big]\mathop{\mathrm{Re}}(u_1u_2^*\phi_b)
\end{align}
and
\begin{align}
    &\tilde{J}_{22}^k-\tilde{J}_{11}^k
    \notag\\
    &=\Big[\xi_k(1-2|u_1|^2)+\eta_k|u_1|^2-\eta_{-k}|u_2|^2\Big]u_1^*u_2\phi_a\notag\\
    &+\Big[\xi_k(1-2|u_1|^2)^2+2(\eta_k+\eta_{-k})|u_1u_2|^2\Big]\phi_b,
\end{align}
where we used $|u_2|^2-|u_1|^2=1-2|u_1|^2$. 
The linear equation for $\phi$ is obtained by summing them up for $k=h,\ell$. 
To this end, we define
\begin{align}
    \Xi\coloneqq \xi_h+\xi_\ell,\quad
    \eta_\pm\coloneqq\eta_{\pm h}+\eta_{\pm \ell}. 
\end{align}
We remark that $\Xi = L_\mathrm{cl}(u)$ and, if $\rho_{eg}=\rho_{ge}=0$, $\eta_+ = L_{\mathrm{cl}}(\rho)$,
where $L_\mathrm{cl}$ is defined in Eq.~\eqref{eq:onsager} and $u=I_\hilb/2$ is the uniform distribution. 
Then, combining Eqs.~\eqref{eq-app:E-dissipator} and \eqref{eq-app:final0}, we finally obtain the equation
\begin{align}
    A\phi_a + 2\mathop{\mathrm{Re}}(B^*\phi_b)
    &=-J_\mathrm{cl}(\rho)\label{eq-app:lineq1}\\
    B\phi_a + D\phi_b &= \act_\mathrm{cl}(u)\rho_{eg}, \label{eq-app:lineq2}
\end{align}
where 
\begin{align}
    A &= 2\Xi|u_1u_2|^2+\eta_+|u_1|^4-\eta_{-}|u_2|^4,\\
    B &= \Big[\Xi(1-2|u_1|^2)+\eta_+|u_1|^2-\eta_{-}|u_2|^2\Big]u_1^*u_2,\\
    D &= \Xi(1-2|u_1|^2)^2+2(\eta_++\eta_{-})|u_1u_2|^2, 
\end{align}
$J_\mathrm{cl}(\rho)=\Gamma_+\rho_{ee}-\Gamma_-\rho_{gg}$, and $\act_\mathrm{cl}(\rho)=\Gamma_+\rho_{ee}+\Gamma_-\rho_{gg}$. 
We use bold-styled $\act_\mathrm{cl}$ to distinguish the dynamical activity from the coefficient $A$. 

Let us consider the phase of $B$ in more depth. It is determined by $u_1^*u_2$. Since the phase of $\ket{u}$ is arbitrary, we fix it so that $u_1$ is a positive real number. 
Moreover, the eigenvalue equation $\rho\ket{u}=p\ket{u}$ reads
\begin{align}
    \rho_{gg}u_1+\rho_{ge}u_2 = pu_1.
\end{align}
Because $\rho_{gg}u_1$ and $pu_1$ are real, $\rho_{ge}u_2$ must be real. 
Thus, $B^*\rho_{eg}$ is also real as it includes $u_1u_2^*\rho_{eg}$ (note $\rho_{eg}=\rho_{ge}^*$). 
Therefore, if we multiply $B^*$ to Eq.~\eqref{eq-app:lineq2}, we realize that $B^*\phi_b$ is a real number since $|B|^2\phi_a$, $D$, and $\act_\mathrm{cl}(u)B^*\rho_{eg}$ are real, which means $\mathop{\mathrm{Re}}(B^*\phi_b)=B^*\phi_b$ in Eq.~\eqref{eq-app:lineq1}.

Consequently, we find that the equation that we need to solve is
\begin{align}
    A\phi_a + 2B^*\phi_b
    &=-J_\mathrm{cl}(\rho)\label{eq-app:lineq11}\\
    B\phi_a + D\phi_b &= \act_\mathrm{cl}(u)\rho_{eg}. \label{eq-app:lineq22}
\end{align}
It is solved by
\begin{align}
    \phi_a 
    &=\frac{-DJ_\mathrm{cl}(\rho)-2\act_\mathrm{cl}(u)B^*\rho_{eg}}{AD-2|B|^2}\\
    \phi_b
    &=\frac{A\act_\mathrm{cl}(u)\rho_{eg}+BJ_\mathrm{cl}(\rho)}{AD-2|B|^2},
\end{align}
which provides an exact solution for Eq.~\eqref{eq-app:GFE}.

When $\rho_{ge}=0$, we have $u_1=1$ and $u_2=0$. 
Then, the coefficients read $A=L_\mathrm{cl}(\rho)$, $B=0$, and $D=L_\mathrm{cl}(u)$. 
As a result, we obtain the classical result [Eq.~\eqref{eq:phiclassical}]
\begin{align}
    \phi_a = -\frac{J_\mathrm{cl}(\rho)}{L_{\mathrm{cl}}(\rho)},\quad
    \phi_b = 0. 
\end{align}

Finally, we briefly examine how the potential is modified when there is a small coherence ($\varepsilon,\theta\in\mathbb{R}$)
\begin{align}
    \rho_{eg}=\varepsilon e^{i\theta}. 
\end{align}
Assuming the perturbation $p=\rho_{gg}+\epsilon_p$ and $u_1 = \sqrt{1 -\epsilon_u}$, we solve the eigenvalue equation
\begin{align}
    \mat{\rho_{gg}&\varepsilon e^{-i\theta}\\
    \varepsilon e^{i\theta}&1-\rho_{gg}}
    \mat{\sqrt{1-\epsilon_u}\\\sqrt{\epsilon_u}e^{i\theta}}
    =(\rho_{gg}+\epsilon_p)
    \mat{\sqrt{1-\epsilon_u}\\\sqrt{\epsilon_u}e^{i\theta}}
\end{align}
to obtain the following relations to the leading order
\begin{align}
    &\varepsilon\sqrt{\epsilon_u} = \epsilon_p \\
    &\varepsilon=(2\rho_{gg}-1)\sqrt{\epsilon_u}.
\end{align}
The second equation implies $\varepsilon$ and $2\rho_{gg}-1 = \rho_{gg}-\rho_{ee}$ have the same sign (this is because we chose $\theta$ such that $u_2e^{-i\theta}=\sqrt{\epsilon_u}$ is positive). Then, we obtain the relation
\begin{align}
    \sqrt{\epsilon_u}=\frac{\varepsilon}{\delta\rho},\quad
    \epsilon_p = \frac{\varepsilon^2}{\delta\rho},
\end{align}
where $\delta\rho\coloneqq\rho_{gg}-\rho_{ee}$. 
After a straightforward calculation, we find the relevant expansion coefficients as
\begin{align}
    A_0&= L_\mathrm{cl}(\rho)
    ,\quad D_0 = L_\mathrm{cl}(u),\\
    B_1 &= \frac{L_\mathrm{cl}(\rho)-L_\mathrm{cl}(u)}{\delta\rho}e^{i\theta},\\
    A_2 &= \eta' - \frac{2}{\delta\rho^2}(L_\mathrm{cl}(\rho)-L_\mathrm{cl}(u)),\\
    D_2 &= \frac{2}{\delta\rho^2}(L_\mathrm{cl}(\rho)+R(\rho)-2L_\mathrm{cl}(u)),
\end{align}
where $X_n$ expands a quantity $X$ as $X=\sum_n X_n\varepsilon^n$. 
Here, $\eta'$ is defined by
\begin{align}
    &\eta' \coloneqq \frac{1}{\rho_{gg}\rho_{ee}\delta\rho}
    \sum_k\frac{j_k(\rho)}{f_k(\rho)}
    \left(\frac{1}{f_k(\rho)}-\frac{2\rho_{gg}\rho_{ee}a_k(u)}{j_k(\rho)}\right)\\
    &\text{with}\quad f_k(\rho)=\ln\frac{\gamma_k\rho_{ee}}{\gamma_{-k}\rho_{gg}},\;\; j_k(\rho)=\gamma_k\rho_{ee}-\gamma_{-k}\rho_{gg},\notag\\
    &\phantom{\text{with}\quad}a_k(u)=(\gamma_k+\gamma_{-k})/2,
\end{align}
and it satisfies $\eta_+=L_\mathrm{cl}(\rho)+\eta'\varepsilon^2$~\footnote{Seemingly, $\eta'$ is ill-defined when $\gamma_k\rho_{ee}=\gamma_{-k}\rho_{gg}$; however, if $\gamma_k\rho_{ee}=\gamma_{-k}\rho_{gg}(1+r)$ with small $r$, it can be shown $1/f_k-2\rho_{ee}\rho_{gg}a_k/j_k= -1/2+r/4+o(r)$. }. 
In addition, $R(\rho)$ is $L_\mathrm{cl}(\rho')$ where $\rho'$ is $\rho$ with its diagonal elements swapped (i.e., $\rho'_{gg}=\rho_{ee}$ and $\rho_{ee}'=\rho_{gg}$). 
Consequently, we can expand the exact solution as $\phi_a=\phi_{a,0}+\varepsilon^2\phi_{a,2}+o(\varepsilon^2)$ and $\phi_b=\varepsilon\phi_{b,1}+o(\varepsilon)$ with
\begin{align}
    \phi_{a,0}
    &= -\frac{J_\mathrm{cl}(\rho)}{L_\mathrm{cl}(\rho)},\\
    \phi_{a,2}
    &= \frac{J_\mathrm{cl}(\rho)A_2}{A_0^2}
    -\frac{2B_1^*}{A_0D_0}\bigg(\frac{J_\mathrm{cl}(\rho)B_1}{A_0}+\act_\mathrm{cl}(u)e^{i\theta}\bigg),\\
    \phi_{b,1}
    &= \frac{\act_\mathrm{cl}(u)\rho_{eg}}{D_0}
    +\frac{J_\mathrm{cl}(\rho)B_1}{A_0D_0}.
\end{align}

Furthermore, we additionally assume $\rho$ is close to the steady state. $J_\mathrm{cl}(\rho)$ vanishes at the steady state, which can be confirmed by plugging Eq.~\eqref{eq:ex1ss} into the definition of $J_\mathrm{cl}(\rho)$. 
Thus, we can neglect second-order terms related to $J_\mathrm{cl}(\rho)$ to get the approximation formulas
\begin{align}
    \phi_a 
    &=-\frac{J_\mathrm{cl}(\rho)}{L_\mathrm{cl}(\rho)}
    -\frac{2B_1^*}{A_0D_0}\act_\mathrm{cl}(u)\rho_{eg}\\
    &=-\frac{J_\mathrm{cl}(\rho)}{L_\mathrm{cl}(\rho)}
    -\frac{1}{\delta\rho}\bigg(\frac{1}{L_\mathrm{cl}(u)}-\frac{1}{L_\mathrm{cl}(\rho)}\bigg)|\rho_{eg}|^2\\
    \phi_b
    &=\frac{\act_\mathrm{cl}(u)}{L_\mathrm{cl}(u)}\rho_{eg}.
\end{align}
We can recover Eqs.~\eqref{eq:approx1} and \eqref{eq:approx2} by $\tilde{\Delta}=-\phi_a$ and $\tilde{\psi}=\phi_b^*$. 

\section{Numerical evidence of convergence}
\label{app:convergence}

\begin{figure*}
    \centering
    \includegraphics[width=\linewidth]{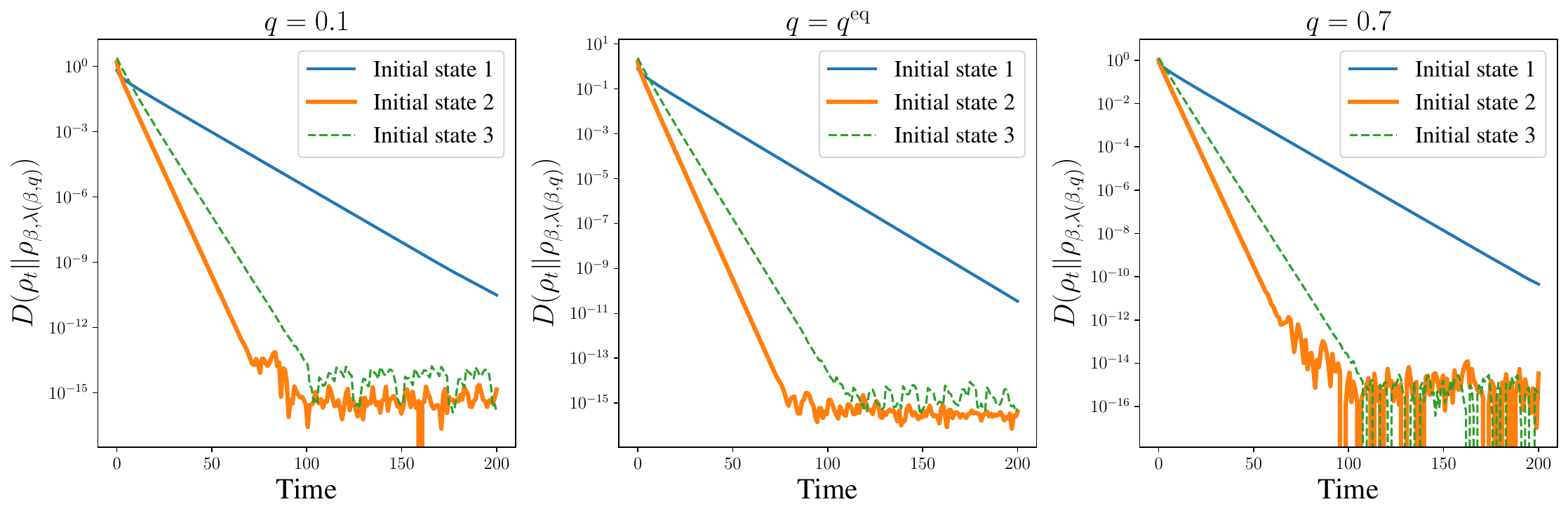}
    \caption{We compute the relative entropy between $\rho_t$ and the generalized Gibbs ensemble (GGE). 
    The panels correspond to three values of the conserved quantity $Q$ (defined in Eq.~\eqref{eq:qdef}): $\langle Q\rangle = 0.1,q^\mathrm{eq},$ and $0.7$, where $q^\mathrm{eq}=\tr(Qe^{-\beta H}/Z)\simeq 0.197$. 
    The GGE $\rho_{\beta,\lambda(\beta,q)}$ is chosen to have the same value of $q=\langle Q\rangle$ as the initial states. 
    For each value of $q$, we prepare three initial states that have the designated value of $q$ and respectively close to $\ket{gg}$, $\ket{+}=(\ket{ge}+\ket{eg})/\sqrt{2}$, or $\ket{ee}$. 
    We observe exponential decay to the corresponding GGE in any case (though the convergence speed is diverse). 
    The irregular fluctuations at $10^{-15}$ are caused by floating-point arithmetic.
    }
    \label{fig:convergence}
\end{figure*}

In Sec.~\ref{subsec:TF}, we discussed that the model has a conserved quantity $Q$ and the system does not generally converge to the thermal equilibrium $\rhoeq = e^{-\beta H}/Z$.
Rather, we suggested that the system relaxes to the generalized Gibbs ensemble (GGE)
\begin{align}
    \rho_{\beta,\lambda}=\frac{e^{-\beta H-\lambda Q}}{\tr(e^{-\beta H-\lambda Q})}
\end{align}
and 
\begin{align}
    \lambda(\beta,q)=-\ln\left[\frac{q}{1-q}(e^{\beta\omega}+e^{-\beta\omega}+1)\right] \label{eq-app:lambda}
\end{align}
is the appropriate value of $\lambda$ when the expectation value of $Q$ is $q$. 
In this section, we show analytically that $\rho_{\beta,\lambda}$ is a steady state, i.e., 
\begin{align}
    -i[H,\rho_{\beta,\lambda}]+\mathcal{D}(\rho_{\beta,\lambda})=0 \label{eq-app:GGEsteady}
\end{align}
holds, and $\lambda(\beta,q)$ is the unique solution to
\begin{align}
    \tr(Q\rho_{\beta,\lambda})=q, \label{eq-app:F1}
\end{align}
and numerically
\begin{align}
    \rho_t\to \rho_{\beta,\lambda(\beta, q)} \label{eq-app:conv}
\end{align}
if $\tr(Q\rho_0)=q$. 

First, we show Eq.~\eqref{eq-app:GGEsteady}. 
The first term vanishes because $Q$ commutes with $H$. 
This commutativity also implies $e^{-\beta H-\lambda Q}=e^{-\beta H}e^{-\lambda Q}$. In the series expansion of $e^{-\lambda Q}$, we only need to care about $1$ because the other terms, which includes $Q^n$, disappear on multiplied by $L_\pm$ in $\mathcal{D}$ ($L_{\pm}Q=QL_\pm = 0$). 
Therefore, all we need is to prove $\mathcal{D}(e^{-\beta H})=0$. 
It immediately follows from the fact that $H$ and $L_\pm$ satisfy the thermodynamic consistency conditions~\eqref{eq:TC} with a single temperature $\beta$, i.e., the system is detailed balanced (cf. Eq.~\eqref{eq:detailedbalance}). 

Next, to find the solution to Eq.~\eqref{eq-app:F1}, it is convenient to express operators in the energy eigenbasis. 
By definition, we have
\begin{align}
    \beta H + \lambda Q
    =\begin{pmatrix}
        0 &&&\\
        &\beta\omega +\lambda/2 & -\lambda/2 & \\
        &-\lambda/2 & \beta\omega +\lambda/2 & \\
        &&& 2\beta\omega
    \end{pmatrix}.
\end{align}
Then, it is a basic task to obtain the exponential 
\begin{align}
    &e^{-\beta H-\lambda Q}
    =\begin{pmatrix}
        1 &&&\\
        &0 && \\
        && 0 & \\
        &&& e^{-2\beta\omega}\\
    \end{pmatrix}\notag\\
    &+\frac{e^{-\beta\omega}}{2}
    \begin{pmatrix}
        0 &&&\\
        &1&1& \\
        &1&1& \\
        &&& 0
    \end{pmatrix}
    +\frac{e^{-\beta\omega-\lambda}}{2}
    \begin{pmatrix}
        0 &&&\\
        &1&-1& \\
        &-1&1& \\
        &&& 0
    \end{pmatrix}.
\end{align}
As a result, we see
\begin{align}
    \tr(e^{-\beta H-\lambda Q})
    =1+e^{-\beta\omega}+e^{-2\beta\omega}
    +e^{-\beta\omega-\lambda}
\end{align}
and
\begin{align}
    \tr(Qe^{-\beta H-\lambda Q})
    =e^{-\beta\omega-\lambda}. 
\end{align}
Therefore, we can obtain $\lambda$ such that
\begin{align}
    q = \tr(Q\rho_{\beta,\lambda})
\end{align}
by solving
\begin{align}
    e^{-\beta\omega-\lambda}
    =q(1+e^{-\beta\omega}+e^{-2\beta\omega}
    +e^{-\beta\omega-\lambda}).
\end{align}
It is a linear equation for $e^{-\lambda}$ and we can easily solve it and obtain Eq.~\eqref{eq-app:lambda}. 
This discussion means not only that $\lambda(\beta, q)$ satisfies Eq.~\eqref{eq-app:F1}, but also that it is the unique solution to $\tr(Q\rho_{\beta,\lambda})=q$. 

With the same parameters as Sec.~\ref{subsec:TF}, we numerically solve the quantum master equation under several conditions of initial conditions. 
The three panels of Fig.~\ref{fig:convergence} show the results for the three values of $q$: $0.1$, $q^\mathrm{eq}=\tr(Q\rhoeq)\simeq 0.197$, and $0.7$. 
For each $q$, we prepare three initial states in the following way:
First, we generate a pure state in the three ways:
\begin{align}
    \ket{\psi_1}&=\sqrt{1-c(q)^2}\ket{gg}+c(q)\ket{-}\\
    \ket{\psi_2}&=\sqrt{1-c(q)^2}\ket{+}+c(q)\ket{-}\\
    \ket{\psi_3}&=\sqrt{1-c(q)^2}\ket{ee}+c(q)\ket{-}
\end{align}
where $\ket{\pm}\coloneqq(\ket{ge}\pm\ket{eg})/\sqrt{2}$ are eigenstates of $Q$ ($Q\ket{+}=0$, $Q\ket{-}=\ket{-}$) and $c(q)$ is a positive coefficient to guarantee $\langle Q\rangle = q$, which is given soon later. 
We prepare $i$th initial state by adding a small noise to avoid the divergence of $\ln\rho$ as
\begin{align}
    \rho_0^{(i)}=(1-4\epsilon)\ketbras{\psi_i} + \epsilon I_\hilb. 
\end{align}
The initial condition $\tr(Q\rho_0^{(i)})=q$ is achieved if we define
\begin{align}
    c(q) = \sqrt{\frac{q-\epsilon}{1-4\epsilon}},
\end{align}
which is possible only if $\epsilon\leq q \leq 1-3\epsilon$; otherwise $\ket{\psi_i}$ will not be a proper state vector. 
If $q$ is small, the initial state $\rho_0^{(i)}$ becomes very close to one of the pure states, $\ket{gg}$, $\ket{+}$, and $\ket{ee}$. 

In Fig.~\ref{fig:convergence}, we compute the relative entropy between $\rho_t$ and the corresponding GGE $\rho_{\beta,\lambda}$
\begin{align}
    D(\rho_t\Vert\rho_{\beta,\lambda(\beta,q)})
    =\tr\big(\rho_t(\ln\rho_t-\ln\rho_{\beta,\lambda(\beta,q)})\big). 
\end{align}
In the simulation, we set $\epsilon = 10^{-3}$. 
We can observe the relative entropy decaying exponentially, which confirms the convergence~\eqref{eq-app:conv} because the relative entropy is zero only if the two states coincide. 

\section{Derivation of Eqs.~\eqref{eq:106} and \eqref{eq:107}}
\label{app:HO}
First, we show the basic relations
\begin{align}
    &\mathcal{D}^*(a)=-\frac{\Gamma}{2}a,\quad
    \mathcal{D}^*(a^\dagger)=-\frac{\Gamma}{2}a^\dagger,\\
    &\mathcal{D}^*(a^2)=-\Gamma a^2,\quad
    \mathcal{D}^*((a^\dagger)^2)=-\Gamma(a^\dagger)^2,\\
    &\mathcal{D}^*(a^\dagger a)=-\Gamma (a^\dagger a-\nth),\quad
    \mathcal{D}^*(1)=0. \label{eq-app:F3}
\end{align}
Because we have $(\mathcal{D}^*(X))^\dagger = \mathcal{D}^*(X^\dagger)$ for any $X\in\opr(\hilb)$, the second equations in the first and second lines follow from the first ones. 
$\mathcal{D}^*(a)=-\frac{\Gamma}{2}a$ and $\mathcal{D}^*(a^2)=-\Gamma a^2$ are shown as
\begin{align}
    &\mathcal{D}^*(a)\notag\\
    &=\gamma_+
    \Big(a^\dagger aa-\frac{1}{2}\{a^\dagger a,a\}\Big)
    +\gamma_-
    \Big(a aa^\dagger-\frac{1}{2}\{a a^\dagger,a\}\Big)\\
    &=\gamma_+
    \frac{a^\dagger a^2-aa^\dagger a}{2}
    +\gamma_-
    \frac{a^2a^\dagger-aa^\dagger a}{2}\\
    &=\gamma_+
    \frac{[a^\dagger,a]a}{2}
    +\gamma_-
    \frac{a[a,a^\dagger]}{2}\\
    &=\frac{1}{2}(\gamma_--\gamma_+)a
    =-\frac{\Gamma}{2}a
\end{align}
and 
\begin{align}
    &\mathcal{D}^*(a^2)\notag\\
    &=\gamma_+
    \Big(a^\dagger a^2a-\frac{1}{2}\{a^\dagger a,a^2\}\Big)
    +\gamma_-
    \Big(a a^2a^\dagger-\frac{1}{2}\{a a^\dagger,a^2\}\Big)\\
    &=\gamma_+
    \frac{a^\dagger a^3-a^2a^\dagger a}{2}
    +\gamma_-
    \frac{a^3a^\dagger-aa^\dagger a^2}{2}\\
    &=\gamma_+
    \frac{[a^\dagger,a^2]a}{2}
    +\gamma_-
    \frac{a[a^2,a^\dagger]}{2}\\
    &=(\gamma_--\gamma_+)a^2
    =-\Gamma a^2,
\end{align}
where we used
\begin{align}
    [a^2,a^\dagger] &= a^2a^\dagger -aa^\dagger a +aa^\dagger a - a^\dagger a^2\\
    &=a[a,a^\dagger]+[a,a^\dagger]a=2a. 
\end{align}
The second equality of Eq.~\eqref{eq-app:F3} is a standard fact of $\mathcal{D}^*$ that can be checked easily. 
The first one is proved as follows:
\begin{align}
    &\mathcal{D}^*(a^\dagger a)\notag\\
    &=\gamma_+
    \Big(a^\dagger a^\dagger aa-\frac{1}{2}\{a^\dagger a,a^\dagger a\}\Big)\notag\\
    &\phantom{=}
    +\gamma_-
    \Big(a a^\dagger aa^\dagger-\frac{1}{2}\{a a^\dagger,a^\dagger a\}\Big)\\
    &=\gamma_+
    a^\dagger [a^\dagger, a]a
    +\gamma_-\frac{aa^\dagger[a,a^\dagger]+[a,a^\dagger]aa^\dagger}{2}\\
    &=-\gamma_+
    a^\dagger a
    +\gamma_-
    aa^\dagger\\
    &=(\gamma_--\gamma_+)a^\dagger a + \gamma_-\\
    &=-\Gamma (a^\dagger a-\nth),
\end{align}
where we used $aa^\dagger = a^\dagger a + 1$. 

Thus, if we apply $\mathcal{D}^*$ to
\begin{align}
    Q&=\sqrt{\frac{\hbar}{2m\omega}}(a+a^\dagger),\\
    Q^2&=\frac{\hbar}{2m\omega}(a^2+(a^\dagger)^2+2a^\dagger a+1),
\end{align}
we will obtain Eqs.~\eqref{eq:106} and \eqref{eq:107}. 


\begin{thebibliography}{80}%
\makeatletter
\providecommand \@ifxundefined [1]{%
 \@ifx{#1\undefined}
}%
\providecommand \@ifnum [1]{%
 \ifnum #1\expandafter \@firstoftwo
 \else \expandafter \@secondoftwo
 \fi
}%
\providecommand \@ifx [1]{%
 \ifx #1\expandafter \@firstoftwo
 \else \expandafter \@secondoftwo
 \fi
}%
\providecommand \natexlab [1]{#1}%
\providecommand \enquote  [1]{``#1''}%
\providecommand \bibnamefont  [1]{#1}%
\providecommand \bibfnamefont [1]{#1}%
\providecommand \citenamefont [1]{#1}%
\providecommand \href@noop [0]{\@secondoftwo}%
\providecommand \href [0]{\begingroup \@sanitize@url \@href}%
\providecommand \@href[1]{\@@startlink{#1}\@@href}%
\providecommand \@@href[1]{\endgroup#1\@@endlink}%
\providecommand \@sanitize@url [0]{\catcode `\\12\catcode `\$12\catcode `\&12\catcode `\#12\catcode `\^12\catcode `\_12\catcode `\%12\relax}%
\providecommand \@@startlink[1]{}%
\providecommand \@@endlink[0]{}%
\providecommand \url  [0]{\begingroup\@sanitize@url \@url }%
\providecommand \@url [1]{\endgroup\@href {#1}{\urlprefix }}%
\providecommand \urlprefix  [0]{URL }%
\providecommand \Eprint [0]{\href }%
\providecommand \doibase [0]{https://doi.org/}%
\providecommand \selectlanguage [0]{\@gobble}%
\providecommand \bibinfo  [0]{\@secondoftwo}%
\providecommand \bibfield  [0]{\@secondoftwo}%
\providecommand \translation [1]{[#1]}%
\providecommand \BibitemOpen [0]{}%
\providecommand \bibitemStop [0]{}%
\providecommand \bibitemNoStop [0]{.\EOS\space}%
\providecommand \EOS [0]{\spacefactor3000\relax}%
\providecommand \BibitemShut  [1]{\csname bibitem#1\endcsname}%
\let\auto@bib@innerbib\@empty
\bibitem [{\citenamefont {de~Groot}\ and\ \citenamefont {Mazur}(1984)}]{de1962non}%
  \BibitemOpen
  \bibfield  {author} {\bibinfo {author} {\bibfnamefont {S.~R.}\ \bibnamefont {de~Groot}}\ and\ \bibinfo {author} {\bibfnamefont {P.}~\bibnamefont {Mazur}},\ }\href@noop {} {\emph {\bibinfo {title} {Non-equilibrium thermodynamics}}}\ (\bibinfo  {publisher} {Dover, New York},\ \bibinfo {year} {1984})\BibitemShut {NoStop}%
\bibitem [{\citenamefont {Breuer}\ and\ \citenamefont {Petruccione}(2002)}]{breuer2002theory}%
  \BibitemOpen
  \bibfield  {author} {\bibinfo {author} {\bibfnamefont {H.-P.}\ \bibnamefont {Breuer}}\ and\ \bibinfo {author} {\bibfnamefont {F.}~\bibnamefont {Petruccione}},\ }\href@noop {} {\emph {\bibinfo {title} {The theory of open quantum systems}}}\ (\bibinfo  {publisher} {Oxford University Press, New York},\ \bibinfo {year} {2002})\BibitemShut {NoStop}%
\bibitem [{\citenamefont {Strasberg}(2022)}]{strasberg2022quantum}%
  \BibitemOpen
  \bibfield  {author} {\bibinfo {author} {\bibfnamefont {P.}~\bibnamefont {Strasberg}},\ }\href@noop {} {\emph {\bibinfo {title} {Quantum Stochastic Thermodynamics: Foundations and Selected Applications}}}\ (\bibinfo  {publisher} {Oxford University Press},\ \bibinfo {year} {2022})\BibitemShut {NoStop}%
\bibitem [{\citenamefont {Seifert}(2012)}]{seifert2012stochastic}%
  \BibitemOpen
  \bibfield  {author} {\bibinfo {author} {\bibfnamefont {U.}~\bibnamefont {Seifert}},\ }\bibfield  {title} {\bibinfo {title} {Stochastic thermodynamics, fluctuation theorems and molecular machines},\ }\href {https://iopscience.iop.org/article/10.1088/0034-4885/75/12/126001/meta} {\bibfield  {journal} {\bibinfo  {journal} {Rep. Prog. Phys.}\ }\textbf {\bibinfo {volume} {75}},\ \bibinfo {pages} {126001} (\bibinfo {year} {2012})}\BibitemShut {NoStop}%
\bibitem [{\citenamefont {Shiraishi}(2023)}]{shiraishi2023introduction}%
  \BibitemOpen
  \bibfield  {author} {\bibinfo {author} {\bibfnamefont {N.}~\bibnamefont {Shiraishi}},\ }\href@noop {} {\emph {\bibinfo {title} {An Introduction to Stochastic Thermodynamics}}},\ \bibinfo {series} {Fundamental Theories of Physics}, Vol.\ \bibinfo {volume} {212}\ (\bibinfo  {publisher} {Springer, Singapore},\ \bibinfo {year} {2023})\BibitemShut {NoStop}%
\bibitem [{\citenamefont {Campisi}\ \emph {et~al.}(2011)\citenamefont {Campisi}, \citenamefont {H{\"a}nggi},\ and\ \citenamefont {Talkner}}]{campisi2011colloquium}%
  \BibitemOpen
  \bibfield  {author} {\bibinfo {author} {\bibfnamefont {M.}~\bibnamefont {Campisi}}, \bibinfo {author} {\bibfnamefont {P.}~\bibnamefont {H{\"a}nggi}},\ and\ \bibinfo {author} {\bibfnamefont {P.}~\bibnamefont {Talkner}},\ }\bibfield  {title} {\bibinfo {title} {Colloquium: Quantum fluctuation relations: Foundations and applications},\ }\href {https://journals.aps.org/rmp/abstract/10.1103/RevModPhys.83.771} {\bibfield  {journal} {\bibinfo  {journal} {Rev. Mod. Phys.}\ }\textbf {\bibinfo {volume} {83}},\ \bibinfo {pages} {771} (\bibinfo {year} {2011})}\BibitemShut {NoStop}%
\bibitem [{\citenamefont {Landi}\ and\ \citenamefont {Paternostro}(2021)}]{landi2021irreversible}%
  \BibitemOpen
  \bibfield  {author} {\bibinfo {author} {\bibfnamefont {G.~T.}\ \bibnamefont {Landi}}\ and\ \bibinfo {author} {\bibfnamefont {M.}~\bibnamefont {Paternostro}},\ }\bibfield  {title} {\bibinfo {title} {Irreversible entropy production: From classical to quantum},\ }\href {https://journals.aps.org/rmp/abstract/10.1103/RevModPhys.93.035008} {\bibfield  {journal} {\bibinfo  {journal} {Rev. Mod. Phys.}\ }\textbf {\bibinfo {volume} {93}},\ \bibinfo {pages} {035008} (\bibinfo {year} {2021})}\BibitemShut {NoStop}%
\bibitem [{\citenamefont {Jarzynski}(1997)}]{jarzynski1997nonequilibrium}%
  \BibitemOpen
  \bibfield  {author} {\bibinfo {author} {\bibfnamefont {C.}~\bibnamefont {Jarzynski}},\ }\bibfield  {title} {\bibinfo {title} {Nonequilibrium equality for free energy differences},\ }\href {https://journals.aps.org/prl/abstract/10.1103/PhysRevLett.78.2690} {\bibfield  {journal} {\bibinfo  {journal} {Phys. Rev. Lett.}\ }\textbf {\bibinfo {volume} {78}},\ \bibinfo {pages} {2690} (\bibinfo {year} {1997})}\BibitemShut {NoStop}%
\bibitem [{\citenamefont {Barato}\ and\ \citenamefont {Seifert}(2015)}]{barato2015thermodynamic}%
  \BibitemOpen
  \bibfield  {author} {\bibinfo {author} {\bibfnamefont {A.~C.}\ \bibnamefont {Barato}}\ and\ \bibinfo {author} {\bibfnamefont {U.}~\bibnamefont {Seifert}},\ }\bibfield  {title} {\bibinfo {title} {Thermodynamic uncertainty relation for biomolecular processes},\ }\href {https://journals.aps.org/prl/abstract/10.1103/PhysRevLett.114.158101} {\bibfield  {journal} {\bibinfo  {journal} {Phys. Rev. Lett.}\ }\textbf {\bibinfo {volume} {114}},\ \bibinfo {pages} {158101} (\bibinfo {year} {2015})}\BibitemShut {NoStop}%
\bibitem [{\citenamefont {Hasegawa}(2020)}]{hasegawa2020quantum}%
  \BibitemOpen
  \bibfield  {author} {\bibinfo {author} {\bibfnamefont {Y.}~\bibnamefont {Hasegawa}},\ }\bibfield  {title} {\bibinfo {title} {Quantum thermodynamic uncertainty relation for continuous measurement},\ }\href {https://journals.aps.org/prl/abstract/10.1103/PhysRevLett.125.050601} {\bibfield  {journal} {\bibinfo  {journal} {Phys. Rev. Lett.}\ }\textbf {\bibinfo {volume} {125}},\ \bibinfo {pages} {050601} (\bibinfo {year} {2020})}\BibitemShut {NoStop}%
\bibitem [{\citenamefont {Hasegawa}(2021)}]{hasegawa2021thermodynamic}%
  \BibitemOpen
  \bibfield  {author} {\bibinfo {author} {\bibfnamefont {Y.}~\bibnamefont {Hasegawa}},\ }\bibfield  {title} {\bibinfo {title} {Thermodynamic uncertainty relation for general open quantum systems},\ }\href {https://journals.aps.org/prl/abstract/10.1103/PhysRevLett.126.010602} {\bibfield  {journal} {\bibinfo  {journal} {Phys. Rev. Lett.}\ }\textbf {\bibinfo {volume} {126}},\ \bibinfo {pages} {010602} (\bibinfo {year} {2021})}\BibitemShut {NoStop}%
\bibitem [{\citenamefont {Onsager}(1931{\natexlab{a}})}]{onsager1931reciprocal1}%
  \BibitemOpen
  \bibfield  {author} {\bibinfo {author} {\bibfnamefont {L.}~\bibnamefont {Onsager}},\ }\bibfield  {title} {\bibinfo {title} {Reciprocal relations in irreversible processes. i.},\ }\href {https://journals.aps.org/pr/abstract/10.1103/PhysRev.37.405} {\bibfield  {journal} {\bibinfo  {journal} {Phys. Rev.}\ }\textbf {\bibinfo {volume} {37}},\ \bibinfo {pages} {405} (\bibinfo {year} {1931}{\natexlab{a}})}\BibitemShut {NoStop}%
\bibitem [{\citenamefont {Onsager}(1931{\natexlab{b}})}]{onsager1931reciprocal2}%
  \BibitemOpen
  \bibfield  {author} {\bibinfo {author} {\bibfnamefont {L.}~\bibnamefont {Onsager}},\ }\bibfield  {title} {\bibinfo {title} {Reciprocal relations in irreversible processes. ii.},\ }\href {https://journals.aps.org/pr/abstract/10.1103/PhysRev.38.2265} {\bibfield  {journal} {\bibinfo  {journal} {Phys. Rev.}\ }\textbf {\bibinfo {volume} {38}},\ \bibinfo {pages} {2265} (\bibinfo {year} {1931}{\natexlab{b}})}\BibitemShut {NoStop}%
\bibitem [{\citenamefont {Schnakenberg}(1976)}]{schnakenberg1976network}%
  \BibitemOpen
  \bibfield  {author} {\bibinfo {author} {\bibfnamefont {J.}~\bibnamefont {Schnakenberg}},\ }\bibfield  {title} {\bibinfo {title} {Network theory of microscopic and macroscopic behavior of master equation systems},\ }\href {https://journals.aps.org/rmp/abstract/10.1103/RevModPhys.48.571} {\bibfield  {journal} {\bibinfo  {journal} {Rev. Mod. Phys.}\ }\textbf {\bibinfo {volume} {48}},\ \bibinfo {pages} {571} (\bibinfo {year} {1976})}\BibitemShut {NoStop}%
\bibitem [{\citenamefont {Esposito}\ and\ \citenamefont {Mukamel}(2006)}]{esposito2006fluctuation}%
  \BibitemOpen
  \bibfield  {author} {\bibinfo {author} {\bibfnamefont {M.}~\bibnamefont {Esposito}}\ and\ \bibinfo {author} {\bibfnamefont {S.}~\bibnamefont {Mukamel}},\ }\bibfield  {title} {\bibinfo {title} {Fluctuation theorems for quantum master equations},\ }\href {https://journals.aps.org/pre/abstract/10.1103/PhysRevE.73.046129} {\bibfield  {journal} {\bibinfo  {journal} {Physical Review E—Statistical, Nonlinear, and Soft Matter Physics}\ }\textbf {\bibinfo {volume} {73}},\ \bibinfo {pages} {046129} (\bibinfo {year} {2006})}\BibitemShut {NoStop}%
\bibitem [{\citenamefont {Horowitz}(2012)}]{horowitz2012quantum}%
  \BibitemOpen
  \bibfield  {author} {\bibinfo {author} {\bibfnamefont {J.~M.}\ \bibnamefont {Horowitz}},\ }\bibfield  {title} {\bibinfo {title} {Quantum-trajectory approach to the stochastic thermodynamics of a forced harmonic oscillator},\ }\href {https://journals.aps.org/pre/abstract/10.1103/PhysRevE.85.031110} {\bibfield  {journal} {\bibinfo  {journal} {Physical Review E—Statistical, Nonlinear, and Soft Matter Physics}\ }\textbf {\bibinfo {volume} {85}},\ \bibinfo {pages} {031110} (\bibinfo {year} {2012})}\BibitemShut {NoStop}%
\bibitem [{\citenamefont {Horowitz}\ and\ \citenamefont {Parrondo}(2013)}]{horowitz2013entropy}%
  \BibitemOpen
  \bibfield  {author} {\bibinfo {author} {\bibfnamefont {J.~M.}\ \bibnamefont {Horowitz}}\ and\ \bibinfo {author} {\bibfnamefont {J.~M.}\ \bibnamefont {Parrondo}},\ }\bibfield  {title} {\bibinfo {title} {Entropy production along nonequilibrium quantum jump trajectories},\ }\href {https://iopscience.iop.org/article/10.1088/1367-2630/15/8/085028/meta} {\bibfield  {journal} {\bibinfo  {journal} {New J. Phys.}\ }\textbf {\bibinfo {volume} {15}},\ \bibinfo {pages} {085028} (\bibinfo {year} {2013})}\BibitemShut {NoStop}%
\bibitem [{\citenamefont {Funo}\ \emph {et~al.}(2019)\citenamefont {Funo}, \citenamefont {Shiraishi},\ and\ \citenamefont {Saito}}]{funo2019speed}%
  \BibitemOpen
  \bibfield  {author} {\bibinfo {author} {\bibfnamefont {K.}~\bibnamefont {Funo}}, \bibinfo {author} {\bibfnamefont {N.}~\bibnamefont {Shiraishi}},\ and\ \bibinfo {author} {\bibfnamefont {K.}~\bibnamefont {Saito}},\ }\bibfield  {title} {\bibinfo {title} {Speed limit for open quantum systems},\ }\href {https://iopscience.iop.org/article/10.1088/1367-2630/aaf9f5/meta} {\bibfield  {journal} {\bibinfo  {journal} {New J. Phys.}\ }\textbf {\bibinfo {volume} {21}},\ \bibinfo {pages} {013006} (\bibinfo {year} {2019})}\BibitemShut {NoStop}%
\bibitem [{\citenamefont {Gorini}\ \emph {et~al.}(1976)\citenamefont {Gorini}, \citenamefont {Kossakowski},\ and\ \citenamefont {Sudarshan}}]{gorini1976completely}%
  \BibitemOpen
  \bibfield  {author} {\bibinfo {author} {\bibfnamefont {V.}~\bibnamefont {Gorini}}, \bibinfo {author} {\bibfnamefont {A.}~\bibnamefont {Kossakowski}},\ and\ \bibinfo {author} {\bibfnamefont {E.~C.~G.}\ \bibnamefont {Sudarshan}},\ }\bibfield  {title} {\bibinfo {title} {Completely positive dynamical semigroups of n-level systems},\ }\href@noop {} {\bibfield  {journal} {\bibinfo  {journal} {Journal of Mathematical Physics}\ }\textbf {\bibinfo {volume} {17}},\ \bibinfo {pages} {821} (\bibinfo {year} {1976})}\BibitemShut {NoStop}%
\bibitem [{\citenamefont {Lindblad}(1976)}]{lindblad1976generators}%
  \BibitemOpen
  \bibfield  {author} {\bibinfo {author} {\bibfnamefont {G.}~\bibnamefont {Lindblad}},\ }\bibfield  {title} {\bibinfo {title} {On the generators of quantum dynamical semigroups},\ }\href {https://link.springer.com/article/10.1007/BF01608499} {\bibfield  {journal} {\bibinfo  {journal} {Commun. Math. Phys.}\ }\textbf {\bibinfo {volume} {48}},\ \bibinfo {pages} {119} (\bibinfo {year} {1976})}\BibitemShut {NoStop}%
\bibitem [{\citenamefont {Oono}\ and\ \citenamefont {Paniconi}(1998)}]{oono1998steady}%
  \BibitemOpen
  \bibfield  {author} {\bibinfo {author} {\bibfnamefont {Y.}~\bibnamefont {Oono}}\ and\ \bibinfo {author} {\bibfnamefont {M.}~\bibnamefont {Paniconi}},\ }\bibfield  {title} {\bibinfo {title} {Steady state thermodynamics},\ }\href {https://academic.oup.com/ptps/article-abstract/doi/10.1143/PTPS.130.29/1842398} {\bibfield  {journal} {\bibinfo  {journal} {Prog. Theor. Phys. Suppl.}\ }\textbf {\bibinfo {volume} {130}},\ \bibinfo {pages} {29} (\bibinfo {year} {1998})}\BibitemShut {NoStop}%
\bibitem [{\citenamefont {Hatano}\ and\ \citenamefont {Sasa}(2001)}]{hatano2001steady}%
  \BibitemOpen
  \bibfield  {author} {\bibinfo {author} {\bibfnamefont {T.}~\bibnamefont {Hatano}}\ and\ \bibinfo {author} {\bibfnamefont {S.-i.}\ \bibnamefont {Sasa}},\ }\bibfield  {title} {\bibinfo {title} {Steady-state thermodynamics of {Langevin} systems},\ }\href {https://journals.aps.org/prl/abstract/10.1103/PhysRevLett.86.3463} {\bibfield  {journal} {\bibinfo  {journal} {Phys. Rev. Lett.}\ }\textbf {\bibinfo {volume} {86}},\ \bibinfo {pages} {3463} (\bibinfo {year} {2001})}\BibitemShut {NoStop}%
\bibitem [{\citenamefont {Esposito}\ and\ \citenamefont {Van~den Broeck}(2010)}]{esposito2010three}%
  \BibitemOpen
  \bibfield  {author} {\bibinfo {author} {\bibfnamefont {M.}~\bibnamefont {Esposito}}\ and\ \bibinfo {author} {\bibfnamefont {C.}~\bibnamefont {Van~den Broeck}},\ }\bibfield  {title} {\bibinfo {title} {Three faces of the second law. {I}. master equation formulation},\ }\href {https://journals.aps.org/pre/abstract/10.1103/PhysRevE.82.011143} {\bibfield  {journal} {\bibinfo  {journal} {Phys. Rev. E}\ }\textbf {\bibinfo {volume} {82}},\ \bibinfo {pages} {011143} (\bibinfo {year} {2010})}\BibitemShut {NoStop}%
\bibitem [{\citenamefont {Spinney}\ and\ \citenamefont {Ford}(2012{\natexlab{a}})}]{spinney2012nonequilibrium}%
  \BibitemOpen
  \bibfield  {author} {\bibinfo {author} {\bibfnamefont {R.~E.}\ \bibnamefont {Spinney}}\ and\ \bibinfo {author} {\bibfnamefont {I.~J.}\ \bibnamefont {Ford}},\ }\bibfield  {title} {\bibinfo {title} {Nonequilibrium thermodynamics of stochastic systems with odd and even variables},\ }\href {https://journals.aps.org/prl/abstract/10.1103/PhysRevLett.108.170603} {\bibfield  {journal} {\bibinfo  {journal} {Phys. Rev. Lett.}\ }\textbf {\bibinfo {volume} {108}},\ \bibinfo {pages} {170603} (\bibinfo {year} {2012}{\natexlab{a}})}\BibitemShut {NoStop}%
\bibitem [{\citenamefont {Spinney}\ and\ \citenamefont {Ford}(2012{\natexlab{b}})}]{spinney2012entropy}%
  \BibitemOpen
  \bibfield  {author} {\bibinfo {author} {\bibfnamefont {R.~E.}\ \bibnamefont {Spinney}}\ and\ \bibinfo {author} {\bibfnamefont {I.~J.}\ \bibnamefont {Ford}},\ }\bibfield  {title} {\bibinfo {title} {Entropy production in full phase space for continuous stochastic dynamics},\ }\href {https://journals.aps.org/pre/abstract/10.1103/PhysRevE.85.051113} {\bibfield  {journal} {\bibinfo  {journal} {Phys. Rev. E}\ }\textbf {\bibinfo {volume} {85}},\ \bibinfo {pages} {051113} (\bibinfo {year} {2012}{\natexlab{b}})}\BibitemShut {NoStop}%
\bibitem [{\citenamefont {Ge}\ and\ \citenamefont {Qian}(2016)}]{ge2016nonequilibrium}%
  \BibitemOpen
  \bibfield  {author} {\bibinfo {author} {\bibfnamefont {H.}~\bibnamefont {Ge}}\ and\ \bibinfo {author} {\bibfnamefont {H.}~\bibnamefont {Qian}},\ }\bibfield  {title} {\bibinfo {title} {Nonequilibrium thermodynamic formalism of nonlinear chemical reaction systems with {Waage}--{Guldberg}'s law of mass action},\ }\href {https://www.sciencedirect.com/science/article/pii/S0301010416300866} {\bibfield  {journal} {\bibinfo  {journal} {Chem. Phys.}\ }\textbf {\bibinfo {volume} {472}},\ \bibinfo {pages} {241} (\bibinfo {year} {2016})}\BibitemShut {NoStop}%
\bibitem [{\citenamefont {Rao}\ and\ \citenamefont {Esposito}(2016)}]{rao2016nonequilibrium}%
  \BibitemOpen
  \bibfield  {author} {\bibinfo {author} {\bibfnamefont {R.}~\bibnamefont {Rao}}\ and\ \bibinfo {author} {\bibfnamefont {M.}~\bibnamefont {Esposito}},\ }\bibfield  {title} {\bibinfo {title} {Nonequilibrium thermodynamics of chemical reaction networks: wisdom from stochastic thermodynamics},\ }\href {https://journals.aps.org/prx/abstract/10.1103/PhysRevX.6.041064} {\bibfield  {journal} {\bibinfo  {journal} {Phys. Rev. X}\ }\textbf {\bibinfo {volume} {6}},\ \bibinfo {pages} {041064} (\bibinfo {year} {2016})}\BibitemShut {NoStop}%
\bibitem [{\citenamefont {Maes}\ and\ \citenamefont {Neto{\v{c}}n{\`y}}(2014)}]{maes2014nonequilibrium}%
  \BibitemOpen
  \bibfield  {author} {\bibinfo {author} {\bibfnamefont {C.}~\bibnamefont {Maes}}\ and\ \bibinfo {author} {\bibfnamefont {K.}~\bibnamefont {Neto{\v{c}}n{\`y}}},\ }\bibfield  {title} {\bibinfo {title} {A nonequilibrium extension of the {C}lausius heat theorem},\ }\href {https://link.springer.com/article/10.1007/s10955-013-0822-9} {\bibfield  {journal} {\bibinfo  {journal} {J. Stat. Phys.}\ }\textbf {\bibinfo {volume} {154}},\ \bibinfo {pages} {188} (\bibinfo {year} {2014})}\BibitemShut {NoStop}%
\bibitem [{\citenamefont {Dechant}\ \emph {et~al.}(2022{\natexlab{a}})\citenamefont {Dechant}, \citenamefont {Sasa},\ and\ \citenamefont {Ito}}]{dechant2022geometric1}%
  \BibitemOpen
  \bibfield  {author} {\bibinfo {author} {\bibfnamefont {A.}~\bibnamefont {Dechant}}, \bibinfo {author} {\bibfnamefont {S.-i.}\ \bibnamefont {Sasa}},\ and\ \bibinfo {author} {\bibfnamefont {S.}~\bibnamefont {Ito}},\ }\bibfield  {title} {\bibinfo {title} {Geometric decomposition of entropy production in out-of-equilibrium systems},\ }\href {https://journals.aps.org/prresearch/abstract/10.1103/PhysRevResearch.4.L012034} {\bibfield  {journal} {\bibinfo  {journal} {Phys. Rev. Res.}\ }\textbf {\bibinfo {volume} {4}},\ \bibinfo {pages} {L012034} (\bibinfo {year} {2022}{\natexlab{a}})}\BibitemShut {NoStop}%
\bibitem [{\citenamefont {Dechant}\ \emph {et~al.}(2022{\natexlab{b}})\citenamefont {Dechant}, \citenamefont {Sasa},\ and\ \citenamefont {Ito}}]{dechant2022geometric2}%
  \BibitemOpen
  \bibfield  {author} {\bibinfo {author} {\bibfnamefont {A.}~\bibnamefont {Dechant}}, \bibinfo {author} {\bibfnamefont {S.-i.}\ \bibnamefont {Sasa}},\ and\ \bibinfo {author} {\bibfnamefont {S.}~\bibnamefont {Ito}},\ }\bibfield  {title} {\bibinfo {title} {Geometric decomposition of entropy production into excess, housekeeping, and coupling parts},\ }\href {https://doi.org/10.1103/PhysRevE.106.024125} {\bibfield  {journal} {\bibinfo  {journal} {Phys. Rev. E}\ }\textbf {\bibinfo {volume} {106}},\ \bibinfo {pages} {024125} (\bibinfo {year} {2022}{\natexlab{b}})}\BibitemShut {NoStop}%
\bibitem [{\citenamefont {Yoshimura}\ \emph {et~al.}(2023)\citenamefont {Yoshimura}, \citenamefont {Kolchinsky}, \citenamefont {Dechant},\ and\ \citenamefont {Ito}}]{yoshimura2023housekeeping}%
  \BibitemOpen
  \bibfield  {author} {\bibinfo {author} {\bibfnamefont {K.}~\bibnamefont {Yoshimura}}, \bibinfo {author} {\bibfnamefont {A.}~\bibnamefont {Kolchinsky}}, \bibinfo {author} {\bibfnamefont {A.}~\bibnamefont {Dechant}},\ and\ \bibinfo {author} {\bibfnamefont {S.}~\bibnamefont {Ito}},\ }\bibfield  {title} {\bibinfo {title} {Housekeeping and excess entropy production for general nonlinear dynamics},\ }\href {https://journals.aps.org/prresearch/abstract/10.1103/PhysRevResearch.5.013017} {\bibfield  {journal} {\bibinfo  {journal} {Phys. Rev. Res.}\ }\textbf {\bibinfo {volume} {5}},\ \bibinfo {pages} {013017} (\bibinfo {year} {2023})}\BibitemShut {NoStop}%
\bibitem [{\citenamefont {Kolchinsky}\ \emph {et~al.}(2022)\citenamefont {Kolchinsky}, \citenamefont {Dechant}, \citenamefont {Yoshimura},\ and\ \citenamefont {Ito}}]{kolchinsky2022information}%
  \BibitemOpen
  \bibfield  {author} {\bibinfo {author} {\bibfnamefont {A.}~\bibnamefont {Kolchinsky}}, \bibinfo {author} {\bibfnamefont {A.}~\bibnamefont {Dechant}}, \bibinfo {author} {\bibfnamefont {K.}~\bibnamefont {Yoshimura}},\ and\ \bibinfo {author} {\bibfnamefont {S.}~\bibnamefont {Ito}},\ }\bibfield  {title} {\bibinfo {title} {Information geometry of excess and housekeeping entropy production},\ }\href {https://arxiv.org/abs/2206.14599} {\bibfield  {journal} {\bibinfo  {journal} {arXiv preprint arXiv:2206.14599}\ } (\bibinfo {year} {2022})}\BibitemShut {NoStop}%
\bibitem [{\citenamefont {Kolchinsky}\ \emph {et~al.}(2024)\citenamefont {Kolchinsky}, \citenamefont {Dechant}, \citenamefont {Yoshimura},\ and\ \citenamefont {Ito}}]{kolchinsky2024generalized}%
  \BibitemOpen
  \bibfield  {author} {\bibinfo {author} {\bibfnamefont {A.}~\bibnamefont {Kolchinsky}}, \bibinfo {author} {\bibfnamefont {A.}~\bibnamefont {Dechant}}, \bibinfo {author} {\bibfnamefont {K.}~\bibnamefont {Yoshimura}},\ and\ \bibinfo {author} {\bibfnamefont {S.}~\bibnamefont {Ito}},\ }\bibfield  {title} {\bibinfo {title} {Generalized free energy and excess entropy production for active systems},\ }\href {https://arxiv.org/abs/2412.08432} {\bibfield  {journal} {\bibinfo  {journal} {arXiv preprint arXiv:2412.08432}\ } (\bibinfo {year} {2024})}\BibitemShut {NoStop}%
\bibitem [{\citenamefont {Nagayama}\ \emph {et~al.}(2023)\citenamefont {Nagayama}, \citenamefont {Yoshimura}, \citenamefont {Kolchinsky},\ and\ \citenamefont {Ito}}]{nagayama2023geometric}%
  \BibitemOpen
  \bibfield  {author} {\bibinfo {author} {\bibfnamefont {R.}~\bibnamefont {Nagayama}}, \bibinfo {author} {\bibfnamefont {K.}~\bibnamefont {Yoshimura}}, \bibinfo {author} {\bibfnamefont {A.}~\bibnamefont {Kolchinsky}},\ and\ \bibinfo {author} {\bibfnamefont {S.}~\bibnamefont {Ito}},\ }\bibfield  {title} {\bibinfo {title} {Geometric thermodynamics of reaction-diffusion systems: Thermodynamic trade-off relations and optimal transport for pattern formation},\ }\href {https://arxiv.org/abs/2311.16569} {\bibfield  {journal} {\bibinfo  {journal} {arXiv preprint arXiv:2311.16569}\ } (\bibinfo {year} {2023})}\BibitemShut {NoStop}%
\bibitem [{\citenamefont {Yoshimura}\ and\ \citenamefont {Ito}(2024)}]{yoshimura2024two}%
  \BibitemOpen
  \bibfield  {author} {\bibinfo {author} {\bibfnamefont {K.}~\bibnamefont {Yoshimura}}\ and\ \bibinfo {author} {\bibfnamefont {S.}~\bibnamefont {Ito}},\ }\bibfield  {title} {\bibinfo {title} {Two applications of stochastic thermodynamics to hydrodynamics},\ }\href {https://journals.aps.org/prresearch/abstract/10.1103/PhysRevResearch.6.L022057} {\bibfield  {journal} {\bibinfo  {journal} {Phys. Rev. Res.}\ }\textbf {\bibinfo {volume} {6}},\ \bibinfo {pages} {L022057} (\bibinfo {year} {2024})}\BibitemShut {NoStop}%
\bibitem [{\citenamefont {Ito}(2023)}]{ito2023geometric}%
  \BibitemOpen
  \bibfield  {author} {\bibinfo {author} {\bibfnamefont {S.}~\bibnamefont {Ito}},\ }\bibfield  {title} {\bibinfo {title} {Geometric thermodynamics for the {Fokker}--{Planck} equation: stochastic thermodynamic links between information geometry and optimal transport},\ }\href {https://link.springer.com/article/10.1007/s41884-023-00102-3} {\bibfield  {journal} {\bibinfo  {journal} {Information Geometry}\ ,\ \bibinfo {pages} {1}} (\bibinfo {year} {2023})}\BibitemShut {NoStop}%
\bibitem [{\citenamefont {Kobayashi}\ \emph {et~al.}(2022)\citenamefont {Kobayashi}, \citenamefont {Loutchko}, \citenamefont {Kamimura},\ and\ \citenamefont {Sughiyama}}]{kobayashi2022hessian}%
  \BibitemOpen
  \bibfield  {author} {\bibinfo {author} {\bibfnamefont {T.~J.}\ \bibnamefont {Kobayashi}}, \bibinfo {author} {\bibfnamefont {D.}~\bibnamefont {Loutchko}}, \bibinfo {author} {\bibfnamefont {A.}~\bibnamefont {Kamimura}},\ and\ \bibinfo {author} {\bibfnamefont {Y.}~\bibnamefont {Sughiyama}},\ }\bibfield  {title} {\bibinfo {title} {Hessian geometry of nonequilibrium chemical reaction networks and entropy production decompositions},\ }\href {https://journals.aps.org/prresearch/abstract/10.1103/PhysRevResearch.4.033208} {\bibfield  {journal} {\bibinfo  {journal} {Phys. Rev. Res.}\ }\textbf {\bibinfo {volume} {4}},\ \bibinfo {pages} {033208} (\bibinfo {year} {2022})}\BibitemShut {NoStop}%
\bibitem [{\citenamefont {Horowitz}\ and\ \citenamefont {Sagawa}(2014)}]{horowitz2014equivalent}%
  \BibitemOpen
  \bibfield  {author} {\bibinfo {author} {\bibfnamefont {J.~M.}\ \bibnamefont {Horowitz}}\ and\ \bibinfo {author} {\bibfnamefont {T.}~\bibnamefont {Sagawa}},\ }\bibfield  {title} {\bibinfo {title} {Equivalent definitions of the quantum nonadiabatic entropy production},\ }\href {https://link.springer.com/article/10.1007/s10955-014-0991-1} {\bibfield  {journal} {\bibinfo  {journal} {Journal of Statistical Physics}\ }\textbf {\bibinfo {volume} {156}},\ \bibinfo {pages} {55} (\bibinfo {year} {2014})}\BibitemShut {NoStop}%
\bibitem [{\citenamefont {Manzano}\ \emph {et~al.}(2018)\citenamefont {Manzano}, \citenamefont {Horowitz},\ and\ \citenamefont {Parrondo}}]{manzano2018quantum}%
  \BibitemOpen
  \bibfield  {author} {\bibinfo {author} {\bibfnamefont {G.}~\bibnamefont {Manzano}}, \bibinfo {author} {\bibfnamefont {J.~M.}\ \bibnamefont {Horowitz}},\ and\ \bibinfo {author} {\bibfnamefont {J.~M.}\ \bibnamefont {Parrondo}},\ }\bibfield  {title} {\bibinfo {title} {Quantum fluctuation theorems for arbitrary environments: adiabatic and nonadiabatic entropy production},\ }\href {https://journals.aps.org/prx/abstract/10.1103/PhysRevX.8.031037} {\bibfield  {journal} {\bibinfo  {journal} {Phys. Rev. X}\ }\textbf {\bibinfo {volume} {8}},\ \bibinfo {pages} {031037} (\bibinfo {year} {2018})}\BibitemShut {NoStop}%
\bibitem [{\citenamefont {Shiraishi}\ \emph {et~al.}(2018)\citenamefont {Shiraishi}, \citenamefont {Funo},\ and\ \citenamefont {Saito}}]{shiraishi2018speed}%
  \BibitemOpen
  \bibfield  {author} {\bibinfo {author} {\bibfnamefont {N.}~\bibnamefont {Shiraishi}}, \bibinfo {author} {\bibfnamefont {K.}~\bibnamefont {Funo}},\ and\ \bibinfo {author} {\bibfnamefont {K.}~\bibnamefont {Saito}},\ }\bibfield  {title} {\bibinfo {title} {Speed limit for classical stochastic processes},\ }\href {https://journals.aps.org/prl/abstract/10.1103/PhysRevLett.121.070601} {\bibfield  {journal} {\bibinfo  {journal} {Phys. Rev. Lett.}\ }\textbf {\bibinfo {volume} {121}},\ \bibinfo {pages} {070601} (\bibinfo {year} {2018})}\BibitemShut {NoStop}%
\bibitem [{\citenamefont {Tuan~Vo}\ \emph {et~al.}(2020)\citenamefont {Tuan~Vo}, \citenamefont {Van~Vu},\ and\ \citenamefont {Hasegawa}}]{tuan2020unified}%
  \BibitemOpen
  \bibfield  {author} {\bibinfo {author} {\bibfnamefont {V.}~\bibnamefont {Tuan~Vo}}, \bibinfo {author} {\bibfnamefont {T.}~\bibnamefont {Van~Vu}},\ and\ \bibinfo {author} {\bibfnamefont {Y.}~\bibnamefont {Hasegawa}},\ }\bibfield  {title} {\bibinfo {title} {Unified approach to classical speed limit and thermodynamic uncertainty relation},\ }\href {https://journals.aps.org/pre/abstract/10.1103/PhysRevE.102.062132} {\bibfield  {journal} {\bibinfo  {journal} {Phys. Rev. E}\ }\textbf {\bibinfo {volume} {102}},\ \bibinfo {pages} {062132} (\bibinfo {year} {2020})}\BibitemShut {NoStop}%
\bibitem [{\citenamefont {Pietzonka}\ \emph {et~al.}(2016)\citenamefont {Pietzonka}, \citenamefont {Barato},\ and\ \citenamefont {Seifert}}]{pietzonka2016universal}%
  \BibitemOpen
  \bibfield  {author} {\bibinfo {author} {\bibfnamefont {P.}~\bibnamefont {Pietzonka}}, \bibinfo {author} {\bibfnamefont {A.~C.}\ \bibnamefont {Barato}},\ and\ \bibinfo {author} {\bibfnamefont {U.}~\bibnamefont {Seifert}},\ }\bibfield  {title} {\bibinfo {title} {Universal bound on the efficiency of molecular motors},\ }\href {https://iopscience.iop.org/article/10.1088/1742-5468/2016/12/124004/meta} {\bibfield  {journal} {\bibinfo  {journal} {J. Stat. Mech.}\ }\textbf {\bibinfo {volume} {2016}},\ \bibinfo {pages} {124004} (\bibinfo {year} {2016})}\BibitemShut {NoStop}%
\bibitem [{\citenamefont {Dechant}(2018)}]{dechant2018multidimensional}%
  \BibitemOpen
  \bibfield  {author} {\bibinfo {author} {\bibfnamefont {A.}~\bibnamefont {Dechant}},\ }\bibfield  {title} {\bibinfo {title} {Multidimensional thermodynamic uncertainty relations},\ }\href {https://iopscience.iop.org/article/10.1088/1751-8121/aaf3ff/meta} {\bibfield  {journal} {\bibinfo  {journal} {J. Phys. A}\ }\textbf {\bibinfo {volume} {52}},\ \bibinfo {pages} {035001} (\bibinfo {year} {2018})}\BibitemShut {NoStop}%
\bibitem [{\citenamefont {Liu}\ \emph {et~al.}(2020)\citenamefont {Liu}, \citenamefont {Gong},\ and\ \citenamefont {Ueda}}]{liu2020thermodynamic}%
  \BibitemOpen
  \bibfield  {author} {\bibinfo {author} {\bibfnamefont {K.}~\bibnamefont {Liu}}, \bibinfo {author} {\bibfnamefont {Z.}~\bibnamefont {Gong}},\ and\ \bibinfo {author} {\bibfnamefont {M.}~\bibnamefont {Ueda}},\ }\bibfield  {title} {\bibinfo {title} {Thermodynamic uncertainty relation for arbitrary initial states},\ }\href {https://journals.aps.org/prl/abstract/10.1103/PhysRevLett.125.140602} {\bibfield  {journal} {\bibinfo  {journal} {Phys. Rev. Lett.}\ }\textbf {\bibinfo {volume} {125}},\ \bibinfo {pages} {140602} (\bibinfo {year} {2020})}\BibitemShut {NoStop}%
\bibitem [{\citenamefont {Horowitz}\ and\ \citenamefont {Gingrich}(2020)}]{horowitz2020thermodynamic}%
  \BibitemOpen
  \bibfield  {author} {\bibinfo {author} {\bibfnamefont {J.~M.}\ \bibnamefont {Horowitz}}\ and\ \bibinfo {author} {\bibfnamefont {T.~R.}\ \bibnamefont {Gingrich}},\ }\bibfield  {title} {\bibinfo {title} {Thermodynamic uncertainty relations constrain non-equilibrium fluctuations},\ }\href {https://www.nature.com/articles/s41567-019-0702-6} {\bibfield  {journal} {\bibinfo  {journal} {Nat. Phys.}\ }\textbf {\bibinfo {volume} {16}},\ \bibinfo {pages} {15} (\bibinfo {year} {2020})}\BibitemShut {NoStop}%
\bibitem [{\citenamefont {Yoshimura}\ and\ \citenamefont {Ito}(2021)}]{yoshimura2021thermodynamic}%
  \BibitemOpen
  \bibfield  {author} {\bibinfo {author} {\bibfnamefont {K.}~\bibnamefont {Yoshimura}}\ and\ \bibinfo {author} {\bibfnamefont {S.}~\bibnamefont {Ito}},\ }\bibfield  {title} {\bibinfo {title} {Thermodynamic uncertainty relation and thermodynamic speed limit in deterministic chemical reaction networks},\ }\href {https://journals.aps.org/prl/abstract/10.1103/PhysRevLett.127.160601} {\bibfield  {journal} {\bibinfo  {journal} {Phys. Rev. Lett.}\ }\textbf {\bibinfo {volume} {127}},\ \bibinfo {pages} {160601} (\bibinfo {year} {2021})}\BibitemShut {NoStop}%
\bibitem [{\citenamefont {Van~Vu}\ and\ \citenamefont {Saito}(2023)}]{van2023thermodynamic}%
  \BibitemOpen
  \bibfield  {author} {\bibinfo {author} {\bibfnamefont {T.}~\bibnamefont {Van~Vu}}\ and\ \bibinfo {author} {\bibfnamefont {K.}~\bibnamefont {Saito}},\ }\bibfield  {title} {\bibinfo {title} {Thermodynamic unification of optimal transport: thermodynamic uncertainty relation, minimum dissipation, and thermodynamic speed limits},\ }\href {https://journals.aps.org/prx/abstract/10.1103/PhysRevX.13.011013} {\bibfield  {journal} {\bibinfo  {journal} {Phys. Rev. X}\ }\textbf {\bibinfo {volume} {13}},\ \bibinfo {pages} {011013} (\bibinfo {year} {2023})}\BibitemShut {NoStop}%
\bibitem [{\citenamefont {Van~Vu}\ and\ \citenamefont {Hasegawa}(2021)}]{van2021geometrical}%
  \BibitemOpen
  \bibfield  {author} {\bibinfo {author} {\bibfnamefont {T.}~\bibnamefont {Van~Vu}}\ and\ \bibinfo {author} {\bibfnamefont {Y.}~\bibnamefont {Hasegawa}},\ }\bibfield  {title} {\bibinfo {title} {Geometrical bounds of the irreversibility in {M}arkovian systems},\ }\href {https://journals.aps.org/prl/abstract/10.1103/PhysRevLett.126.010601} {\bibfield  {journal} {\bibinfo  {journal} {Phys. Rev. Lett.}\ }\textbf {\bibinfo {volume} {126}},\ \bibinfo {pages} {010601} (\bibinfo {year} {2021})}\BibitemShut {NoStop}%
\bibitem [{\citenamefont {Dechant}(2022)}]{dechant2022minimum}%
  \BibitemOpen
  \bibfield  {author} {\bibinfo {author} {\bibfnamefont {A.}~\bibnamefont {Dechant}},\ }\bibfield  {title} {\bibinfo {title} {Minimum entropy production, detailed balance and {W}asserstein distance for continuous-time {M}arkov processes},\ }\href {https://iopscience.iop.org/article/10.1088/1751-8121/ac4ac0} {\bibfield  {journal} {\bibinfo  {journal} {J. Phys. A}\ } (\bibinfo {year} {2022})}\BibitemShut {NoStop}%
\bibitem [{\citenamefont {Otsubo}\ \emph {et~al.}(2020)\citenamefont {Otsubo}, \citenamefont {Ito}, \citenamefont {Dechant},\ and\ \citenamefont {Sagawa}}]{otsubo2020estimating}%
  \BibitemOpen
  \bibfield  {author} {\bibinfo {author} {\bibfnamefont {S.}~\bibnamefont {Otsubo}}, \bibinfo {author} {\bibfnamefont {S.}~\bibnamefont {Ito}}, \bibinfo {author} {\bibfnamefont {A.}~\bibnamefont {Dechant}},\ and\ \bibinfo {author} {\bibfnamefont {T.}~\bibnamefont {Sagawa}},\ }\bibfield  {title} {\bibinfo {title} {Estimating entropy production by machine learning of short-time fluctuating currents},\ }\href {https://journals.aps.org/pre/abstract/10.1103/PhysRevE.101.062106} {\bibfield  {journal} {\bibinfo  {journal} {Phys. Rev. E}\ }\textbf {\bibinfo {volume} {101}},\ \bibinfo {pages} {062106} (\bibinfo {year} {2020})}\BibitemShut {NoStop}%
\bibitem [{\citenamefont {Tajima}\ and\ \citenamefont {Funo}(2021)}]{tajima2021superconducting}%
  \BibitemOpen
  \bibfield  {author} {\bibinfo {author} {\bibfnamefont {H.}~\bibnamefont {Tajima}}\ and\ \bibinfo {author} {\bibfnamefont {K.}~\bibnamefont {Funo}},\ }\bibfield  {title} {\bibinfo {title} {Superconducting-like heat current: Effective cancellation of current-dissipation trade-off by quantum coherence},\ }\href {https://journals.aps.org/prl/abstract/10.1103/PhysRevLett.127.190604} {\bibfield  {journal} {\bibinfo  {journal} {Physical Review Letters}\ }\textbf {\bibinfo {volume} {127}},\ \bibinfo {pages} {190604} (\bibinfo {year} {2021})}\BibitemShut {NoStop}%
\bibitem [{\citenamefont {Falasco}\ and\ \citenamefont {Esposito}(2020)}]{falasco2020dissipation}%
  \BibitemOpen
  \bibfield  {author} {\bibinfo {author} {\bibfnamefont {G.}~\bibnamefont {Falasco}}\ and\ \bibinfo {author} {\bibfnamefont {M.}~\bibnamefont {Esposito}},\ }\bibfield  {title} {\bibinfo {title} {Dissipation-time uncertainty relation},\ }\href {https://journals.aps.org/prl/abstract/10.1103/PhysRevLett.125.120604} {\bibfield  {journal} {\bibinfo  {journal} {Phys. Rev. Lett.}\ }\textbf {\bibinfo {volume} {125}},\ \bibinfo {pages} {120604} (\bibinfo {year} {2020})}\BibitemShut {NoStop}%
\bibitem [{\citenamefont {Mandelstam}\ and\ \citenamefont {Tamm}(1945)}]{mandelstam1945uncertainty}%
  \BibitemOpen
  \bibfield  {author} {\bibinfo {author} {\bibfnamefont {L.}~\bibnamefont {Mandelstam}}\ and\ \bibinfo {author} {\bibfnamefont {I.}~\bibnamefont {Tamm}},\ }\bibfield  {title} {\bibinfo {title} {The uncertainty relation between energy and time in non-relativistic quantum mechanics},\ }\href {https://link.springer.com/chapter/10.1007/978-3-642-74626-0} {\bibfield  {journal} {\bibinfo  {journal} {J. Phys. USSR}\ }\textbf {\bibinfo {volume} {9}},\ \bibinfo {pages} {249} (\bibinfo {year} {1945})}\BibitemShut {NoStop}%
\bibitem [{\citenamefont {Schuster}\ and\ \citenamefont {Schuster}(1989)}]{schuster1989generalization}%
  \BibitemOpen
  \bibfield  {author} {\bibinfo {author} {\bibfnamefont {S.}~\bibnamefont {Schuster}}\ and\ \bibinfo {author} {\bibfnamefont {R.}~\bibnamefont {Schuster}},\ }\bibfield  {title} {\bibinfo {title} {A generalization of {W}egscheider's condition. implications for properties of steady states and for quasi-steady-state approximation},\ }\href {https://link.springer.com/article/10.1007/BF01171883} {\bibfield  {journal} {\bibinfo  {journal} {J. Math. Chem.}\ }\textbf {\bibinfo {volume} {3}},\ \bibinfo {pages} {25} (\bibinfo {year} {1989})}\BibitemShut {NoStop}%
\bibitem [{\citenamefont {Kondepudi}\ and\ \citenamefont {Prigogine}(2015)}]{kondepudi2014modern}%
  \BibitemOpen
  \bibfield  {author} {\bibinfo {author} {\bibfnamefont {D.}~\bibnamefont {Kondepudi}}\ and\ \bibinfo {author} {\bibfnamefont {I.}~\bibnamefont {Prigogine}},\ }\href@noop {} {\emph {\bibinfo {title} {Modern thermodynamics: from heat engines to dissipative structures}}}\ (\bibinfo  {publisher} {John Wiley \& Sons, Chichester, West Sussex},\ \bibinfo {year} {2015})\BibitemShut {NoStop}%
\bibitem [{\citenamefont {Carlson}(1972)}]{carlson1972logarithmic}%
  \BibitemOpen
  \bibfield  {author} {\bibinfo {author} {\bibfnamefont {B.~C.}\ \bibnamefont {Carlson}},\ }\bibfield  {title} {\bibinfo {title} {The logarithmic mean},\ }\href {https://www.tandfonline.com/doi/pdf/10.1080/00029890.1972.11993095} {\bibfield  {journal} {\bibinfo  {journal} {Am. Math. Mon.}\ }\textbf {\bibinfo {volume} {79}},\ \bibinfo {pages} {615} (\bibinfo {year} {1972})}\BibitemShut {NoStop}%
\bibitem [{\citenamefont {de~Oliveira}(2023)}]{de2023quantum}%
  \BibitemOpen
  \bibfield  {author} {\bibinfo {author} {\bibfnamefont {M.~J.}\ \bibnamefont {de~Oliveira}},\ }\bibfield  {title} {\bibinfo {title} {Quantum fokker-planck structure of the lindblad equation},\ }\href {https://link.springer.com/article/10.1007/s13538-023-01330-9} {\bibfield  {journal} {\bibinfo  {journal} {Braz. J. Phys.}\ }\textbf {\bibinfo {volume} {53}},\ \bibinfo {pages} {121} (\bibinfo {year} {2023})}\BibitemShut {NoStop}%
\bibitem [{\citenamefont {{\"O}ttinger}(2010)}]{ottinger2010nonlinear}%
  \BibitemOpen
  \bibfield  {author} {\bibinfo {author} {\bibfnamefont {H.~C.}\ \bibnamefont {{\"O}ttinger}},\ }\bibfield  {title} {\bibinfo {title} {Nonlinear thermodynamic quantum master equation: Properties and examples},\ }\href {https://journals.aps.org/pra/abstract/10.1103/PhysRevA.82.052119} {\bibfield  {journal} {\bibinfo  {journal} {Phys. Rev. A}\ }\textbf {\bibinfo {volume} {82}},\ \bibinfo {pages} {052119} (\bibinfo {year} {2010})}\BibitemShut {NoStop}%
\bibitem [{\citenamefont {Green}(1954)}]{green1954markoff}%
  \BibitemOpen
  \bibfield  {author} {\bibinfo {author} {\bibfnamefont {M.~S.}\ \bibnamefont {Green}},\ }\bibfield  {title} {\bibinfo {title} {Markoff random processes and the statistical mechanics of time-dependent phenomena. ii. irreversible processes in fluids},\ }\href {https://pubs.aip.org/aip/jcp/article-abstract/22/3/398/203406/Markoff-Random-Processes-and-the-Statistical} {\bibfield  {journal} {\bibinfo  {journal} {J. Chem. Phys.}\ }\textbf {\bibinfo {volume} {22}},\ \bibinfo {pages} {398} (\bibinfo {year} {1954})}\BibitemShut {NoStop}%
\bibitem [{\citenamefont {Kubo}\ \emph {et~al.}(2012)\citenamefont {Kubo}, \citenamefont {Toda},\ and\ \citenamefont {Hashitsume}}]{kubo2012statistical}%
  \BibitemOpen
  \bibfield  {author} {\bibinfo {author} {\bibfnamefont {R.}~\bibnamefont {Kubo}}, \bibinfo {author} {\bibfnamefont {M.}~\bibnamefont {Toda}},\ and\ \bibinfo {author} {\bibfnamefont {N.}~\bibnamefont {Hashitsume}},\ }\href@noop {} {\emph {\bibinfo {title} {Statistical physics II: nonequilibrium statistical mechanics}}},\ \bibinfo {series} {SSSOL}, Vol.~\bibinfo {volume} {31}\ (\bibinfo  {publisher} {Springer Berlin, Heidelberg},\ \bibinfo {year} {2012})\BibitemShut {NoStop}%
\bibitem [{\citenamefont {Vroylandt}\ \emph {et~al.}(2018)\citenamefont {Vroylandt}, \citenamefont {Lacoste},\ and\ \citenamefont {Verley}}]{vroylandt2018degree}%
  \BibitemOpen
  \bibfield  {author} {\bibinfo {author} {\bibfnamefont {H.}~\bibnamefont {Vroylandt}}, \bibinfo {author} {\bibfnamefont {D.}~\bibnamefont {Lacoste}},\ and\ \bibinfo {author} {\bibfnamefont {G.}~\bibnamefont {Verley}},\ }\bibfield  {title} {\bibinfo {title} {Degree of coupling and efficiency of energy converters far-from-equilibrium},\ }\href {https://iopscience.iop.org/article/10.1088/1742-5468/aaa8fe/meta?casa_token=bU5FYgDmzLkAAAAA:YD0xrEkqeov3t2eeRjZk_P_gwkL2LXkZpudmv3zV8fECstLqh2iYrwo8omwJa8ANHIOLzYU7nksfUi_OB35aU1H8kg} {\bibfield  {journal} {\bibinfo  {journal} {Journal of Statistical Mechanics: Theory and Experiment}\ }\textbf {\bibinfo {volume} {2018}},\ \bibinfo {pages} {023205} (\bibinfo {year} {2018})}\BibitemShut {NoStop}%
\bibitem [{\citenamefont {Mittnenzweig}\ and\ \citenamefont {Mielke}(2017)}]{mittnenzweig2017entropic}%
  \BibitemOpen
  \bibfield  {author} {\bibinfo {author} {\bibfnamefont {M.}~\bibnamefont {Mittnenzweig}}\ and\ \bibinfo {author} {\bibfnamefont {A.}~\bibnamefont {Mielke}},\ }\bibfield  {title} {\bibinfo {title} {An entropic gradient structure for lindblad equations and couplings of quantum systems to macroscopic models},\ }\href {https://link.springer.com/article/10.1007/s10955-017-1756-4} {\bibfield  {journal} {\bibinfo  {journal} {J. Stat. Phys.}\ }\textbf {\bibinfo {volume} {167}},\ \bibinfo {pages} {205} (\bibinfo {year} {2017})}\BibitemShut {NoStop}%
\bibitem [{Note1()}]{Note1}%
  \BibitemOpen
  \bibinfo {note} {The first equality is proved by the mathematical induction; $L_kH = (H+\omega _k)L_k$ holds and assuming $L_kH^n=(H+\omega _k)^nL_k$ leads to $L_kH^{n+1}=(H+\omega _k)^nL_kH =(H+\omega _k)^{n+1}L_k$. The second one is derived by applying this equality to each term of the Taylor expansion of $e^{-\beta H}$.}\BibitemShut {Stop}%
\bibitem [{\citenamefont {Polettini}\ and\ \citenamefont {Esposito}(2014)}]{polettini2014irreversible}%
  \BibitemOpen
  \bibfield  {author} {\bibinfo {author} {\bibfnamefont {M.}~\bibnamefont {Polettini}}\ and\ \bibinfo {author} {\bibfnamefont {M.}~\bibnamefont {Esposito}},\ }\bibfield  {title} {\bibinfo {title} {Irreversible thermodynamics of open chemical networks. {I.} {Emergent} cycles and broken conservation laws},\ }\href {https://aip.scitation.org/doi/full/10.1063/1.4886396} {\bibfield  {journal} {\bibinfo  {journal} {J. Chem. Phys.}\ }\textbf {\bibinfo {volume} {141}},\ \bibinfo {pages} {07B610\_1} (\bibinfo {year} {2014})}\BibitemShut {NoStop}%
\bibitem [{\citenamefont {Komatsu}\ \emph {et~al.}(2008)\citenamefont {Komatsu}, \citenamefont {Nakagawa}, \citenamefont {Sasa},\ and\ \citenamefont {Tasaki}}]{komatsu2008steady}%
  \BibitemOpen
  \bibfield  {author} {\bibinfo {author} {\bibfnamefont {T.~S.}\ \bibnamefont {Komatsu}}, \bibinfo {author} {\bibfnamefont {N.}~\bibnamefont {Nakagawa}}, \bibinfo {author} {\bibfnamefont {S.-i.}\ \bibnamefont {Sasa}},\ and\ \bibinfo {author} {\bibfnamefont {H.}~\bibnamefont {Tasaki}},\ }\bibfield  {title} {\bibinfo {title} {Steady-state thermodynamics for heat conduction: microscopic derivation},\ }\href {https://journals.aps.org/prl/abstract/10.1103/PhysRevLett.100.230602} {\bibfield  {journal} {\bibinfo  {journal} {Phys. Rev. Lett.}\ }\textbf {\bibinfo {volume} {100}},\ \bibinfo {pages} {230602} (\bibinfo {year} {2008})}\BibitemShut {NoStop}%
\bibitem [{Note2()}]{Note2}%
  \BibitemOpen
  \bibinfo {note} {Though we defined the decomposition for autonomous systems, where $H$ and $\gamma _k$ are constant, the decomposition can be applied to non-autonomous cases with a slight modification.}\BibitemShut {Stop}%
\bibitem [{\citenamefont {Puntanen}\ \emph {et~al.}(2011)\citenamefont {Puntanen}, \citenamefont {Styan},\ and\ \citenamefont {Isotalo}}]{puntanen2011matrix}%
  \BibitemOpen
  \bibfield  {author} {\bibinfo {author} {\bibfnamefont {S.}~\bibnamefont {Puntanen}}, \bibinfo {author} {\bibfnamefont {G.~P.}\ \bibnamefont {Styan}},\ and\ \bibinfo {author} {\bibfnamefont {J.}~\bibnamefont {Isotalo}},\ }\href@noop {} {\emph {\bibinfo {title} {Matrix tricks for linear statistical models: our personal top twenty}}}\ (\bibinfo  {publisher} {Springer Berlin, Heidelberg},\ \bibinfo {year} {2011})\BibitemShut {NoStop}%
\bibitem [{\citenamefont {Villani}(2009)}]{villani2009optimal}%
  \BibitemOpen
  \bibfield  {author} {\bibinfo {author} {\bibfnamefont {C.}~\bibnamefont {Villani}},\ }\href@noop {} {\emph {\bibinfo {title} {Optimal transport: old and new}}},\ \bibinfo {series} {GL}, Vol.\ \bibinfo {volume} {338}\ (\bibinfo  {publisher} {Springer Berlin, Heidelberg},\ \bibinfo {year} {2009})\BibitemShut {NoStop}%
\bibitem [{\citenamefont {Maas}(2011)}]{maas2011gradient}%
  \BibitemOpen
  \bibfield  {author} {\bibinfo {author} {\bibfnamefont {J.}~\bibnamefont {Maas}},\ }\bibfield  {title} {\bibinfo {title} {Gradient flows of the entropy for finite {M}arkov chains},\ }\href {https://www.sciencedirect.com/science/article/pii/S0022123611002278} {\bibfield  {journal} {\bibinfo  {journal} {J. Funct. Anal.}\ }\textbf {\bibinfo {volume} {261}},\ \bibinfo {pages} {2250} (\bibinfo {year} {2011})}\BibitemShut {NoStop}%
\bibitem [{\citenamefont {Liero}\ and\ \citenamefont {Mielke}(2013)}]{liero2013gradient}%
  \BibitemOpen
  \bibfield  {author} {\bibinfo {author} {\bibfnamefont {M.}~\bibnamefont {Liero}}\ and\ \bibinfo {author} {\bibfnamefont {A.}~\bibnamefont {Mielke}},\ }\bibfield  {title} {\bibinfo {title} {Gradient structures and geodesic convexity for reaction--diffusion systems},\ }\href {https://royalsocietypublishing.org/doi/full/10.1098/rsta.2012.0346} {\bibfield  {journal} {\bibinfo  {journal} {Philos. Trans. R. Soc. A}\ }\textbf {\bibinfo {volume} {371}},\ \bibinfo {pages} {20120346} (\bibinfo {year} {2013})}\BibitemShut {NoStop}%
\bibitem [{\citenamefont {Benamou}\ and\ \citenamefont {Brenier}(2000)}]{benamou2000computational}%
  \BibitemOpen
  \bibfield  {author} {\bibinfo {author} {\bibfnamefont {J.-D.}\ \bibnamefont {Benamou}}\ and\ \bibinfo {author} {\bibfnamefont {Y.}~\bibnamefont {Brenier}},\ }\bibfield  {title} {\bibinfo {title} {A computational fluid mechanics solution to the monge-kantorovich mass transfer problem},\ }\href {https://link.springer.com/article/10.1007/s002110050002} {\bibfield  {journal} {\bibinfo  {journal} {Numer. Math.}\ }\textbf {\bibinfo {volume} {84}},\ \bibinfo {pages} {375} (\bibinfo {year} {2000})}\BibitemShut {NoStop}%
\bibitem [{\citenamefont {Parrondo}\ \emph {et~al.}(2015)\citenamefont {Parrondo}, \citenamefont {Horowitz},\ and\ \citenamefont {Sagawa}}]{parrondo2015thermodynamics}%
  \BibitemOpen
  \bibfield  {author} {\bibinfo {author} {\bibfnamefont {J.~M.}\ \bibnamefont {Parrondo}}, \bibinfo {author} {\bibfnamefont {J.~M.}\ \bibnamefont {Horowitz}},\ and\ \bibinfo {author} {\bibfnamefont {T.}~\bibnamefont {Sagawa}},\ }\bibfield  {title} {\bibinfo {title} {Thermodynamics of information},\ }\href {https://www.nature.com/articles/nphys3230} {\bibfield  {journal} {\bibinfo  {journal} {Nat. Phys.}\ }\textbf {\bibinfo {volume} {11}},\ \bibinfo {pages} {131} (\bibinfo {year} {2015})}\BibitemShut {NoStop}%
\bibitem [{\citenamefont {Maes}\ and\ \citenamefont {Neto{\v{c}}n{\`y}}(2008)}]{maes2008canonical}%
  \BibitemOpen
  \bibfield  {author} {\bibinfo {author} {\bibfnamefont {C.}~\bibnamefont {Maes}}\ and\ \bibinfo {author} {\bibfnamefont {K.}~\bibnamefont {Neto{\v{c}}n{\`y}}},\ }\bibfield  {title} {\bibinfo {title} {Canonical structure of dynamical fluctuations in mesoscopic nonequilibrium steady states},\ }\href {https://iopscience.iop.org/article/10.1209/0295-5075/82/30003/meta} {\bibfield  {journal} {\bibinfo  {journal} {Europhys. Lett.}\ }\textbf {\bibinfo {volume} {82}},\ \bibinfo {pages} {30003} (\bibinfo {year} {2008})}\BibitemShut {NoStop}%
\bibitem [{\citenamefont {Pires}\ \emph {et~al.}(2016)\citenamefont {Pires}, \citenamefont {Cianciaruso}, \citenamefont {C{\'e}leri}, \citenamefont {Adesso},\ and\ \citenamefont {Soares-Pinto}}]{pires2016generalized}%
  \BibitemOpen
  \bibfield  {author} {\bibinfo {author} {\bibfnamefont {D.~P.}\ \bibnamefont {Pires}}, \bibinfo {author} {\bibfnamefont {M.}~\bibnamefont {Cianciaruso}}, \bibinfo {author} {\bibfnamefont {L.~C.}\ \bibnamefont {C{\'e}leri}}, \bibinfo {author} {\bibfnamefont {G.}~\bibnamefont {Adesso}},\ and\ \bibinfo {author} {\bibfnamefont {D.~O.}\ \bibnamefont {Soares-Pinto}},\ }\bibfield  {title} {\bibinfo {title} {Generalized geometric quantum speed limits},\ }\href {https://journals.aps.org/prx/abstract/10.1103/PhysRevX.6.021031} {\bibfield  {journal} {\bibinfo  {journal} {Phys. Rev. X}\ }\textbf {\bibinfo {volume} {6}},\ \bibinfo {pages} {021031} (\bibinfo {year} {2016})}\BibitemShut {NoStop}%
\bibitem [{\citenamefont {Johansson}\ \emph {et~al.}(2012)\citenamefont {Johansson}, \citenamefont {Nation},\ and\ \citenamefont {Nori}}]{johansson2012qutip}%
  \BibitemOpen
  \bibfield  {author} {\bibinfo {author} {\bibfnamefont {J.~R.}\ \bibnamefont {Johansson}}, \bibinfo {author} {\bibfnamefont {P.~D.}\ \bibnamefont {Nation}},\ and\ \bibinfo {author} {\bibfnamefont {F.}~\bibnamefont {Nori}},\ }\bibfield  {title} {\bibinfo {title} {Qutip: An open-source python framework for the dynamics of open quantum systems},\ }\href {https://www.sciencedirect.com/science/article/pii/S0010465512000835?casa_token=7PgMV9h4Qm4AAAAA:OY4kghYuzAmsPP2NqbGz37iMPwDnJspHX7VNgIy2UZcMBH2pUb0W9Puuf_jVTIP-NplkfXKvCQ} {\bibfield  {journal} {\bibinfo  {journal} {Comput. Phys. Commun.}\ }\textbf {\bibinfo {volume} {183}},\ \bibinfo {pages} {1760} (\bibinfo {year} {2012})}\BibitemShut {NoStop}%
\bibitem [{\citenamefont {Rigol}\ \emph {et~al.}(2007)\citenamefont {Rigol}, \citenamefont {Dunjko}, \citenamefont {Yurovsky},\ and\ \citenamefont {Olshanii}}]{rigol2007relaxation}%
  \BibitemOpen
  \bibfield  {author} {\bibinfo {author} {\bibfnamefont {M.}~\bibnamefont {Rigol}}, \bibinfo {author} {\bibfnamefont {V.}~\bibnamefont {Dunjko}}, \bibinfo {author} {\bibfnamefont {V.}~\bibnamefont {Yurovsky}},\ and\ \bibinfo {author} {\bibfnamefont {M.}~\bibnamefont {Olshanii}},\ }\bibfield  {title} {\bibinfo {title} {Relaxation in a completely integrable many-body quantum system: an ab initio study of the dynamics of the highly excited states of 1d lattice hard-core bosons},\ }\href {https://journals.aps.org/prl/abstract/10.1103/PhysRevLett.98.050405} {\bibfield  {journal} {\bibinfo  {journal} {Phys. Rev. Lett.}\ }\textbf {\bibinfo {volume} {98}},\ \bibinfo {pages} {050405} (\bibinfo {year} {2007})}\BibitemShut {NoStop}%
\bibitem [{\citenamefont {Isar}\ \emph {et~al.}(1994)\citenamefont {Isar}, \citenamefont {Sandulescu}, \citenamefont {Scutaru}, \citenamefont {Stefanescu},\ and\ \citenamefont {Scheid}}]{isar1994open}%
  \BibitemOpen
  \bibfield  {author} {\bibinfo {author} {\bibfnamefont {A.}~\bibnamefont {Isar}}, \bibinfo {author} {\bibfnamefont {A.}~\bibnamefont {Sandulescu}}, \bibinfo {author} {\bibfnamefont {H.}~\bibnamefont {Scutaru}}, \bibinfo {author} {\bibfnamefont {E.}~\bibnamefont {Stefanescu}},\ and\ \bibinfo {author} {\bibfnamefont {W.}~\bibnamefont {Scheid}},\ }\bibfield  {title} {\bibinfo {title} {Open quantum systems},\ }\href {https://www.worldscientific.com/doi/abs/10.1142/S0218301394000164} {\bibfield  {journal} {\bibinfo  {journal} {Int. J. Mod. Phys. E}\ }\textbf {\bibinfo {volume} {3}},\ \bibinfo {pages} {635} (\bibinfo {year} {1994})}\BibitemShut {NoStop}%
\bibitem [{\citenamefont {Weiderpass}\ and\ \citenamefont {Caldeira}(2020)}]{weiderpass2020neumann}%
  \BibitemOpen
  \bibfield  {author} {\bibinfo {author} {\bibfnamefont {G.~A.}\ \bibnamefont {Weiderpass}}\ and\ \bibinfo {author} {\bibfnamefont {A.~O.}\ \bibnamefont {Caldeira}},\ }\bibfield  {title} {\bibinfo {title} {von neumann entropy and entropy production of a damped harmonic oscillator},\ }\href {https://journals.aps.org/pre/abstract/10.1103/PhysRevE.102.032102} {\bibfield  {journal} {\bibinfo  {journal} {Phys. Rev. E}\ }\textbf {\bibinfo {volume} {102}},\ \bibinfo {pages} {032102} (\bibinfo {year} {2020})}\BibitemShut {NoStop}%
\bibitem [{\citenamefont {Manikandan}\ \emph {et~al.}(2020)\citenamefont {Manikandan}, \citenamefont {Gupta},\ and\ \citenamefont {Krishnamurthy}}]{manikandan2020inferring}%
  \BibitemOpen
  \bibfield  {author} {\bibinfo {author} {\bibfnamefont {S.~K.}\ \bibnamefont {Manikandan}}, \bibinfo {author} {\bibfnamefont {D.}~\bibnamefont {Gupta}},\ and\ \bibinfo {author} {\bibfnamefont {S.}~\bibnamefont {Krishnamurthy}},\ }\bibfield  {title} {\bibinfo {title} {Inferring entropy production from short experiments},\ }\href {https://journals.aps.org/prl/abstract/10.1103/PhysRevLett.124.120603} {\bibfield  {journal} {\bibinfo  {journal} {Phys. Rev. Lett.}\ }\textbf {\bibinfo {volume} {124}},\ \bibinfo {pages} {120603} (\bibinfo {year} {2020})}\BibitemShut {NoStop}%
\bibitem [{Note3()}]{Note3}%
  \BibitemOpen
  \bibinfo {note} {Seemingly, $\eta '$ is ill-defined when $\gamma _k\rho _{ee}=\gamma _{-k}\rho _{gg}$; however, if $\gamma _k\rho _{ee}=\gamma _{-k}\rho _{gg}(1+r)$ with small $r$, it can be shown $1/f_k-2\rho _{ee}\rho _{gg}a_k/j_k= -1/2+r/4+o(r)$.}\BibitemShut {Stop}%
\end{thebibliography}
%

\end{document}